\documentclass[11pt,a4paper]{article}

\usepackage[title]{appendix}

\usepackage{jheppub}
\usepackage{graphicx} 
\usepackage{color}
\usepackage{amsmath}
\usepackage{listings}
\usepackage{exercise}
\usepackage{amssymb}
\usepackage{cases}
\usepackage{graphicx}
\usepackage{hyperref}
\hypersetup{
	colorlinks=true,
	linkcolor=blue,
	filecolor=blue,      
	urlcolor=blue,
	citecolor=blue
}

\usepackage{fancyhdr}
\usepackage{mdframed}
\mdfdefinestyle{exampledefault}{
	outerlinewidth=5pt,innerlinewidth=0pt,
	outerlinecolor=red,roundcorner=5pt
}
\usepackage{empheq}
\usepackage{xcolor}
\usepackage[compat=1.0.0]{tikz-feynman}
\usepackage{tcolorbox}
\usepackage{cancel}
\usepackage{subcaption}

\usepackage[compat=1.0.0]{tikz-feynman}
\usepackage{tcolorbox}
\usepackage{soul}

\newcommand{\Tr}{\text{Tr}}
\newcommand{\ket}[1]{\vert #1\rangle}
\newcommand{\bra}[1]{\langle #1\vert}
\newcommand{\MFP}{\lambda_{\rm MFP}}

\newcommand{\checked}[1]{\textcolor{red}{\checkmark}}
\definecolor{darkgreen}{RGB}{50,150,0}

\newcommand{\secref}[1]{Sec.(\ref{#1})}

\title{Chern-Simons Induced Thermal Friction on Axion Domain Walls}
\author[a,b]{Saquib Hassan,}
\author[a]{Gaurang Ramakant Kane,}
\author[a]{John March-Russell,}
\author[a]{and Georges Obied}
\affiliation[a]{Rudolf Peierls Centre for Theoretical Physics, 
University of Oxford\\Parks Road,
Oxford, OX1 3PU, United Kingdom}
\affiliation[b]{Christ Church College, 
University of Oxford\\St Aldate's,
Oxford, OX1 1DP, United Kingdom}
\emailAdd{saquib.hassan@chch.ox.ac.uk}
\emailAdd{gaurang.kane@physics.ox.ac.uk}
\emailAdd{john.march-russell@physics.ox.ac.uk}
\emailAdd{georges.obied@physics.ox.ac.uk}

\abstract{We study the dynamics and interactions of the solitonic domain walls that occur in realistic axion electrodynamics models including the Chern-Simons interaction, $a\epsilon_{\mu\nu\lambda\sigma}F^{\mu\nu} F^{\lambda\sigma}$, between an axion $a(x)$ of mass $m_a$, and a massless U(1) gauge field, e.g. EM, interacting with strength $\alpha=e^2/4\pi$ with charged matter, e.g. electron-positron pairs.  In particular, in the presence of a U(1) gauge-and-matter relativistic thermal plasma we study the friction experienced by the walls due to the Chern-Simons term. Utilizing the linear response method we include the collective effects of the plasma, as opposed to purely particle scattering across the wall (as is done in previous treatments) which is valid only in the thin wall regime that is rarely applicable in realistic cases.  We show that the friction depends on the Lorentz-$\gamma$-factor-dependent inverse thickness of the wall in the plasma frame, $\ell^{-1} \sim \gamma m_a$, compared to the three different plasma scales, the temperature $T$, the Debye mass $m_D\sim\sqrt{\alpha} T$, and the damping rate $\Gamma \sim \alpha^2 T$, and elucidate the underlying physical intuition for this behavior. (For friction in the thin-wall-limit we correct previous expressions in the literature.)  We further consider the effects of long-range coherent magnetic fields that are possibly present in the early universe and compare their effect with that of thermal magnetic fields. We comment on the changes to our results that likely apply in the thermal deconfined phase of a non-Abelian gauge theory. Finally, we briefly discuss the possible early universe consequences of our results for domain wall motion and network decay, stochastic gravitational wave production from domain wall networks, and possible primordial black hole production from domain wall collapse, though a more complete discussion of these topics is reserved for a companion paper.
}

\begin{document}

\maketitle

\section{Introduction}
\label{introduction}
The Peccei-Quinn solution to the strong CP problem \cite{Peccei:1977hh, Peccei:2006as} implies the existence of a light pseudo-Nambu-Goldstone boson,
the QCD axion \cite{Weinberg:1977ma,Wilczek:1977pj}, $a(x)$, which has gone on to become one of the most studied and searched for new physics states. 
Although the axion concept originated out of a dynamical solution to the QCD strong CP problem, it is now the case that generalized axions, ``axion-like particles" (ALPs), are well motivated new physics candidates with potential consequences for some of the most interesting problems in physics, in particular as an attractive dark matter candidate~\cite{Preskill:1982cy,Abbott:1982af,Dine:1982ah,Sikivie:2006ni,DiLuzio:2020wdo,Adams:2022pbo}.  Moreover, a wide variety of axions appear in motivated UV completions of the Standard Model (SM), such as string theory constructions~\cite{Svrcek:2006yi, Arvanitaki:2009fg}.
 
Two important properties of all axions are their compact field range, $a(x)\sim a(x) +2\pi f_a$ (here $f_a$ is the mass-dimension one axion decay constant), and a non-zero periodic potential, $V(a)$, necessarily generated by non-perturbative effects (QCD non-perturbative effects in the case of the QCD axion) which can have a sub-periodicity $V(a+2\pi f_a/N)=V(a)$, depending on an integer-valued anomaly coefficient, $N\geq 1$. These properties imply that axion theories possess global
cosmic string solutions as well as domain walls that can be either bounded by the cosmic strings or form closed surfaces with no boundary. In many cases, such as, for the
QCD axion in much of its parameter space, the early universe dynamics and eventual decay of these cosmic string and domain wall networks play a vital role in setting the final axion
dark matter density \cite{Kawasaki:2014sqa,Gorghetto:2018myk,Gorghetto:2020qws,Buschmann:2021sdq}, as well as possibly generating other striking and potentially observable phenomena, such as a gravitational wave background \cite{Hiramatsu:2013qaa,Ramberg:2019dgi,ZambujalFerreira:2021cte,Kitajima:2023cek,Li:2024psa}, or a population
of primordial black holes \cite{Vachaspati:2017hjw, Ferrer:2018uiu, Chen:2021wcf,PhysRevD.109.123507, Gouttenoire:2023ftk, Dunsky:2024zdo, Ferreira:2024eru, Escriva:2022duf, Khlopov:2008qy}. Thus, a careful study of the
evolution of these defects is an important ingredient in
assessing the phenomenological and observational consequences of axion models.  

The details of the underlying axion model affect the properties of cosmic strings and domain walls: For example, $N$ determines the number of domain walls that end on a cosmic string, leading to a possibly everlasting domain wall newtork for $N>1$ (unless other, $\mathbb{Z}_N$ breaking effects are included), while the domain wall tension $\sigma \simeq 8 f_a^2 m_a^2$ is set by $f_a$ along with the axion mass $m_a^2 = d^2 V(a)/da^2 |_{\rm min}$. The resulting cosmic string--domain wall network interacts
with and produces a population of free axion particles during its evolution and decay (as well as some amount of heavier states, such as the Peccei-Quinn radial mode in field theoretic UV completions of the low-energy axion theory), and such effects are taken into account by modern numerical simulations \cite{Kawasaki:2014sqa,Gorghetto:2018myk,Gorghetto:2020qws,Buschmann:2021sdq}, though only Hubble ``friction" is typically taken into account.

Most interesting for our purposes is the well-motivated possibility of further interactions of the domain walls with SM particles.\footnote{Our main interest in this work is axion domain walls and their physics, rather than the interactions of axion cosmic strings, though there are of course relations between these two topics. We emphasize that though our primary interest is where the U(1)-gauge-fermion plasma is the SM plasma, it is possible to consider an ALP having Chern-Simons interactions with a \emph{hidden} sector U(1)$_{h}$-gauge (and possibly charged matter) system at finite temperature, and our results apply to this situation too with suitable modifications. Moreover, in Section~\ref{Discussions} we outline the changes to the friction calculation that would apply to the case of a thermal, deconfined phase, non-Abelian plasma interacting with an axion domain wall via the appropriate ``$a\Tr[G\wedge G]"$ Chern-Simons term. Note that the mass of the axion can change substantially in this case as non-perturbative non-Abelian processes can contribute to the axion mass significantly, and differently, in the confined, low-$T$, and deconfined, high-$T$ phases. Relatedly, it is amusing to consider the possibility of an axion domain/bubble wall separating a deconfined region from a confined one.} In fact, as we will explain in detail in the following sections, axions and thus axion domain walls
are expected to possess a particular class of Chern-Simons(-Pontryagin) quasi-topological interactions with gauge fields.
In the case that the relevant gauge field is a U(1), such as electromagnetism, the interaction term in the Lagrangian density is
\begin{equation}
    \mathcal{L}_{int}=\dfrac{\alpha \kappa_{a\gamma\gamma}}{8\pi f_{a}}a F\wedge F ~~~,
    \label{eqn:chern_simons_interaction_lagrangian}
\end{equation}
where here we have used form notation so $F$ is the 
2-form field strength tensor of the U(1) gauge field, and $\alpha$ is the fine-structure constant of the U(1).  The dimensionless coupling $\kappa_{a\gamma\gamma}$ sets the strength of this interaction and is equal to the PQ-U(1)$^2$
anomaly coefficient $K\in \mathbb{Z}$ (in the normalization we use here) up to corrections due to mixing with other neutral pseudoscalar states at low energy, such as $\pi^0,\eta$ and $\eta'$ in the QCD axion case.

The existence of such a Chern-Simons coupling is motivated by its consistency with the symmetries of the theory, its appearance in the
majority of UV completions of axion theories, and, especially for the QCD axion, its natural appearance because of the mixing
with the neutral pseudoscalar mesons which have such couplings with U(1)$_{em}$. Additionally, for the QCD axion where the QCD version of the term Eqn.\eqref{eqn:chern_simons_interaction_lagrangian} is necessarily present with integer $K_{QCD}\neq 0$, any grand unified structure at high energy implies that it is accompanied by its U(1) counterpart \cite{Agrawal:2022lsp}.

The Chern-Simons interaction of an axion, whether the QCD axion or a generalized ALP, with U(1)$_{em}$ is also by far the most utilized coupling in searches for these new states over a broad range of parameter space, with many experiments across the globe trying to detect axions using this interaction \cite{Carosi:2013rla,ParticleDataGroup:2024cfk}.  This system of a massive neutral axion coupled to a massless (massive in some generalizations) U(1) gauge theory through the Chern-Simons term is known as \emph{axion electrodynamics} and possesses a surprisingly rich phenomenology, as well as a theoretical structure with many hidden
generalized higher-form and non-invertible symmetries (see, e.g., \cite{Brennan:2020ehu, Hidaka:2020izy, Hidaka:2020iaz, Yokokura:2022alv, Cordova:2023her}), as well as surprising dynamics \cite{Hassan:2024nbl}.

At the non-perturbative level, the QCD axion and generalised ALPs always acquire a potential that explicitly violates the leading order continuous non-linearly-realized U(1)$_{PQ}$ shift symmetry of the axion, $a(x) \sim a(x)+ c$, $c\in \mathbb{R}$, while respecting the compactness of the field range  $a(x)\sim a(x) + 2\pi k  f_a$ for $k\in \mathbb{Z}$ (equivalently this can be thought of as an exact gauged discrete shift symmetry).\footnote{We expect that, inevitably, there are non-perturbative gravitational effects, which may be due to wormholes, or virtual black holes, or mixed gravitational-U(1)$_{PQ}$ anomalies, or other dynamics of the gravitational UV completion, that generate a potential, possibly the dominant one, for the axion~\cite{Giddings:1987cg,Lee:1988ge,Coleman:1989zu,Kamionkowski:1992mf,Holman:1992us,Kallosh:1995hi,Hebecker:2019vyf}.  This is enshrined and extended in the ``no global symmetries" conjecture which has significant theoretical support \cite{Banks:1988yz,Abbott:1989jw,Arkani-Hamed:2006emk,Banks:2010zn,Harlow:2018tng,Fichet:2019ugl,Daus:2020vtf}.} 
 In particular, in the simple ``single-instanton" limit, that applies to non-perturbative effects in weakly coupled theories, the potential takes the simple form
\begin{equation}
    V(a)\sim \Lambda^{4}\left(1-\cos\left(N\dfrac{a}{f_{a}}\right)\right),~~~~\Lambda^{2}= \dfrac{m_{a}f_{a}}{N}~~.
    \label{eqn:simple_potential}
\end{equation}
 Here the mass of the axion, $m_{a}$, around any one of the minima and the effective period $f_{a}/N$ are related to the dynamical scale, $\Lambda$, generated by the relevant non-perturbative process explicitly breaking the continuous $a(x)\sim a(x)+ c$ shift symmetry. One also explicitly sees that $N$ tells us the number of distinct degenerate vacua between $[a,a+2\pi f_{a})$. This degeneracy in vacua for $N>1$ leads to the existence of $N$ distinct topologically stable domain walls which interpolate between adjacent vacua $a_k = 2\pi f_a k/N$ and $a_{k+1} = 2\pi f_a (k+1)/N$, for $k=0,1,2,...,N-1$. 
 
 In the early universe context, these domain walls form when the axion mass becomes significant compared to the decreasing Hubble rate, and they can either be topologically closed, e.g. usually of $S^2$ topology, or part of a complicated wall network with strings as boundaries. In the case $N=1$ a domain wall solution still exists, as we recall in detail in Section~\ref{Section Axion Domain Walls}, but since it can be bounded by a single axion string it is no longer absolutely stable, just metastable. (Despite this lack of absolute stability in the $N=1$ case, the evolution and eventual decay of the axion domain walls can still dominate the production of ALPs and gravitational waves, as well as the other phenomena that are affected by domain wall friction.)  The $N=1$ walls annihilate\footnote{As already mentioned, in the $N>1$ case with an exact $\mathbb{Z}_N$ global symmetry the domain walls are topologically stable and, in addition, a basic axion cosmic string which winds once through the compact field range, $a\in [0,2\pi f_a)$, bounds $N$ walls. These two features lead to the existence of a scaling solution for the string-wall network in which the domain walls never fully disappear (unlike the $N=1$ case). This in turn leads to a phenomenological disaster for much of the $\{ m_a,f_a\}$ parameter space: The domain wall tension of the remaining scaling-solution walls, $\sigma\simeq 8m_a f_a^2$, can be high enough to either cause unacceptable gravitational distortions in the cosmic microwave background radiation, or even to come to dominate the energy budget of the universe during an epoch before dark energy domination, thus leading to an unacceptable equation of state for the BBN- and post-BBN universe. The solution to this is that the $\mathbb{Z}_N$ global symmetry must be explicitly broken by yet further small non-perturbative effects (which could be gravitational in origin; cf. the "no global symmetries" conjecture), which in turn leads to a pressure difference across the domain walls (see, e.g. \cite{Beyer:2022ywc} for a recent study). This pressure difference can lead to relativistic or even ultra-relativistic wall velocities, and certainly, for large enough pressure difference, the eventual destruction of the wall-string network consistent with cosmological bounds. Additional frictional forces of the kind we have calculated can play a role in the dynamics of this destructive process, and thus the gravitational wave background production and other signatures of wall-string-network demise.} and in this process, they move and oscillate emitting gravitational waves \cite{Huang:1985tt,Vilenkin:2000jqa}. Studies of the evolution of these domain walls (without the additional friction terms that are the subject of this work) show that they can in principle become relativistic, or even ultra-relativistic, and dominate the scaling factor \cite{Vilenkin:2000jqa,Dunsky:2024zdo}. 
 In the sections to follow, we assume $N=1$ unless otherwise stated, but generalizing our results on the friction experienced by the walls (\emph{not} necessarily other features of the wall evolution!) to cases with $N>1$ can be easily achieved by adjusting a few obvious factors. 

As mentioned above, one of the more important potential windows into axion physics in the early universe is \emph{gravitational wave signals}. Indeed, the domain wall network dynamics exhibits the emission of such gravitational wave signatures, and with upcoming detectors enabling increasing sensitivity and frequency coverage in gravitational wave detection, this tool will become an ever more important technique for studying early universe axion physics. 

With this in mind, we now come to the topic studied in this work: In the early universe, the walls are moving in a hot, usually relativistic plasma consisting of particles that interact, directly or indirectly, with the axions via the Chern-Simons term. Hence, the wall experiences \emph{friction} when in motion. This friction can dominate over the Hubble friction caused by the expansion of the universe for certain parameter values and change the overall dynamics of the wall, making it an important feature to study even from an observational perspective (see, for example, appendix D of \cite{ZambujalFerreira:2021cte}). Taking motivation from past literature \cite{McLerran:1990de, Khlebnikov:1992bx, Bodeker:2022ihg}, we calculate the friction experienced by the walls due to the hitherto mainly neglected but key Chern-Simons interaction~\eqref{eqn:chern_simons_interaction_lagrangian} using \emph{linear response theory}. We also summarize previous work on axion domain wall friction and its relation to our results. 

The remainder of this paper is organized as follows. Since the details of our calculations are rather involved and likely mainly of interest to specialists, we first provide a substantial introduction summarizing our setup and main results (important variations and applications, as well as possible fruitful follow-up directions, are discussed in Section~\ref{Discussions}). Specifically, in Section~\ref{The Physical setup} we define the physical system we study and establish notation, while the associated domain wall solutions are described in Section~\ref{Section Axion Domain Walls}.  In Section~\ref{sec:thin_wall_overview} we provide a high-level overview of the calculation of friction on an axion domain wall in the relatively straightforward (but rarely appropriate) \emph{thin-wall-limit}. This has been the subject of previous study in the literature, but we disagree with these earlier results as written, so, before passing on to the most physically relevant thick-wall case, we present what we believe to be the correct thin-wall friction formulae in Eqn.\eqref{eqn:scatteringFriction}.

Section~\ref{subsec:thick_walls_qualitative_picture} gives an overview of the physical intuition behind the somewhat involved calculation of domain wall friction in the physically most interesting \emph{thick-wall limit}.  For convenience of the reader, we provide quick summaries of the main results of our paper in Eqn.\eqref{eqn:results_summary} and in Figure~\ref{thermal friction_ranges} and Table~\ref{tab:my_label}. Together with the discussion in Section \ref{Discussions} of the physical consequences of our results and variations, Section~\ref{subsec:thick_walls_qualitative_picture} provides a compact summary of the present work.

In Section~\ref{The Linear Response} we commence the detailed study of friction on axion domain walls by outlining the methods of linear response theory and their application to the problems at hand, including in Section~\ref{Average Force on the Wall} a discussion of how to evaluate the average force on the wall in a situation dominated by fluctuations, often at a parametrically small scale compared to the wall thickness.
Section~\ref{Calculating the Thermal Averages} contains the calculation of the
thermal averages that determine the domain wall friction in the thick wall limit, first in the simpler case of a long-range coherent background magnetic field, and secondly for a relativistic
thermal plasma with only the standard thermally-produced and fluctuating magnetic and electric fields. In Section~\ref{Comparison_with_Hubble} we turn to a comparison, in various cosmological scenarios, of the magnitude of the Chern-Simons produced friction with the standard cosmological Hubble friction that effectively damps wall motion (and is included in the leading numerical simulations) -- see also the discussion at the end of Section~\ref{sec:thin_wall_overview}.  

Finally in Section~\ref{Discussions} we both summarise our work and the main results, and briefly discuss a variety of interesting variations of our basic calculation.  This includes the case of a high-temperature deconfined non-Abelian plasma; the physics of wall collapse when there is a long-range coherent magnetic field present (as sometimes occurs in variant early universe scenarios) and the associated probability of primordial black hole production; the case where the axion mass and/or decay constant changes on either side of the domain wall (possibly due to explicit $\mathbb{Z}_N$ breaking effects) where previous work \cite{GarciaGarcia:2022yqb,GarciaGarcia:2024dfx} has shown that surprising, and large, frictional forces can occur at high Lorentz-$\gamma$ factors; and finally the issue of a numerically accurate friction calculation for fully realistic QCD axion domain walls with their extra structure due to the mixing with the $\pi^0,\eta,\eta'$ mesons of QCD.  The paper also has five substantial appendices covering various details, or pedagogical asides not suitable for the main text.

 \subsection{The physical setup}
 \label{The Physical setup}
Our system has three components: a U(1) gauge field $A_{\mu}$, fermions $\psi$ charged under the U(1), and the axion field $a(x)$ associated to the domain wall solution. The Lagrangian we consider is
\begin{equation}
\mathcal{L}=\dfrac{1}{2}\partial_{\mu}\partial^{\mu}a-V(a)-\dfrac{1}{4}F_{\mu\nu}F^{\mu\nu}+\Bar{\psi}(i\cancel{D}-m_{f})\psi+\dfrac{\alpha}{8\pi f_{a}}a\epsilon^{\mu\nu\sigma\tau}F_{\mu\nu}F_{\sigma\tau}~~,
\label{eqn:main_lagrangian}
\end{equation}
with an axion potential
\begin{equation}
    V(a)=m_{a}^{2}f_{a}^{2}\Big(1-\cos\Big(\dfrac{a}{f_{a}}\Big)\Big)~~.
    \label{eqn:simple_potential2}
\end{equation}
In the potential, $m_{a}$ is the axion mass and $2\pi f_{a}$ is the fundamental period of the axion field.
Note that in Eqns.\eqref{eqn:main_lagrangian} and \eqref{eqn:simple_potential2} we have taken $N=1$ and $\kappa_{a\gamma\gamma}=1$.
It is straightforward to reinstate these factors in our final results. More importantly, Eqns.\eqref{eqn:main_lagrangian} and \eqref{eqn:simple_potential2} are sufficient to describe an ALP domain wall assuming that: 1) there are not further pseudoscalar states
which meaningfully mix with $a(x)$, and 2) there do not exist wall-localised degrees of freedom apart from the obvious Nambu-Goldstone modes of the spontaneously broken wall-transverse translation symmetry.  We will take the second of these assumptions 
to hold for the entirety of this work, just briefly commenting in Section~\ref{Discussions} on the possible effects of such states (which are well-motivated in certain cases -- see e.g. section 6 of \cite{Alford:1990ur} and the discussion of QHE and FQHE domain wall states in e.g. \cite{Tong:2016kpv}). 

The final $a F\wedge F$ term in Eqn.\eqref{eqn:main_lagrangian} is the interaction term between the wall and the plasma in the environment. In differential form notation this can also be written as $da\wedge A\wedge dA$ by using integration by parts. In an EFT where the momentum cutoff $k$ is less than $m_{a}$, then, 
the term ``$da$" is like a Dirac delta function localised where the gradient of $a$ is greatest, and hence $da\wedge A\wedge dA$ is a boundary term lying completely on the domain wall surface. The term $A\wedge dA$ is the well-known Chern-Simons term in the Quantum Hall Effect context. Hence, the term $a F\wedge F$ is also called the Chern-Simons interaction term. 

Regarding the environment, the plasma temperature we will consider will be substantially greater than the fermion mass, so our plasma is highly relativistic. Furthermore, we assume that the ratio of the chemical potential of the fermion to the temperature is $\mu_{f}/T \ll 1$. Hence, there are almost an equal number of fermions and antifermions. This is true for electrons in the pre-Big-Bang-Nucleosynthesis universe where the ratio is $\mu_{e}/T\sim 10^{-10}$.

From the Lagrangian written above, the equation of motion for the ALP field is
\begin{equation*}
    \dfrac{\partial ^{2}a}{\partial t^{2}}=\nabla^{2}a-m_{a}^{2}f_{a}\sin\Big(\dfrac{a}{f}\Big)+\dfrac{\alpha}{\pi f_{a}} \mathbf{E}\cdot\mathbf{B}~~.
\end{equation*}
Note that here we have simplified $F\wedge F= 8\mathbf{E}\cdot \mathbf{B}$, where $\mathbf{E},\mathbf{B}$ are the usual electric and magnetic fields. In the calculations to follow, we express the axion in terms of a dimensionless angular variable $\varphi$ which is related to the ALP field by
\begin{equation*}
    a=f_{a}\varphi~~.
\end{equation*}
Hence, re-writing the equation of motion for the field $\varphi$ gives
\begin{equation}
      \dfrac{\partial ^{2}\varphi}{\partial t^{2}}=\nabla^{2}\varphi-m_{a}^{2}\sin\varphi+\dfrac{\alpha}{\pi f^{2}_{a}} \mathbf{E}\cdot\mathbf{B}~~. 
      \label{eqn:full_eom}
\end{equation}

\subsection{Brief recap of axion domain walls}\label{Section Axion Domain Walls}

We now quickly revisit the simplest domain wall solution that applies in the case that: a) the potential is as given in Eqn.\eqref{eqn:simple_potential}; b) where the Chern-Simons interaction part in Eqn.\eqref{eqn:full_eom} is absent, and c) where there is no mixing of the field $a(x)$ with other pseudo-scalar states (such as the neutral spin zero mesons of QCD). This then maps our axion equation of motion on to the well-studied Sine-Gordon model. In a (1+1) dimensional theory, this has a non-trivial stationary solution of the form \cite{Weinberg},
\begin{equation}
\varphi(z)=4\arctan\left(\exp(m_{a}z)\right)~~.
\label{eqn:static_planar_pure_wall}
\end{equation}
 The energy density of the walls is defined as
\begin{equation}
    \mathcal{E}(z)=f_{a}^{2}\Big(\dfrac{1}{2}\Big(\dfrac{\partial\varphi}{\partial t}\Big)^{2}+\dfrac{1}{2}\Big(\dfrac{\partial\varphi}{\partial z}\Big)^{2}+m_{a}^{2}(1-\cos\varphi)\Big)~~.
\end{equation}
 
\begin{figure}[h!]
\begin{minipage}{0.5\linewidth}
    \centering
    \includegraphics[width=7.2cm, height=5cm]{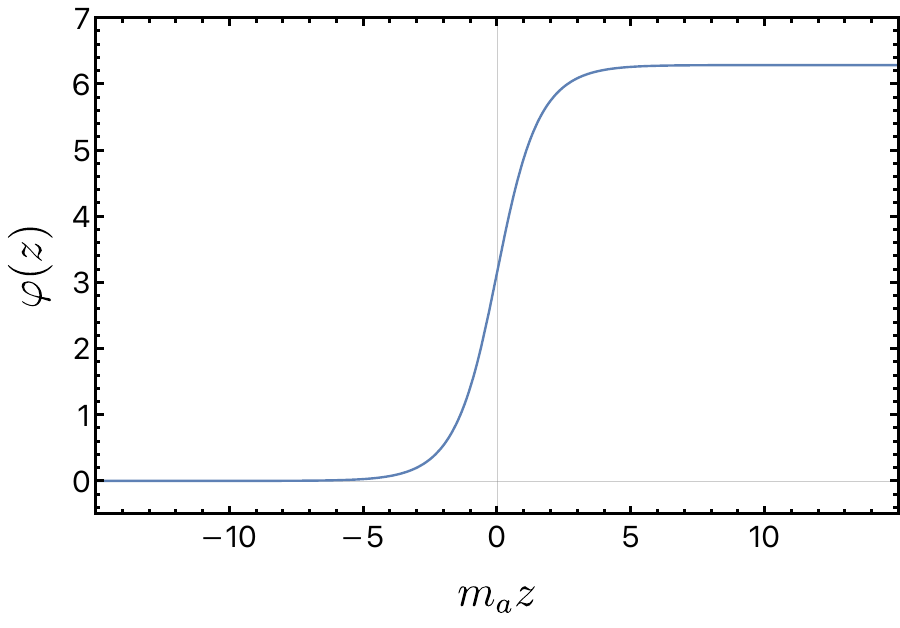}
    \end{minipage}
    \begin{minipage}{0.5\linewidth}
    \centering
    \includegraphics[width=7.5cm, height=5cm]{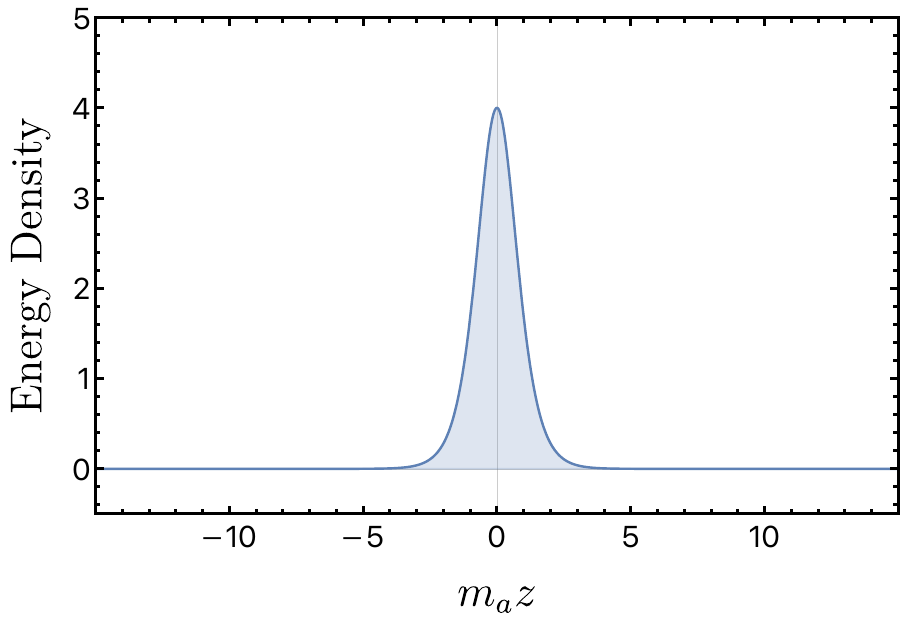}
    \end{minipage}
    \caption{\textbf{Left Panel}: Domain wall field profile interpolating between $a/f_a \equiv \varphi = 0, 2\pi$. \textbf{Right Panel}: Energy density of the domain wall configuration. The shaded region is what we consider as the \emph{wall}. This region has a width of the order $1/\gamma m_{a}$ in case the wall is moving with Lorentz factor $\gamma=1/\sqrt{(1-v^{2})}$.}
\label{fig:sine_gordon_pure_wall_profiles}
\end{figure}

 For us, the solution of Eqn.\eqref{eqn:static_planar_pure_wall} corresponds to a static planar wall configuration oriented in the $x-y$ plane that, along the transverse $z$ direction, interpolates between 0 and $2\pi$. In this $N=1$ case these are the same vacua, but the solution still has a non-trivial energy density as can be seen in Fig.(\ref{fig:sine_gordon_pure_wall_profiles}) and acts as a domain wall in the cosmological context. Under a Lorentz boost normal to the wall, the solution \eqref{eqn:static_planar_pure_wall} simply becomes
\begin{equation}
\varphi(z,t)=4\arctan\left(\exp(\gamma m_{a}(z-vt))\right)~,
\label{eqn:moving_flat_wall}
\end{equation}
where, $v$ is the wall velocity and $\gamma$ is the corresponding boost factor. Of course, when the wall moves relative to the plasma the effective thickness of the wall in the plasma frame is reduced by the usual Lorentz contraction factor $1/\gamma$. For slow walls, this is not a big change, but for walls that are relativistic, this is significant.

Although the above description is valid for planar walls, one can use this approximation for large walls as well, with the radius of curvature of the wall being much larger than the thickness of the wall. Hence, we can approximate the local configuration of the wall to be of the form,
\begin{equation}
\varphi(x_{\perp},t)\approx 4\arctan(\exp(\gamma m_{a}(x_{\perp}-vt)))~~,
\end{equation}
$x_{\perp}$ being the direction normal to the wall locally. 
\subsection{Overview of friction on thin axion domain walls}
\label{sec:thin_wall_overview}

Phase transitions in the early universe are expected to produce strings and domain walls through the Kibble-Zurek mechanism \cite{Kibble:1976sj, Kibble:1980mv}. These have complicated dynamics and can interact with the thermal bath in the early universe. These interactions exchange energy and momentum between the wall and the plasma, inducing friction as the wall evolves over time. This friction can affect the phenomenologically important signatures of these domain walls. For instance, oscillating domain walls produce a strong gravitational wave signal that is a target for many detectors but a large frictional force can remove energy from the coupled gravitational wave-wall system and dump it into the plasma. Even if this does not happen, the friction may delay the onset of an ultra-relativistic regime where gravitational wave emission is dominant. Collapsing domain walls also provide a production mechanism for primordial black holes~\cite{Dunsky:2024zdo} but if the wall collapse is slowed down or if it loses enough energy to its surroundings, this process may be inhibited. Finally, the decay of domain walls into Beyond the Standard Model (BSM) relics is also sensitive to the total energy of the wall and is thereby affected by the friction. This can have a major effect on the bounds derived by assuming the existence of these relics~\cite{Beyer:2022ywc}. Understanding the effect of friction on these phenomena is essential for axion searches.

Frictional forces on ALP domain walls can be classified according to their physical origin including particle scattering, changing masses across the wall, or collective dissipative plasma effects. The former two cases have been studied extensively in the literature (such as \cite{Blaizot:1996az, GarciaGarcia:2024dfx, Vilenkin:2000jqa} and references within), and they apply in the thin-wall limit, where the particle mean-free-path is longer than the wall thickness. The latter case has been discussed in a general context in~\cite{Khlebnikov:1992bx}, without an application to ALP domain walls, which is the subject of this work. We will discuss this in detail in the following sections where we study the friction experienced by (thin and thick) ALP domain walls due to the Chern-Simons interaction term. In the thin wall limit, this interaction can also be treated along the lines described above, i.e. by studying particle scattering off the domain wall.

In the thin wall limit, the friction calculation is done in two steps: (a) calculating the reflection coefficient of the particle interaction with the wall using the semi-classical WKB approximation and (b) calculating the pressure difference on the two sides of the wall in the wall frame. In the first step, one only considers the particle-wall interaction and ignores any interactions between the particle and the plasma. In the domain wall frame, the problem reduces to finding the reflection probability of the plasma particle in an effective potential induced by the wall, which can be treated using the WKB approximation. The force from one particle reflection is then thermally averaged to give the total friction (more details are given below around eq.~\eqref{eqn:pressureThermalAverage}). Heuristically, one can think of this as integrating across the wall first followed by thermal averaging. This procedure can be carried out for any domain wall and any particle that interacts with it.

In the context of axion domain walls with a Chern-Simons interaction term, the friction has been first described by~\cite{Huang:1985tt}. Their focus was on superhozion, moderately-relativistic QCD axion domain walls. In summary, the friction per unit area due to photons in the thermal bath is given by
\begin{equation}
    P\sim \Big(\dfrac{\alpha}{\pi}\Big)^{2}\begin{cases}
        m_{a}^{3}T e^{-m_{a}/T}~~T \ll m_{a}\\
        m_{a}^{2}T ^{2}~~~~~~~~~~T \gg m_{a}~~.
    \end{cases}
    \label{eqn:Sikivie_result}
\end{equation}
where we have quoted two limits depending on the hierarchy between the temperature and the axion mass. Fermions in the plasma do not contribute to this result as mentioned above.

Recently,~\cite{Blasi:2022ayo} presented a comprehensive study\footnote{The goal of~\cite{Blasi:2022ayo} was to investigate fermion interactions of the form $\mathcal{L}_{\rm int} \supset \partial_\mu a \overline{\psi} \gamma^\mu \gamma^5 \psi$ which are different from the Chern-Simons interaction in which we are interested.} of friction on ALP walls. We will use generic thin-wall formulae derived in that paper for comparison with the our results derived in the following sections. Specifically, the pressure (friction per unit area) difference between the two sides of the wall in the wall frame, and for any interaction, is given by the thermal average \cite{Blasi:2022ayo}:
\begin{equation}
    \label{eqn:pressureThermalAverage}
  P=\dfrac{2}{(2\pi)^{2}}\dfrac{1}{\beta\gamma}\int_{0}^{\infty}dp~p^{2}\mathcal{R}(p)\left(\log\left(\dfrac{f(-v)}{f(v)}\right)-2\beta\gamma vp\right)
\end{equation}
where
\begin{equation}
\label{eq:BEDistribution}
f(v)=\dfrac{1}{e^{\gamma\beta\left(\sqrt{p^{2}+m^{2}}+vp\right)}-1}~~.
\end{equation}
Here, $m$ is the mass of the particle scattering off the wall and $\mathcal{R}(p)$ is the reflection coefficient calculated using the WKB method which encapsulates all the information about the particle-wall interaction\footnote{Readers can also refer to Appendix A for the derivation of this formula}. 

Our interest is the axion-photon coupling described in Section~\ref{The Physical setup}. For this coupling, and when treating thin domain walls, we can use the above formalism to determine the friction from photon scattering. To that end, we take $m=0$ in~\eqref{eq:BEDistribution}. The remaining ingredient is the reflection coefficient on which we comment in Appendix \ref{appendix:thin_wall_wkb} for a simple toy model of a realistic axion domain wall.\footnote{A detailed analysis of similar toy model has been discussed in a recent paper on optical properties in axion-electrodynamics \cite{Favitta:2023hlx}.} Putting everything together, the integrand peaks when the momentum of the photon is of order of the axion mass $m_{a}$ and gives
\begin{equation}
    \label{eqn:scatteringFriction}
    P\approx \dfrac{\alpha^{2}}{2\pi^{2}}  \dfrac{m_{a}^{3}T}{\gamma}\log\left(\dfrac{1+v}{1-v}\right)\approx \begin{cases}
    \dfrac{\alpha^{2}}{\pi^{2}}\dfrac{1}{\gamma} m_{a}^{3}Tv~~~~\gamma \approx 1 \\ 
    \vspace{-1.0em}\\
        \dfrac{\alpha^{2}}{2\pi^{2}}\dfrac{1}{\gamma} m_{a}^{3}T \ln\left(\dfrac{2}{1-v}\right)~~\gamma \geq 2
    \end{cases}.
\end{equation}
where we have the leading contribution in the limit $\gamma m_{a} \ll T$. The first result is when the velocity is such that $\gamma$ is still close to 1. The second case applies when $(1-v)\ll 1$ such that $\gamma\geq 2$. This expression is different from the published~\footnote{We thank Simone Blasi and Alberto Mariotti for bringing to our attention a difference between the preprint and published versions of~\cite{Huang:1985tt}. The parametric dependence we get agrees with expressions that appear in the preprint by Huang and Sikivie and also with those derived in \cite{Blasi:2023sej} when studying friction from gluons scattering on the domain wall.} result of~\cite{Huang:1985tt} which we show in Eqn.\eqref{eqn:Sikivie_result}. As we will see, the dissipative thermal field theory approach matches Eqn.\eqref{eqn:scatteringFriction}. One final comment is in order, the factor of $\gamma^{-1}$ in Eqn.\eqref{eqn:scatteringFriction} is because this result is written in the wall frame. In the plasma frame, this factor is absent.

To complete the discussion on thin walls, we briefly mention how friction is taken into account when numerically studying the evolution. As such, the majority of models trying to understand this evolution work in a thin-wall approximation. Using this approximation, one gets a world-volume action similar to a world-sheet action for fundamental strings (or an EFT for axion strings). One can vary this action to get the motion of embedding coordinates in terms of world-volume coordinates \cite{Vilenkin:2000jqa}. In a simplified setting, these equations can be reduced to coupled differential equations for the wall size $L$ and wall velocity $v$. These models are called ``Velocity-dependent One-Scale" (VOS) models \cite{Avelino:2005kn, Martins:2016ois} and are inspired by string evolution models \cite{Martins:1996jp, Martins:2000cs}. In these models, the friction term appears in the differential equation for the velocity and takes the form \cite{Avelino:2005kn, Blasi:2022ayo},
\begin{equation*}
    F_{thin}\propto -\dfrac{\gamma v}{l_{d}},~~~\dfrac{1}{l_{d}}=3H+\dfrac{1}{l_{f}}~~.
\end{equation*}
Here, $H$ is the Hubble constant and signifies the Hubble friction contribution due to cosmic expansion. The second term given by $1/l_{f}$ is the friction contribution from particles in the environment. The length $l_f$ is determined by an independent friction calculation and used as an input to this formalism. As mentioned above, one crucial approximation in these studies is that the wall is thin.

\subsection{Thick wall friction - qualitative overview \& summary of results}
\label{subsec:thick_walls_qualitative_picture}

A relativistic plasma with charged\footnote{In this work, we will mainly only consider Abelian U(1) gauge groups, though in our conclusions we will comment on the changes that we believe apply in the non-Abelian case.} particles introduces several new scales in the domain wall friction problem. Previously, we commented on the mean-free-path and compared wall thickness to this length scale which is typically on the order of:
\begin{align}
    \lambda_{\rm MFP}^{-1} \sim \alpha^2 T
\end{align}
in QED-like theories where $\alpha$ is the fine-structure constant of the gauge theory and $T$ is the temperature. The relativistic plasma also has other, shorter length scales that are also relevant for understanding domain wall friction. At distances shorter than $\lambda_{\rm MFP}$, we encounter the Debye length:
\begin{align}
    l_D^{-1} = m_D \sim &\sqrt{\alpha}T
\end{align}
which sets the scale of electric shielding effects and is sometimes known simply as the `electric scale'. In the plasma, the photon acquires a thermal mass on the order of $m_D$ which is responsible for this screening effect. The Debye length is the shortest distance where we observe collective behaviour of particles in the plasma~\cite{Bellac:2011kqa}, which are long-lived excitations with lifetimes on the order of $(\alpha^2 T)^{-1}$. At yet shorter distances than the Debye length, there is the thermal length set by the plasma temperature:
\begin{align}
    l_T^{-1} \sim T.
\end{align}
This scale sets thermal fluctuations in the system. For distances between $l_D$ and $l_T$, a description in terms of collective modes is not valid and one must treat the plasma as a collection of relativistic particles. 

These scales, along with the emergence of collective behaviour enrich the domain wall problem beyond the simple scattering picture described in the previous section. Specifically, we have the hierarchy
\begin{align}
    l_T \ll l_D \ll \MFP
\end{align}
and the friction experienced by a domain wall depends on how the Lorentz-contracted wall thickness $(\gamma m_a)^{-1}$ compares to the various plasma scales. In the treatment of \secref{sec:thin_wall_overview}, we considered particle scattering off a thin wall where the notion of thin was defined relative to $\MFP$. However, a natural question arises: what happens for domain walls thicker than $\MFP$? At these length scales, collective plasma effects play an important role and are essential in understanding domain wall friction. Furthermore, as mentioned above, these effects are important even for walls with thickness between $\MFP$ and $\ell_D$. It is only for walls thinner than $l_D$ that the particle scattering picture is sufficient to capture plasma friction. 

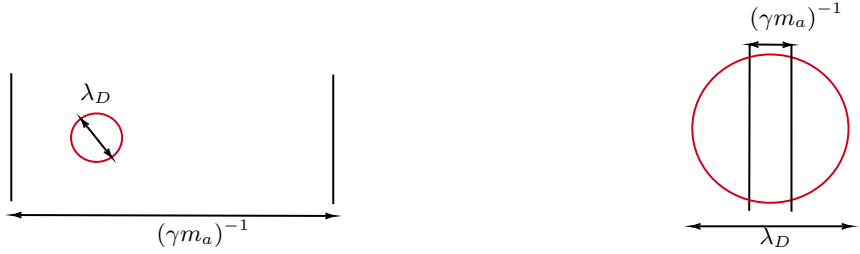
\begin{figure}[h!]
\begin{subfigure}{0.5\linewidth}
\centering
\tikzset{every picture/.style={line width=0.75pt}}         

\begin{tikzpicture}[x=0.75pt,y=0.75pt,yscale=-1,xscale=1]

\draw    (138.32,53.32) -- (138.32,117.75) ;
\draw    (299.04,54.04) -- (299.04,119.19) ;

\draw  [color={rgb, 255:red, 208; green, 2; blue, 27 }  ,draw opacity=1 ] (168.14,85.77) .. controls (168.14,79.05) and (173.85,73.59) .. (180.9,73.59) .. controls (187.94,73.59) and (193.66,79.05) .. (193.66,85.77) .. controls (193.66,92.5) and (187.94,97.96) .. (180.9,97.96) .. controls (173.85,97.96) and (168.14,92.5) .. (168.14,85.77) -- cycle ;

\draw    (141.03,124.68) -- (297.49,125.14) ;
\draw [shift={(299.49,125.15)}, rotate = 180.17] [color={rgb, 255:red, 0; green, 0; blue, 0 }  ][line width=0.75]    (4.37,-1.32) .. controls (2.78,-0.56) and (1.32,-0.12) .. (0,0) .. controls (1.32,0.12) and (2.78,0.56) .. (4.37,1.32)   ;
\draw [shift={(139.03,124.67)}, rotate = 0.17] [color={rgb, 255:red, 0; green, 0; blue, 0 }  ][line width=0.75]    (4.37,-1.32) .. controls (2.78,-0.56) and (1.32,-0.12) .. (0,0) .. controls (1.32,0.12) and (2.78,0.56) .. (4.37,1.32)   ;
\draw    (174.51,78.09) -- (187.11,93.98) ;
\draw [shift={(188.36,95.55)}, rotate = 231.57] [color={rgb, 255:red, 0; green, 0; blue, 0 }  ][line width=0.75]    (4.37,-1.32) .. controls (2.78,-0.56) and (1.32,-0.12) .. (0,0) .. controls (1.32,0.12) and (2.78,0.56) .. (4.37,1.32)   ;
\draw [shift={(173.26,76.52)}, rotate = 51.57] [color={rgb, 255:red, 0; green, 0; blue, 0 }  ][line width=0.75]    (4.37,-1.32) .. controls (2.78,-0.56) and (1.32,-0.12) .. (0,0) .. controls (1.32,0.12) and (2.78,0.56) .. (4.37,1.32)   ;

\draw (209.23,125.34) node [anchor=north west][inner sep=0.75pt]  [font=\footnotesize] [align=left] {$\displaystyle (\gamma m_{a})^{-1}$};
\draw (171.19,57.04) node [anchor=north west][inner sep=0.75pt]  [font=\footnotesize] [align=left] {$\displaystyle \lambda _{D}$};
\label{sfig:a}
\end{tikzpicture}
\end{subfigure}
\hfill
\begin{subfigure}{0.45\linewidth}
\centering
\tikzset{every picture/.style={line width=0.75pt}} 

\begin{tikzpicture}[x=0.75pt,y=0.75pt,yscale=-1,xscale=1]

\draw    (199.73,64.65) -- (199.5,148.34) ;
\draw  [color={rgb, 255:red, 208; green, 2; blue, 27 }  ,draw opacity=1 ] (171.09,106.99) .. controls (171.09,86.38) and (188.54,69.68) .. (210.08,69.68) .. controls (231.61,69.68) and (249.07,86.38) .. (249.07,106.99) .. controls (249.07,127.59) and (231.61,144.29) .. (210.08,144.29) .. controls (188.54,144.29) and (171.09,127.59) .. (171.09,106.99) -- cycle ;
\draw    (220.57,64.85) -- (220.6,148.73) ;
\draw    (172.8,156.12) -- (248.08,156.12) ;
\draw [shift={(250.08,156.12)}, rotate = 180] [color={rgb, 255:red, 0; green, 0; blue, 0 }  ][line width=0.75]    (4.37,-1.32) .. controls (2.78,-0.56) and (1.32,-0.12) .. (0,0) .. controls (1.32,0.12) and (2.78,0.56) .. (4.37,1.32)   ;
\draw [shift={(170.8,156.12)}, rotate = 0] [color={rgb, 255:red, 0; green, 0; blue, 0 }  ][line width=0.75]    (4.37,-1.32) .. controls (2.78,-0.56) and (1.32,-0.12) .. (0,0) .. controls (1.32,0.12) and (2.78,0.56) .. (4.37,1.32)   ;
\draw    (201.73,64.67) -- (218.57,64.83) ;
\draw [shift={(220.57,64.85)}, rotate = 180.57] [color={rgb, 255:red, 0; green, 0; blue, 0 }  ][line width=0.75]    (4.37,-1.32) .. controls (2.78,-0.56) and (1.32,-0.12) .. (0,0) .. controls (1.32,0.12) and (2.78,0.56) .. (4.37,1.32)   ;
\draw [shift={(199.73,64.65)}, rotate = 0.57] [color={rgb, 255:red, 0; green, 0; blue, 0 }  ][line width=0.75]    (4.37,-1.32) .. controls (2.78,-0.56) and (1.32,-0.12) .. (0,0) .. controls (1.32,0.12) and (2.78,0.56) .. (4.37,1.32)   ;

\draw (203.11,155.36) node [anchor=north west][inner sep=0.75pt]  [font=\footnotesize] [align=left] {$\displaystyle \lambda _{D}$};
\draw (198.2,42.29) node [anchor=north west][inner sep=0.75pt]  [font=\footnotesize] [align=left] {$\displaystyle (\gamma m_{a})^{-1}$};
\end{tikzpicture}
\label{sfig:b}
\end{subfigure}
\caption{Cartoon depiction of scales. The domain wall, of plasma-frame Lorentz-contracted width, $L\sim (\gamma m_{a})^{-1}$, is shown between the two vertical lines. The red circles depict the size of the Debye sphere - the scale at which collective modes respond. \textbf{Left Panel}: In this case, the wall thickness is much larger than the Debye scale. \textbf{Right Panel}: The opposite case where the wall thickness as seen in the plasma frame is smaller than the Debye scale.}
\end{figure}

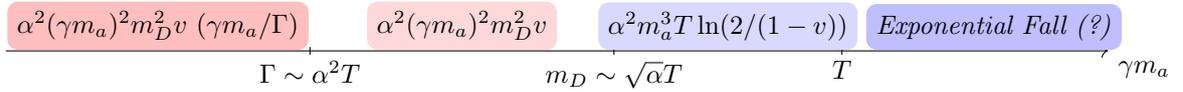
\begin{figure}[h!]
    \centering
    \begin{tikzpicture}
    \draw [->] (0,0)-- (14.5,0) node[anchor=north west] {\small$\gamma m_{a}$};
     \draw (4,0) node [anchor=north] {\small$\Gamma\sim \alpha^{2}T$};
    \draw (8,0) node [anchor=north] {\small$m_{D}\sim \sqrt{\alpha}T$};
    \draw (11,0) node [anchor=north] {\small$T$};
    \draw (4,0.05)-- (4,-0.05);
    \draw (8,0.05)-- (8,-0.05);
    \draw (11,0.05)-- (11,-0.05);
    \draw (2,0) node [shape=rectangle, anchor=south, fill=red! 25, rounded corners=4, thick] {\small$\alpha^{2}(\gamma m_{a})^{2}m_{D}^{2}v~(\gamma m_{a}/\Gamma)$};
    \draw (6,0) node [shape=rectangle, anchor=south, fill=red! 15, rounded corners=4, thick] {\small$\alpha^{2}(\gamma m_{a})^{2}m_{D}^{2}v$};
     \draw (9.5,0) node [shape=rectangle, anchor=south, fill=blue! 15, rounded corners=4, thick] {\small$\alpha^{2}m_{a}^{3}T\ln(2/(1-v))$};
      \draw (13,0) node [shape=rectangle, anchor=south, fill=blue! 25, rounded corners=4, thick] {\small\emph{Exponential Fall (?)}};
\end{tikzpicture}
    \caption{Schematic diagram of the behavior of the friction per area on a moving ALP domain wall as a function of the plasma-frame inverse wall thickness, $\gamma m_a$ (equivalently, for fixed axion mass, as a function of increasing wall velocity reading left-to-right). These expressions, in which $\mathcal{O}(1)$ coefficients have been dropped for simplicity, capture the parametric dependence of the Chern-Simons induced friction upon the wall with velocity $v$ (with associated Lorentz $\gamma$ factor), axion mass $m_{a}$, plasma temperature $T$, and U(1) fine structure constant $\alpha$ assuming that $T$ is high enough that ultra-relativistic charged particles are in equilibrium in the plasma, i.e. $T\gg {\rm all~relevant~particle~masses}$. Note that a major role is played by the three plasma scales, the Abelian plasma damping rate $\Gamma\sim \alpha^2 T$, the Debye (electric) mass $m_D \sim \sqrt{\alpha} T$, and the temperature $T$. These expressions for the friction assume that the axion electrodynamics parameters satisfy $N=1$ and $\kappa_{a\gamma\gamma}=1$. For  wall velocities such that $\gamma m_a \gg T$, the exponential fall-off is a leading-order result and this conclusion can in principle be modified at higher order. For instance, there might exist effects formally sub-dominant in $\alpha$ which are proportional to $\gamma^2$ and so in fact become important as $\gamma\gg 1/\alpha$ (such behaviour was described in~\cite{GarciaGarcia:2022yqb,GarciaGarcia:2024dfx} for a different but related case).  We have not perfomed the higher-order calculations necessary to determine if this applies to Chern-Simons induced friction on axion domain walls.
    }
    \label{thermal friction_ranges}
\end{figure}

Before proceeding with the calculations, it is useful to gain a qualitative understanding of plasma processes that go beyond the scattering picture. To that end, consider a domain wall moving through a plasma where there are local magnetic and electric field fluctuations. In standard axion-electrodynamics, axion field gradients acquire an effective charge density in the presence of a magnetic field according to
\begin{align}
\nabla\cdot\mathbf{E}\sim (\nabla\varphi\cdot\mathbf{B}).
\end{align}
Since $\nabla \varphi \neq 0$ for a domain wall, fluctuations in the magnetic field orthogonal to the wall will induce a local charge density on the wall. This turns the wall into a charged object immersed in the plasma which will respond to restore local charge neutrality. The time it takes to establish neutrality is given by the timescale for plasma rearrangement \cite{Holzer:2016tcg}:
\begin{align}
    t_D \sim \frac{l_D}{v_{th}}
\end{align}
where $v_{th}$ is the thermal velocity of the charged particles. For a relativistic plasma, with temperature $T$ larger than the mass of charged particles, $v_{th} \approx 1$ and the timescale is simply the Debye length defined above. These phenomena that happen at the frequencies $\omega_{p}\sim eT$ (or, equivalently, length scales $(eT)^{-1}$) are the effects of \emph{collective behavior} of the particles in plasma \cite{Bellac:2011kqa}. For walls thicker than $(eT)^{-1}$ (and `slower' than $\omega_p$, as outlined in the next paragraph), we can average these fast plasma processes and include their effects in the axion equation of motion (by replacing $\mathbf{E}\cdot\mathbf{B} \rightarrow \langle \mathbf{E}\cdot\mathbf{B}\rangle$ in Eqn.\eqref{eqn:full_eom}) to find the friction.

This averaging also requires that the wall is `slow' compared to plasma processes. This is easy to justify by comparing the wall self-crossing time to the plasma frequency. The wall thickness is of the order $(\gamma m_{a})^{-1}$. For a wall moving with velocity $v$, the self-crossing time of the wall $(\gamma m_{a})^{-1}/v$ is larger than $(\gamma m_{a})^{-1}$. 
So, as long as $(\gamma m_{a})^{-1}$ is sufficiently larger than the thermal scales in the problem like $\sim T^{-1}$ and $(eT)^{-1}$, we can treat the wall as moving slowly. In this limit, the plasma responds quickly to the motion of the wall and we are interested in effects of the plasma on long\footnote{Later, this will allow us to expand the retarded Green's functions linearly around $\omega=0$.} timescales compared to the thermal scales (since these are the relevant timescales for the domain wall motion).  The outcome of this averaging procedure includes a host of complicated effects that depend on the profile, dynamics and other properties of the domain wall. In this paper, we will focus on plasma friction only and extract this quantity.

The thermal average of the fluctuations of the electromagnetic fields\footnote{In our case, the relevant averaged quantity is $\mathbf{E}\cdot \mathbf{B}$.} at a given time, in the presence of the wall, depends on how they are correlated to similar fluctuations in the past. Because of the effect of fermions in the plasma, which give rise to the screening, such correlations depend on the thermal scales described above and bring various scales into the picture that are smaller than the thickness of the wall. 

If the wall is thinner than the Debye length, then the charge of the wall due to a fluctuation of the magnetic field is screened outside the Debye length. Inside this length scale, the system is almost un-screened and can be viewed as a collection of photons and charges. In this limit, particles scattering off the wall is the dominant effect. In other words, a wall can be considered thin and one can view the friction as purely due to particle scattering off the wall if there is no other smaller length scale at which macroscopic effects are seen in the plasma. This approximation is valid when the wall thickness in the plasma frame is smaller than the Debye length. In this case, the correct procedure is to integrate across the wall and find the microscopic effects of the wall-particle interaction first and then average the result against the thermal particle distributions. 

On the other hand, as mentioned above, if the wall is not thin compared to the Debye length, one needs to first integrate out the thermal effects and then average over the wall. Even if the wall thickness is somewhere between the mean free path and the Debye length, pure scattering doesn't capture the full response.  The friction calculated using this method originates from \emph{dissipative effects} of the plasma rather than pure particle scattering. A similar approach was taken in \cite{Khlebnikov:1992bx} to study the plasma friction on bubble walls in a generic sense.  In this method, we view the wall as a local disturbance to the plasma and, in turn, find the corresponding local response of the plasma to the wall. Further, as we are averaging thermal effects at scales smaller than the wall thickness first and then averaging across the wall, this also takes into account the Lorentz contraction of the wall in the plasma frame. 

The need for this averaging can be seen by considering axion-electrodynamics without a plasma. In this case, one can derive the Lorentz force density acting on the axion field profile as\footnote{Considering the axion profile as some charge distribution with $\rho=-(\alpha/\pi)\nabla\varphi\cdot\mathbf{B}$ and $\mathbf{j}=(\alpha/\pi)(\nabla\times\mathbf{E}+(\partial_{t}\varphi)\mathbf{B})$, the Lorentz force density $f=\rho\mathbf{E}+\mathbf{j}\times\mathbf{B}$ is given by the expression Eqn.\eqref{eqn:Lorentz_force}.}
\begin{equation}
    f=-\dfrac{\alpha}{8\pi}(F\wedge F)\nabla \varphi=-\dfrac{\alpha}{8\pi}(F\wedge F)\dfrac{\partial \varphi}{\partial z}~~
    \label{eqn:Lorentz_force}
\end{equation}
where in the second equality we assumed that the wall profile depends only on the $z$ coordinate. 
To get a net force on the wall, one needs to integrate the above expression along the normal direction to the wall. When we have a wall in a thermal environment, we also need to take the thermal average of the $F\wedge F$ configurations. This is essential when the wall thickness $(\gamma m_{a})^{-1}$ is larger than the scales of thermal fluctuation.

Here, we summarize some important features of the calculations performed in the following sections. For thick walls, the idea is to first calculate thermal effects as a response to the motion of the wall and then integrate the thermal effects across the wall to get the friction. As we know, the classical equation of the wall depends on $F\wedge F$ (or equivalently $\mathbf{E}\cdot\mathbf{B}$) as in Eqn.\eqref{eqn:full_eom} and we need to calculate the thermal average of the quantity $\mathbf{E}\cdot\mathbf{B}$. This is done using linear response theory where the ALP scalar field $\varphi$ is the external disturbance to the plasma. In the linear response method, the thermal average $\mathbf{E}\cdot\mathbf{B}$ will turn out to be some function of $\varphi$ and its derivatives. 

In the calculations, which are done using the real-time formalism of thermal field theory, we include the effects of the fermions in the plasma giving thermal corrections to the photon propagator. Further, to calculate an analytic form for friction, we make certain approximations and use physically motivated toy models. For details, we refer the reader to \secref{integral 2}. After the thermal average is determined, we integrate it across the wall with a weight function to get the effect on the wall. The weight function depends on the profile of the wall. Details about this are presented in \secref{Average Force on the Wall}. The friction depends on the inverse wall thickness $\gamma m_{a}$ with respect to the thermal scales $\Gamma\sim \alpha^{2}T$ (the Abelian damping rate) and $m_{D} \sim \sqrt{\alpha}T$ (the Debye mass, or ``electric scale"). Our results for the friction per unit area (frictional pressure) are:
\begin{equation}
    P\propto \Big(\dfrac{\alpha}{\pi}\Big)^{2}\begin{cases}
        (\gamma m_{a})^{2}m_{D}^{2}(\gamma m_{a}/\Gamma)v, &\gamma m_{a}<\Gamma\sim \alpha^{2}T\\ 
        \vspace{-1.0em}\\
        (\gamma m_{a})^{2}m_{D}^{2}v, &\Gamma<\gamma m_{a}<m_{D}\sim \sqrt{\alpha}T
    \end{cases}
    \label{eqn:results_summary}
\end{equation}
Once $\gamma m_{a}>m_{D}$, we have the purely thin wall case which we have already reviewed in the last subsection.  The schematic diagram, Fig.(\ref{thermal friction_ranges}), summarises these results.  One can further generalise these results to include
the possible presence of an additional long-range coherent magnetic field in the plasma that arises from novel early-universe dynamics. We will describe this in more detail in the following section. 

\begin{table}[h!]
    \centering
    \begin{tabular}{|p{2.8cm}|p{5.1cm}|p{5.1cm}|}
    \hline
        & Thin Wall & Thick Wall \\
        \hline
        Validity & Plasma frame wall thickness $<$ Plasma response scale (Debye length)
         & Plasma frame wall thickness $>$ Plasma response scale (Debye length)\\
         \hline
         Averaging Order & Individual wall-particle interaction computed using WKB followed by averaging using thermal particle distribution & Thermal average first as a local linear response using plasma spectral functions followed by integration across the wall\\
         \hline
        Effects Captured & Individual particle scattering & Collective disipative effects\\
        \hline
    \end{tabular}
    \caption{Qualitative summary of the thin and thick wall regimes and calculational approaches taken.}
    \label{tab:my_label}
\end{table}

\section{Plasma reaction to a moving wall - linear response approach}
\label{The Linear Response}

In the last section, we saw that in the early universe, when we have a relativistic plasma, treating the wall as an infinitesimally thin object is not always a good approximation. To capture the collective effects of the plasma, one needs to take an approach different from particle scattering. To do this, we consider the domain wall field $\varphi$ to be a local perturbation to the plasma which is itself in equilibrium. Furthermore, the motion of the wall is at time scales much longer than the time scales for the plasma to respond, i.e., the inverse of plasma frequency. To look at such a system, we use linear response theory. The idea behind linear response theory is to understand how a system reacts in the presence of a weak local perturbation. In the limit of weak perturbation, one can look at the change in observables at a linear order. For example, in quantum field theory or thermal field theory, the observables are nothing but averages of certain physical quantities. Hence, the linear response tells us how these averages change to linear order in the weak perturbation. The machinery used to study the finite-temperature linear response \cite{Kapusta:1989tk,Bellac:2011kqa,Laine:2016hma} is similar to the usual perturbation theory in quantum field theory. In this case, we can take a thermally averaged equation of motion of $\varphi$, where we average the plasma fields. In this treatment, as mentioned in the last section, what we want is a linear response of the wall in a small frequency limit. In spirit, this is similar to the calculation of the fluctuation-dissipation friction as discussed in \cite{Khlebnikov:1992bx}. 

To calculate the linear response, first, we need to identify the interaction Hamiltonian part that gives the information about how the scalar field $\varphi$ interacts with the plasma. Our calculations are motivated by the formalism used in \cite{McLerran:1990de}. However, after performing the calculations, we will highlight some subtle differences between the system that we are analyzing and the one studied in \cite{McLerran:1990de}. From Eqn.\eqref{eqn:main_lagrangian}, one can derive that the Hamiltonian density takes the form \cite{Huang:1985tt,McLerran:1990de},
\begin{equation}
\mathcal{H}=\dfrac{1}{2}(E^{2}+B^{2})+\bar{\psi}(i\gamma^{i}\partial_{i}+m_{f})\psi+eJ^{\mu}A_{\mu}+\dfrac{f_{a}^{2}}{2}((\partial_{t}\varphi)^{2}+(\partial_{i}\varphi)^{2}+2m_{a}^{2}(1-\cos\varphi))~~.
\label{eqn:hamiltonian_full}
\end{equation}
In this case, just by looking at the expression of the Hamiltonian, one might assume that there is no interaction Hamiltonian present as it seems that the plasma Hamiltonian and the ALP field Hamiltonian are completely separated with no interaction terms. This is because the interaction term is of the form $\varphi F\wedge F= 4\dot{A}_{i}(\varphi\epsilon^{0ijk}\partial_{j}A_{k})$. Terms of this form modify the canonical conjugate momentum but cancel out when calculating the Hamiltonian. Hence, if the Hamiltonian is written in the old canonical momenta (momenta in the absence of the interaction), the form of the Hamiltonian is the same.\footnote{Consider a Lagrangian of the form $L=L_{0}+\dot{q}f(q)$, where $L_{0}$ contains the usual kinetic term and potential of the form $V(q)$. Then, $p_{new}=\partial_{\dot{q}}L=p_{old}+f(q)$. The Hamiltonian is $H=p_{new}\dot{q}-L=p_{old}\dot{q}-L_{0}=H_{0}$. But now, to quantize, one needs to express the $H_{0}\equiv H_{0}(p_{new},q)$. The kinetic term hence will have the form $(p_{new}-f(q))^{2}=p_{new}^{2}-2p_{new}f(q)+f(q)^{2}$. Hence, considering $f(q)$ is small (i.e., it has a coupling factor which is small), then at leading order $H_{int}=-p_{old}f(q)$. From the Hamiltonian equations, one can see that $\dot{q}=\partial_{p_{new}}H=p_{old}$. Hence, $H_{int}=-\dot{q}f(q)$. If we now take $q_{i}=A_{i}$ for $i=1,2,3$, then, one can see that $\mathcal{H}_{int}\propto\epsilon^{0ijk}\dot{A}_{i}\partial_{j}A_{k}~\varphi=F\wedge F~\varphi$.} In this case, one has to note that the electric field ``$-\mathbf{E}$" is not the canonical conjugate momentum of ``$\mathbf{A}$". Instead, the conjugate momenta are,
\begin{equation}
    \Pi_{i}=-E_{i}+\dfrac{\alpha}{\pi}B_{i}\varphi~~.
\end{equation}
Substituting this back in the equation of the Hamiltonian gives:
\begin{equation}
    \mathcal{H}_{int}=\dfrac{\alpha}{\pi}\mathbf{\Pi}\cdot\mathbf{B}~\varphi+\dfrac{1}{2}\Big(\dfrac{\alpha}{\pi}\Big)^{2}\mathbf{B}\cdot\mathbf{B}~\varphi^{2}=-\dfrac{\alpha}{\pi}\mathbf{E}\cdot\mathbf{B}~\varphi+\dfrac{3}{2}\Big(\dfrac{\alpha}{\pi}\Big)^{2}\mathbf{B}\cdot\mathbf{B}~\varphi^{2}~~.
\end{equation}
In the calculations to follow, we take the leading order term, and hence
\begin{equation}
    \mathcal{H}_{int}=-\dfrac{\alpha}{\pi}\mathbf{E}\cdot\mathbf{B}~\varphi~~.
    \label{eqn:interaction_hamiltonian_leading_order}
\end{equation}
This means we are looking at a thermal system with U(1) gauge fields and fermions charged under that field along with an external perturbation mediated to the plasma by the interaction in the equation above. In the linear response approach, we keep the scalar field $\varphi$ to be a background entity (i.e., not to be averaged over the thermal ensemble) as it is the perturbation to the plasma. In this case, the thermally averaged equation of motion of the field $\varphi$ is given by, 
\cite{McLerran:1990de, Bodeker:2022ihg},
\begin{equation}
    \partial_{t}^{2}\varphi=\nabla^{2}\varphi-m_{a}^{2}\sin\varphi+\dfrac{\alpha}{8\pi f_{a}^{2}}\langle F\wedge F\rangle~~.
    \label{eqn:thermally_averaged_phi_eom}
\end{equation}
The thermal average of the operator $F \wedge F$ in the equation above is calculated in the full system where there is the classical field $\varphi$ along with the plasma. This can be expressed in terms of the thermal averages in the free plasma (i.e., the system with no field $\varphi$) up to linear order in the coupling as 
\begin{equation}
  \langle \mathbf{E}\cdot\mathbf{B}\rangle=\langle \mathbf{E}_{I}\cdot\mathbf{B}_{I}\rangle_{0}+i\dfrac{\alpha}{\pi}\int_{-\infty}^{\infty}dt'~\theta(t-t')\int d^{3}x'\langle[\mathbf{E}_{I}\cdot\mathbf{B}_{I}(t,\mathbf{x}),\mathbf{E}_{I}\cdot\mathbf{B}_{I}(t',\mathbf{x}')]\rangle_{0}\varphi(t',\mathbf{x}')~.
\end{equation}
Refer to Appendix (\ref{appendix:Apx_linear_resp}) for the derivation. The first term in the equation above is zero as a consequence of CP conservation when there is no interaction between the wall and plasma. Hence, to leading order, 
\begin{equation}
    \langle \mathbf{E}\cdot\mathbf{B}\rangle=i\dfrac{\alpha}{\pi}\int_{-\infty}^{\infty}dt'\theta(t-t')\int d^{3}x'\langle[\mathbf{E}_{I}\cdot\mathbf{B}_{I}(t,\mathbf{x}),\mathbf{E}_{I}\cdot\mathbf{B}_{I}(t',\mathbf{x}')]\rangle_{0}\varphi(t',\mathbf{x}')~. 
    \label{eqn:thermal_average_eb_1}
\end{equation}
Note that the thermal average in the integral above is going to be some function of $(t-t',\mathbf{x}-\mathbf{x}')$. The averages in the integral are taken with respect to the density matrix of thermal QED in the absence of a wall. This means that after simplification, the above equation should end up as a function of $\varphi$, its derivatives, and some known plasma properties. Before going ahead with the simplification of this expression, we make an important comment on the magnetic fields. In the early universe, along with thermal magnetic fields, there could be long-range coherent magnetic fields present. These long-range magnetic fields in the early universe are envoked to explain the existence of the long-range magnetic fields that we observe today. 
 
Hence, we decompose the magnetic field as, 
\begin{equation}
    \mathbf{B}=\mathbf{B}_{bg}+\mathbf{B}_{th}~~, 
\end{equation}
where $\mathbf{B}_{bg}$ is the long-range background magnetic field and $\mathbf{B}_{th}$ is the magnetic field in the thermal radiation part. If we use this separation of the fields, in Eqn.\eqref{eqn:thermal_average_eb_1}, we can see that we will get terms of the form $(B_{th,i}B_{th,j})$, $(B_{bg,i}B_{th,j})$, and $(B_{bg,i}B_{bg,j})$. In the first term, we take $B_{th}$ as a \emph{thermal operator} and calculate the full average using the techniques of real-time thermal field theory. Now, consider the other two products.  
\begin{equation}
B_{bg,i}(t,\mathbf{x})B_{bg,j}(t',\mathbf{x}'),~~B_{th,i}(t,\mathbf{x})B_{bg,j}(t',\mathbf{x}')~~,
\end{equation}
Simulations trying to understand these primordial magnetic fields show that the coherence scale of these fields is larger than the thermal part and their spectrum is independent, for example, refer to \cite{Diaz-Gil:2007fch} and references within. To model this, we consider $\mathbf{B}_{bg}$ to be a \emph{Gaussian random field} and take the Gaussian random average over such field configuration.
After taking this Gaussian random average, the second term in the equation above of the form $(B_{th}B_{bg})$ vanishes. The random field average  then stores all the information about the fields in their power spectrum. This means, we can write 
\begin{equation}
B_{bg,i}(t,\mathbf{x})B_{bg,j}(t',\mathbf{x}')\rightarrow \langle B_{bg,i}(t,\mathbf{x})B_{bg,j}(t',\mathbf{x}')\rangle_{gaussian}=\xi_{ij}(t-t',\mathbf{x}-\mathbf{x}')
\end{equation}
In the energy-momentum space, this looks like,
\begin{equation}
    \mathsf{FT}(\xi_{ij})=\langle\widetilde{B}_{bg,i}(\mathbf{k})\widetilde{B}_{bg,j}(-\mathbf{k}')\rangle_{gaussian}=(2\pi)^{3}\delta^{(3)}(\mathbf{k}-\mathbf{k}')\Big(\delta_{ij}-\dfrac{k_{i}k_{j}}{k^{2}}\Big)\eta(k),~~k=\vert\mathbf{k}\vert
    \label{eqn:background_field_power_spectrum}
\end{equation}
The symbol $\mathsf{FT}$ in the equation above and everywhere else in the paper stands for Fourier transform. The tensor structure is a consequence of the Maxwell's equation $\nabla\cdot \mathbf{B}_{bg}=0$. Lastly, we also consider that locally, the background magnetic fields are weaker than the thermal fluctuation in fields as also seen in \cite{Diaz-Gil:2007fch}. Due to this, one can immediately see that any corrections to any propagators due to the background magnetic field will be subdominant to the thermal ones and hence we ignore these corrections. Coming back to the thermal average, this gives us two separate contributions in
Eqn.\eqref{eqn:thermal_average_eb_1} as follows\footnote{We loose the subscript $I$ henceforth in the calculations},
\begin{equation}
    \mathcal{I}_{1}=i\dfrac{\alpha}{\pi}\int_{-\infty}^{\infty}dt'\theta(t-t')\int d^{3}x'\langle[E_{i}(t,\mathbf{x}),E_{j}(t',\mathbf{x}')]\rangle_{0}\xi_{ij}(t-t',\mathbf{x}-\mathbf{x}')\varphi(t',\mathbf{x}')~, 
    \label{eqn:I_1_full}
\end{equation}
and
\begin{equation}
   \mathcal{I}_{2}=i\dfrac{\alpha}{\pi}\int_{-\infty}^{\infty}dt'\theta(t-t')\int d^{3}x'\langle[\mathbf{E}\cdot\mathbf{B}_{th}(t,\mathbf{x}),\mathbf{E}\cdot\mathbf{B}_{th}(t',\mathbf{x}')]\rangle_{0}\varphi(t',\mathbf{x}')~.
   \label{eqn:I_2_full}
\end{equation}

The quantity inside the integral is the thermal retarded Green's function. Physically, this tells us how a fluctuation of $\mathbf{E}\cdot\mathbf{B}$ in the past affects fluctuations at later times. As we shall see, we can simplify the above integral further in the small frequency limit.

\subsection{Average force on the wall}
\label{Average Force on the Wall}
We have expressed the thermal average of $\mathbf{E}\cdot\mathbf{B}$, where $\mathbf{B}$ was the sum of a background magnetic field and the thermal magnetic field that we need to calculate. Before calculating these integrals, let us first look at how to calculate the net effect of the thermal average on the wall. As we saw in the last section, the thermal average can be expressed in terms of the ALP field $\varphi$ and some other functions yet to be calculated. Hence, we have everything in terms of $\varphi$ and one can solve the entire equation exactly. On the other hand, we can treat this term as a force acting on the free wall. The motivation behind this is that the term is $(\alpha/\pi)^{2}$ suppressed and hence will not affect the overall structure and behavior of the wall at the leading order. Consider a general equation of the form,
\begin{equation}
    \partial_{t}^{2}\varphi=\partial_{x}^{2}\varphi-\dfrac{\partial V}{\partial\varphi}+F(t,x)
\end{equation}
where $F(t,x)$ comes from the $\langle F \wedge F \rangle$ term. We are interested in extracting terms that have an explicit dependence on velocity ``$v$" (i.e. not just through the $\gamma$ factor). Such terms can be interpreted as friction in a conventional manner. To find these terms, we start by writing the energy in the ALP field $``a"$ (which is nothing by $f_{a}\varphi$):
\begin{equation}
E=f_{a}^{2}A\int dx\Big(\dfrac{1}{2}\Big(\dfrac{\partial\varphi}{\partial t}\Big)^{2}+\dfrac{1}{2}\Big(\dfrac{\partial\varphi}{\partial x}\Big)^{2}+V(\varphi)\Big)~~,
\end{equation}
where $A$ is the area in the transverse direction of the wall considering the fact that $\varphi$ is dependent on a single coordinate. This approximation should be valid as long as the thickness is small compared to the radius of the wall. The change in the energy with respect to time, i.e., the power exchange is,
\begin{equation}
    \dfrac{dE}{dt}=f_{a}^{2}A\int dx\Big(\dfrac{\partial\varphi}{\partial t}\dfrac{\partial^{2}\varphi}{\partial t^{2}}+\dfrac{\partial \varphi}{\partial x}\dfrac{\partial ^{2}\varphi}{\partial t\partial x}+\dfrac{\partial V}{\partial\varphi}\dfrac{\partial \varphi}{\partial t}\Big)~~.
\end{equation}
We can then substitute the equation of motion of the field to get a function in terms of $F(t,x)$. This gives us,
\begin{equation}
    \dfrac{dE}{dt}=f_{a}^{2}A\int dx\Big(\dfrac{\partial\varphi}{\partial t}F(t,x)+\dfrac{\partial \varphi}{\partial x}\dfrac{\partial ^{2}\varphi}{\partial t\partial x}+\dfrac{\partial^{2}\varphi}{\partial x^{2}}\dfrac{\partial \varphi}{\partial t}\Big)~~.
\end{equation}
The last two terms together are derivatives. Hence rearranging the equation gives, 
\begin{equation*}
     \dfrac{dE}{dt}=f_{a}^{2}A\int dx\Big(\dfrac{\partial\varphi}{\partial t}F(t,x)+\dfrac{\partial}{\partial x}\Big(\dfrac{\partial \varphi}{\partial x}\dfrac{\partial\varphi}{\partial t}\Big)\Big)~~.
\end{equation*}
The final term is a pure boundary term. This term can be thought of as the power loss through the boundary of the system. As long as the boundary is far compared to where $\varphi$ is localized, this power loss is zero.  Hence, what we have is,
\begin{equation}
    \dfrac{dE}{dt}=f_{a}^{2}A\int dx~\dfrac{\partial\varphi}{\partial t}F(t,x)~~.
\end{equation}
If the wall is moving with velocity $v$, then, for the zero mode (or the collective coordinate of the wall), we can write $\dfrac{\partial}{\partial t}=-v\dfrac{\partial }{\partial x}$. Then, 
\begin{equation}
    \dfrac{dE}{dt}=-vf_{a}^{2}A\int dx~\dfrac{\partial\varphi}{\partial x}F(t,x)~~.
\end{equation}
Hence, the force acting on the wall is,
\begin{equation}
F_{avg}=-f_{a}^{2}A\int dx~\dfrac{\partial \varphi}{\partial x} F(t,x)
\end{equation}
giving the pressure
\begin{equation}
    P=\dfrac{F_{avg}}{A}=-f_{a}^{2}\int dx~\dfrac{\partial\varphi}{\partial x}F(t,x)~~.
    \label{eqn:pressure_on_the_wall_general}
\end{equation}
We know that $\partial_{x}\varphi$ is a locally peaked function. This function appropriately acts as a weight function to be multiplied by the force to calculate the average force.\footnote{One can note that this expression exactly matches the thermally averaged Lorentz force mentioned in Sec.(\ref{subsec:thick_walls_qualitative_picture}).} A limiting case to see is that if the wall is extremely thin, one can approximate $\partial_{x}\varphi\approx \delta(x-x_{0})$ at some time $t$. In this case, the force is nothing but $F(t,x_{0})$, as expected. An important feature of the friction pressure above is that it is proportional to the zero modes $(\partial\varphi/\partial x)$ of the wall. This feature is one of the common behaviors we have with the expression in \cite{Khlebnikov:1992bx}. A similar expression is also found in literature which tries to compute friction on a phase transition bubble such as \cite{Moore:1995ua, Megevand:2013hwa}. With Eqn.\eqref{eqn:pressure_on_the_wall_general} in mind, we return to the thermal averages to find the final expression for friction pressure.

\section{Calculating the averaged correlation functions}
\label{Calculating the Thermal Averages}
In this section, the aim is to simplify the two thermal average integrals we have in Eqns.\eqref{eqn:I_1_full} and \eqref{eqn:I_2_full}. 
. 
We drop the overall $(\alpha/\pi)$ factor from these equations while simplifying them and include it in the end for notational convenience.

\subsection{Case 1: Coherent background magnetic fields}
Let us first look at the simpler case of coherent, non-thermal, background magnetic fields. In this case, the equation at hand is,
\begin{equation}
    \mathcal{I}_{1}=i\int_{-\infty}^{\infty}dt'\theta(t-t')\int d^{3}x'\langle[E_{i}(t,\mathbf{x}),E_{j}(t',\mathbf{x}')]\rangle_{0}\xi_{ij}(t-t',\mathbf{x}-\mathbf{x}')\varphi(t',\mathbf{x}')~. 
\end{equation}
 One can write the retarded Green's function of the electric fields (i.e. $\theta(t-t')[E_{i}(t,\mathbf{x}),E_{j}(t',\mathbf{x}')]$) in terms of its Fourier transform. Further, a linear approximation of this Fourier transform simplifies the integral and can be written as,
\begin{equation}
    \begin{split}
        \mathcal{I}_{1}=-\int d^{3}x'\int\dfrac{d^{3}k}{(2\pi)^{3}}e^{i\mathbf{k}\cdot(\mathbf{x}-\mathbf{x}')}\Bigg\{\dfrac{k^{2}}{2m_D^{2}+k^{2}}P^{L}_{ij}\xi_{ij}(0,\mathbf{x}-\mathbf{x'})\varphi(t,\mathbf{x}')\\+\dfrac{m_D^{2}\pi k}{(2m_D^{2}+k^{2})^{2}}P^{L}_{ij}\partial_{t}(\xi_{ij}\varphi(t,\mathbf{x}'))\Bigg\}+\xi_{ii}(0,\mathbf{0})\varphi(t,\mathbf{x})~~.
    \end{split}  
    \label{eqn:I_1_full_2}
\end{equation}
where $m_D = eT / \sqrt{6}$. We have performed a linear approximation because the time scale for wall motion (i.e. $(\gamma m_{a})^{-1}$) is much longer than the time scales at which thermal effects such as fluctuations and screening occur. Hence, when understanding the response of the plasma on the wall at the time scales at which wall motion happens, we take the small frequency limit (or, equivalently, the long-time response). We will comment on the last term in the equation above at the end. We direct the reader to Appendix (\ref{breaking the integrals in two parts}) for details of this simplification. 

Next, we can integrate with respect to $\omega$ and integrate with respect to $t'$ to obtain
\begin{equation}
    \begin{split}
        \mathcal{I}_{1}'=-\int d^{3}x'\int\dfrac{d^{3}k}{(2\pi)^{3}}e^{i\mathbf{k}\cdot(\mathbf{x}-\mathbf{x}')}\Big\{\dfrac{k^{2}}{2m_D^{2}+k^{2}}P^{L}_{ij}\xi_{ij}(0,\mathbf{x}-\mathbf{x'})\varphi(t,\mathbf{x}')\\+\dfrac{m_D^{2}\pi k}{(2m_D^{2}+k^{2})^{2}}P^{L}_{ij}\partial_{t}(\xi_{ij}\varphi(t,\mathbf{x}'))\Big\}~~.
    \end{split}  
\end{equation}
To proceed, we assume that the rate at which the background field is changing is much slower than any other physical process, i.e., the wall motion and of course, the local thermal processes. Hence, we can take out $\xi_{ij}$ outside the time derivative. This gives us,
\begin{equation}
    \begin{split}
        \mathcal{I}_{1}'=-\int d^{3}x'\int\dfrac{d^{3}k}{(2\pi)^{3}}e^{i\mathbf{k}\cdot(\mathbf{x}-\mathbf{x}')}\Big\{\dfrac{k^{2}}{2m_D^{2}+k^{2}}P^{L}_{ij}\xi_{ij}(0,\mathbf{x}-\mathbf{x'})\varphi(t,\mathbf{x}')\\+\dfrac{m_D^{2}\pi k}{(2m_D^{2}+k^{2})^{2}}P^{L}_{ij}\xi_{ij}\partial_{t}(\varphi(t,\mathbf{x}'))\Big\}~~.
        \label{eqn:I_1_integral_full}
    \end{split}  
\end{equation}
Here, the prime notation (i.e., $\mathcal{I}_{1}'$) is just to remind that we are simplifying only the integrals in $\mathcal{I}_{1}$ and will be including the final term in Eqn.\eqref{eqn:I_1_full_2} at the end. The above integral can be rewritten as 
\begin{equation}
\mathcal{I}_{1}'=\mathcal{I}_{1,1}+\mathcal{I}_{1,2}
\end{equation}
where $\mathcal{I}_{1,1}$ and $\mathcal{I}_{1,2}$ are in the first and the second integral respectively in Eqn.\eqref{eqn:I_1_integral_full}. These integrals can be written in a simpler manner as,
\begin{equation}
    \mathcal{I}_{1,2}=\dfrac{m_D^{2}}{3\pi}\partial_{t}\partial_{x_{\perp}}^{2}\varphi(t,x_{\perp})\int_{0}^{\infty}dk \dfrac{k}{(2m_D^{2}+k^{2})^{2}}\eta(k)~~.
    \label{eqn:I_1_2_full}
\end{equation}
and,
\begin{equation}
    \mathcal{I}_{1,1}=\dfrac{1}{3\pi^{3}}\partial_{x_{\perp}}^{2}\varphi(t,x_{\perp})\int_{0}^{\infty}\dfrac{k^{2}}{2m_D^{2}+k^{2}}\eta(k)~~.
\end{equation}
Here, $\eta(k)$ is as defined in Eqn.\eqref{eqn:background_field_power_spectrum}. We direct the reader to Appendix (\ref{Appendix:Simplification of integral I_1}) for further details. 

Note that when finding the effective force on the wall, as shown in section~\ref{Average Force on the Wall}, we need to integrate the above quantities with respect to $x_{\perp}$ along with the weight function $(\partial\varphi/\partial x)$. One can check that the odd derivatives of the function $\varphi$ are even whereas the even derivatives are odd functions of $x_{\perp}$. In this case, the integral containing $\mathcal{I}_{1,1}$ vanishes as it is an odd function whereas $\partial_{x}\varphi$ is an even function. Hence, the term of interest in this case is Eqn.\eqref{eqn:I_1_2_full}. Finally, we need to discuss the final term in Eqn.\eqref{eqn:I_1_full} which arises from the $\delta$-function in Eqn.\eqref{eqn:electric_field_thermal_corrected_retarded_green}. Including all the multiplicative factors of $\alpha$ and $f_{a}$ gives the contribution, 
\begin{equation}
    \Big(\dfrac{\alpha}{\pi}\Big)^{2}\dfrac{1}{f_{a}^{2}}\xi_{ii}({0,\mathbf{0}})\varphi(t,x)~~.
\end{equation}
From the tensor structure, we can see that, 
\begin{equation*}
    \xi_{ii}(0,\mathbf{0})=2\int\dfrac{d^{3}k}{(2\pi)^{3}}\eta(k)
\end{equation*}
This is nothing but the Gaussian average $\langle B^{2}_{bg}\rangle_{gaussian}$ which is the local energy density stored in the background magnetic field. Following~\cite{Diaz-Gil:2007fch}, we are considering the limit where this energy density is small compared to the thermal energy density, i.e. $\langle B^{2}_{bg}\rangle_{gaussian}\lesssim T^{4}$.  One can think of this latter expression as a correction to the mass which is small compared to $m_{a}^{2}$. This can be seen as follows: the ratio of $m_{a}^{2}$ to the coefficient in the equation above,
\begin{equation}
   \Big(\dfrac{\alpha}{\pi}\Big)^{2} \dfrac{\xi_{ii}(0,\mathbf{0})}{m_{a}^{2}f_{a}^{2}}< \Big(\dfrac{\alpha}{\pi}\Big)^{2}\dfrac{T^{4}}{\Lambda^{4}}
\end{equation}
where $\Lambda$ is the scale of domain wall formation. This statement assumes the fact the axion potential is derived from a confinement mechanism\footnote{As discussed in the introduction, the potential comes when the axion couples to a gauge group with the term $a\Tr[G\wedge G]$ in the UV}. Hence, we are interested in temperature $T\lesssim\Lambda$ which immediately implies that the ratio above is small. 

We now return to simplifying Eqn.\eqref{eqn:I_1_2_full}. If we substituting the wall profile in that equation, we get
\begin{equation}
\label{eqn:simplified_I_1_2}
\partial_{t}\partial^{2}_{x_{\perp}}\varphi(t,x_{\perp})=-4(\gamma m_{a})^{3}v\Big(\dfrac{8e^{5y}}{(1+e^{2y})^{3}}-\dfrac{8 e^{3y}}{(1+e^{2y})^{2}}+\dfrac{e^{y}}{1+e^{2y}}\Big),~~y=\gamma m_{a}(x-vt)~.
\end{equation}
To calculate the force, we need to integrate the product of the above function with $(\partial\varphi/\partial x)$. This product is plotted below in Fig.(\ref{I_1_2_phi}). The result of this integration is negative.
There are two further negative signs, one coming from $``-v"$ (time derivative of the field) and the second is the negative sign in Eqn.\eqref{eqn:pressure_on_the_wall_general} which cancel each other. The overall force
is negative and hence a frictional force. The magnitude of this force (per unit area) is 
\begin{equation}
    P_{bg}= \Big(\dfrac{\alpha}{\pi}\Big)^{2}\dfrac{1}{18\pi}m_D^{2}(\gamma m_{a})^{3}\int_{0}^{\infty}dk \dfrac{k}{(2m_D^{2}+k^{2})^{2}}\eta(k)~~.
    \label{eqn:pressure_due_to_background_field}
\end{equation}
The expression depends on the power spectrum of the background magnetic fields. Let us look at a simple toy case where
\begin{equation}
    \eta(k)=c_{0}T^{2}\delta(k-k_{0})~~,
\end{equation}
where $k_{0}\ll eT$ and $c_{0}$ is a constant. Here the reason for choosing $k_{0} \ll eT$ is motivated by \cite{Diaz-Gil:2007fch} where it is shown that the coherence length of these magnetic fields is very large compared to the thermal scales. This choice of $\eta(k)$ means that there is rouhgly one dominant spatial mode in the power spectrum of the coherent magnetic fields. The pre-factor $T^{2}$ is chosen to satisfy the dimensionality of $\eta(k)$ and seems to be a natural choice to compare to the results from the thermal fluctuation part. Further, we consider $c_{0}$ to be a small number, again taking motivation from \cite{Diaz-Gil:2007fch} where numerical simulations suggest that the power stored in the background magnetic fields is an order of magnitude smaller than the energy stored in the thermal magnetic fields. 
In this case, 
\begin{equation}
    P_{bg}=\Big(\dfrac{\alpha}{\pi}\Big)^{2}\dfrac{1}{18\pi}c_{0}m_D^{2}(\gamma m_{a})^{3}T^{2}\dfrac{k_{0}}{(2m_D^{2}+k_{0}^{2})^{2}}~~.
\end{equation}
Here, we recall that the Debye mass is $m_D=eT/\sqrt{6}$. In the limit where $k_{0}\ll eT$, i.e., the magnetic field is coherent at scales larger than the Debye length, this gives 
\begin{equation}
    P_{bg}=\alpha c_{0}\dfrac{1}{288 \pi^{4}}(\gamma m_{a})^{3}k_{0}~~.
\end{equation} 
As we will see, this is less than the contribution from thermal magnetic fields $\mathbf{B}_{th}$. This pressure roughly becomes equal to the thermal part (as we shall see in later sections) when the coherence length of the magnetic field is of the order of the Debye length. 

\begin{figure}[h!]
    \centering
    \includegraphics[width=8cm, height=5cm]{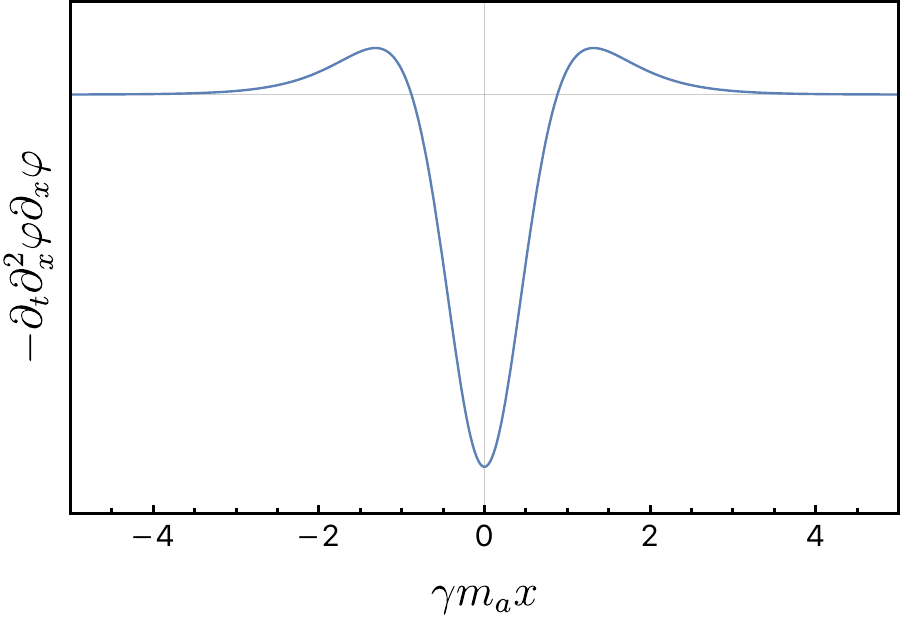}
\caption{The frictional pressure density given by the integrand of Eqn.\eqref{eqn:pressure_on_the_wall_general} when $F(t,x)$ is the contribution from long-range coherent magnetic fields, as in~\eqref{eqn:simplified_I_1_2}. This is plotted against $\gamma m_{a}x$. The distribution is scaled by a constant factor and is plotted at $t=0$. Choosing a different time just shifts the plot along the $x$ axis.}
    \label{I_1_2_phi}
\end{figure}
\subsection{Case 2: Thermal magnetic fields}
\label{integral 2}
Now that we have simplified the expression for the contribution from a background magnetic field, we move on to the thermal magnetic field contribution. In this section, we calculate the integral corresponding to the thermal part of the magnetic field given by:
\begin{equation}
     \mathcal{I}_{2}=i\int_{-\infty}^{\infty}dt'\theta(t-t')\int d^{3}x'\langle[\mathbf{E}\cdot\mathbf{B}_{th}(t,\mathbf{x}),\mathbf{E}\cdot\mathbf{B}_{th}(t',\mathbf{x}')]\rangle_{0}\varphi(t',\mathbf{x}')~.
\end{equation}
To simplify this, let us define the Retarded Green's function
\begin{equation}
    \mathcal{G}^{R}_{th}(t-t',\mathbf{x}-\mathbf{x}')=i\theta(t-t')\langle[\mathbf{E}\cdot\mathbf{B}_{th}(t,\mathbf{x}),\mathbf{E}\cdot\mathbf{B}_{th}(t',\mathbf{x}')]\rangle_{0}~~.
\end{equation}
Then in this notation, the integral of interest takes the form
\begin{equation*}
\mathcal{I}_{2}=\int dt'\int d^{3}x' \mathcal{G}^{R}_{th}~~\varphi(t',\mathbf{x}')~~.
\end{equation*}
We can express the retarded Green's function in terms of its Fourier transform. This can be written as
\begin{equation}
   \mathcal{I}_{2}= \int_{-\infty}^{\infty} dt'\int d^{3}x'\int_{-\infty}^{\infty}\dfrac{d\omega}{2\pi}\int \dfrac{d^{3}k}{(2\pi)^{3}}e^{-i\omega(t-t')}e^{i\mathbf{k}\cdot(\mathbf{x}-\mathbf{x}')}\widetilde{\mathcal{G}}^{R}_{th}(\omega,\mathbf{k})\varphi(t',\mathbf{x}')~~.
\end{equation}
As before, we expand the Fourier transform of the retarded Green's function around $\omega=0$. The integral becomes
\begin{equation}
    \mathcal{I}_{2}= \int_{-\infty}^{\infty} dt'\int d^{3}x'\int_{-\infty}^{\infty}\dfrac{d\omega}{2\pi}\int \dfrac{d^{3}k}{(2\pi)^{3}}e^{-i\omega(t-t')}e^{i\mathbf{k}\cdot(\mathbf{x}-\mathbf{x}')}\Big(\widetilde{\mathcal{G}}^{R}_{th}(0,\mathbf{k})+\omega\dfrac{\partial\widetilde{\mathcal{G}}^{R}_{th}}{\partial\omega}(0,\mathbf{k})\Big)\varphi(t',\mathbf{x}')~~. 
\end{equation}
Taking the inverse Fourier transform with respect to $\omega$ and then integrating with respect to $t'$, we get
\begin{equation}
\begin{split}
   \mathcal{I}_{2}=\int d^{3}x'\int\dfrac{d^{3}k}{(2\pi)^{3}}e^{i\mathbf{k}\cdot(\mathbf{x}-\mathbf{x}')}\Big\{\widetilde{\mathcal{G}}^{R}_{th}(0,\mathbf{k})\varphi(t,\mathbf{x}')+i\dfrac{\partial\widetilde{\mathcal{G}}^{R}_{th}}{\partial\omega}(0,\mathbf{k})\partial_{t}\varphi(t,\mathbf{x}')\Big\} ~~. 
\end{split}
\end{equation}
As we saw in the earlier calculation, the friction comes from that part of the integral that has a time derivative acting on the wall field $\varphi$. The first term in the equation above is like we had in Sec.(\ref{Calculating the Thermal Averages}), a thermal correction to the mass and is very small below the confinement scale of the gauge group that gives the ALP scalar field the periodic potential. That said, we concentrate on the second term of the integral. This can be simplified as
\begin{equation}
\mathcal{I}_{2,2}=-\dfrac{1}{2T}\int d^{3}x'\int\dfrac{d^{3}k}{(2\pi)^{3}}e^{i\mathbf{k}\cdot(\mathbf{x}-\mathbf{x}')}\rho_{0}(0,\mathbf{k})\partial_{t}\varphi(t,\mathbf{x}')~~,
\end{equation}
where 
\begin{equation}
    \rho_{0}(\omega,\mathbf{k})=\mathsf{FT}\Big(\langle \mathbf{E}\cdot\mathbf{B}_{th}(t,\mathbf{x})~\mathbf{E}\cdot\mathbf{B}_{th}(t',\mathbf{x}')\rangle_{0}\Big)~~.
\end{equation}
Refer to Appendix (\ref{appendix:retarded_greens_function_general_operator}) for the derivation. Here, $\rho_{0}$ is the Fourier transform of the correlation between two thermal fluctuations. One can get a physical intuition of the equation above by rewriting it as :
\begin{equation*}
   \mathcal{I}_{2,2}=-\dfrac{1}{2T}\int d^{3}x'\partial_{t}\varphi(t,\mathbf{x'})\int dt''\langle \mathbf{E}\cdot\mathbf{B}_{th}(t'',\mathbf{x}-\mathbf{x}')~\mathbf{E}\cdot\mathbf{B}_{th}(0,0)\rangle_{0} ~~.
\end{equation*}
A thermal fluctuation, including particles, collective excitations, and quasi-particles has a lifetime. Hence, a fluctuation of $\mathbf{E}\cdot\mathbf{B}$ that happens at $t=0$ will be strongly correlated for some small time to the other local fluctuations. Further, the correlation of the two thermal fluctuations depends on their spatial separation. As the electric fields are screened at the Debye scale, one can see that the fluctuations at longer distances will be weakly correlated. On the other hand, fluctuations at scales smaller than the Debye length will tend to have a stronger correlation. Thermal fluctuations also decay which can be because of collisions or other damping thermal processes in plasma. Qualitatively speaking, the second term in the integral will depend on these thermal quantities.

To simplify $\mathcal{I}_{2,2}$, we use the flat wall approximation where we choose the normal coordinate to the wall to be the $z$ direction without loss of generality. This gives us
\begin{equation}
   \mathcal{I}_{2,2}=-\dfrac{1}{2T}\int d^{3}x'\int\dfrac{d^{3}k}{(2\pi)^{3}}e^{i\mathbf{k}\cdot(\mathbf{x}-\mathbf{x}')}\rho_{0}(0,\mathbf{k})\partial_{t}\varphi(t,z')~~.
\end{equation}
Integrating with respect to $(x',y')$ followed by integration with respect to $(k_{x},k_{y})$ simplifies the integral to
\begin{equation}
    \mathcal{I}_{2,2}=-\dfrac{1}{2T}\int dz'\int\dfrac{dk_{z}}{2\pi}e^{i k_{z}(z-z')}\rho_{0}(0,k_{z})\partial_{t}\varphi(t,z')~~.
    \label{eqn:I_2_2}
\end{equation}
Now, the task is to find $\rho_{0}(0,k_{z})$ which we get from calculating:
\begin{equation}
    \rho_{0}(\omega,\mathbf{k})=\dfrac{1}{64}\mathsf{FT}\Big(\epsilon^{\mu\nu\sigma\tau}\epsilon^{\alpha\beta\gamma\delta}\langle \partial_{\mu}A_{\nu}\partial_{\sigma}A_{\tau}\partial'_{\alpha}A_{\beta}\partial'_{\gamma}A_{\delta}\rangle\Big)~~,
\end{equation}
where we have expressed $\mathbf{E}\cdot\mathbf{B}_{th}$ in a covariant way. Now, if we substitute
\begin{equation}
    F(t,x)=-\Big(\dfrac{\alpha}{\pi}\Big)^{2}\dfrac{1}{f_{a}^{2}}\mathcal{I}_{2,2},~~
\end{equation}
in Eqn.\eqref{eqn:pressure_on_the_wall_general}, where $\mathcal{I}_{2,2}$ is given by Eqn.\eqref{eqn:I_2_2}, we see that there is a negative sign coming from $\partial_{t}\varphi=-v\partial_{z'}\varphi$. We therefore have
\begin{equation}
    P=-\dfrac{1}{2T}\Big(\dfrac{\alpha}{\pi}\Big)^{2} v\int_{-\infty}^{\infty}\dfrac{d k_{z}}{2\pi}g^{2}(k_{z}) \rho_{0}(0,k_{z})~~.
\end{equation}
Here, $g(k_{z})$ is the Fourier transform of $\partial_{z}\varphi$ and we shall see that $\rho_{0}(0,k)$ is a positive function of $k$. Hence the integral is positive and the entire expression is negative signaling that this is indeed an opposing force. Keeping this in mind, from now on, we talk only about the magnitude of $P$. Hence,  we have
\begin{equation}
    P=\dfrac{1}{2T}\Big(\dfrac{\alpha}{\pi}\Big)^{2} v\int_{-\infty}^{\infty}\dfrac{d k_{z}}{2\pi}g^{2}(k_{z}) \rho_{0}(0,k_{z})~~.
    \label{eqn:friction_in_terms_of_g}
\end{equation}
In the approximation that the wall is locally flat where we can use Eqn.\eqref{eqn:moving_flat_wall}, the Fourier transform looks like  
\begin{equation}
    g(k_{z})=2\pi\text{Sech}\Big(\dfrac{k_{z}\pi}{2\gamma m_{a}}\Big)~~.
\end{equation}
\begin{figure}[h!]
    \centering
    \includegraphics[width=8cm, height=5cm]{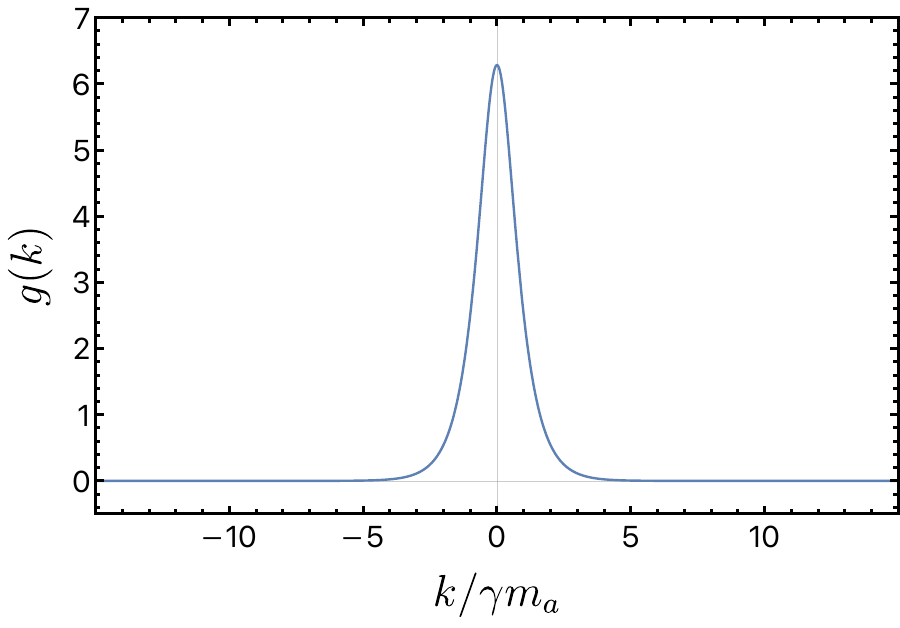}
    \caption{Fourier transform of $(\partial\varphi/\partial x)$, i.e, $g(k)$. This configuration has a characteristic width that is proportional to $\gamma m_{a}$.}
    \label{fig:Wall_derivative_Fourier}
\end{figure}
\\
This gives
\begin{equation}
    P=\Big(\dfrac{\alpha}{\pi}\Big)^{2} \dfrac{\pi v}{T}\int_{-\infty}^{\infty}d k_{z}~ \text{Sech}^{2}\Big(\dfrac{k_{z}\pi}{2\gamma m_{a}}\Big)\rho_{0}(0,k_{z})~~.
    \label{eqn:pressure}
\end{equation}
Hence, the final expression depends on the spectral function $\rho_{0}(0,k)$. This function basically encodes information about the plasma back-reaction on the wall. The details of calculating this function are given in Appendix \ref{Appendix:Calculating the Spectral density}. The expression of $\rho_{0}(0,k)$ where $\mathbf{k}=(0,0,k)$ is given by
\begin{equation}
\begin{split}
 \rho_{0}(0,k)=\dfrac{1}{256\pi^{3}}k^{2}\int_{0}^{\pi}d\theta\sin\theta\int_{0}^{\infty}dp~p^{2}\int_{-\infty}^{\infty}dp_{0}\Big(2\cos^{2}\theta p_{0}^{2}\Delta^{T}(p)\Delta^{T}(k-p)\\+\sin^{2}\theta (p_{0}^{2}-p^{2})(\Delta^{T}(p)\Delta^{L}(k-p)+\Delta^{T}(k-p)\Delta^{L}(p))\Big)
\end{split}
\label{eqn:spectral_function_rho_0}
\end{equation}
In the remaining part of this section, we are going to see how we get pressure for different choices of the $\Delta(p)$ propagator. In particular, the function $\Delta(p)$ depends on the thermal corrections through the spectral function of the photon. 
\subsubsection{Purely U(1) fields}
\begin{figure}[h!]
        \centering
        \includegraphics[width=10cm,height=5cm]{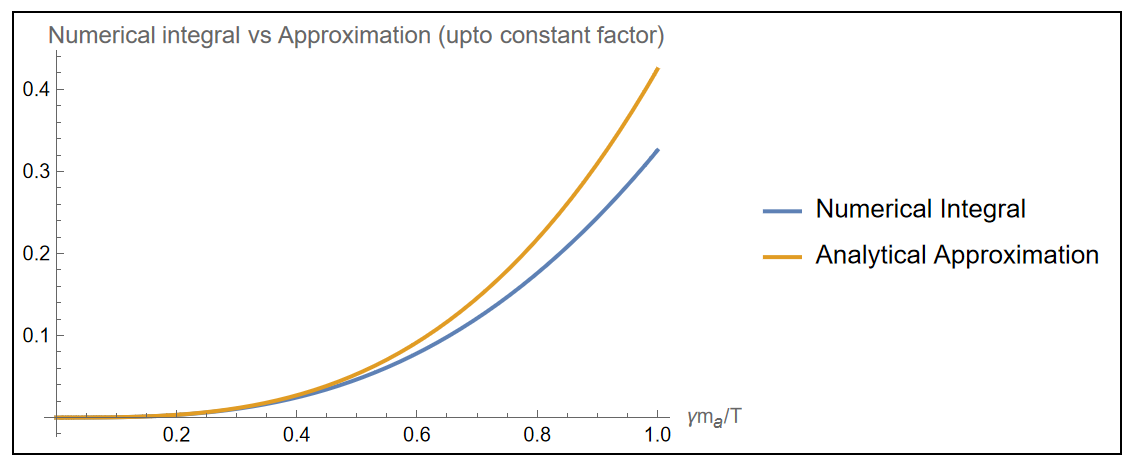}
        \caption{Comparision between the full integral performed numerically and the approximation in the limit $\beta \gamma m_{a}<1$. We can see that the approximation is of the same order as the full numerical integral when $\gamma m_{a}<T$.}
        \label{full_integral_vs_approximation}
\end{figure}

Let us first look at the easiest case where we have bare photon propagators. This is a warm-up example and we will do a more realistic case later. Looking at this simple case serves a dual purpose: (a) It gives us an idea of how the friction calculation is done and (b) this calculation will also show why the order in which effects are averaged is important. In a more realistic case, one needs to understand the effects of fermion thermal loops that give photons a thermal mass and an associated damping factor. The bare photon propagators take the form:
\begin{equation}
\Delta^{T}(p)=\Delta^{L}(p)=\Delta(p)=(1+f(p_{0}))\rho(p_{0})=(2\pi)(\Theta(p_{0})+n_{B}(p_{0}))\delta(p_{0}^{2}-p^{2})~~.
\label{eqn:D_12_free}
\end{equation}
Here, $f(p_{0})=1/(\exp(\beta p_{0})-1)$, $\Theta$ is the Heaviside function and $\rho(p)$ is the spectral density of the photon\footnote{Not to be confused with $\rho_{0}$ which is given in Eqn.\eqref{eqn:spectral_function_rho_0}. More comments on this function will be made in the next subsection}. As mentioned before, this is nothing but the $D_{21}$ type propagator in the real-time thermal field theory\cite{Bellac:2011kqa} \footnote{The $D_{21}$ type propagators are also referred as $D^{>}(t-t'-i\sigma)$ propagators where $\sigma\in [0,\beta]$ is the displacement of the contour along which the path integral is done in the Schwinger-Keldysh formalism. Here, as we want the Fourier transform $\mathsf{FT}(\langle A(t) A(t')\rangle)$, this corresponds to $\mathsf{FT}(D^{>}(t-t'))$ and hence $\sigma=0$ is the right choice}. Using this, we can write 
\begin{equation}
  \begin{split}
 \rho_{0}(0,k)=\dfrac{1}{128\pi^{3}}k^{2}\int_{0}^{\pi}d\theta\sin\theta\int_{0}^{\infty}dp~p^{2}\int_{-\infty}^{\infty}dp_{0}\Delta(p)\Delta(k-p)\Big( p_{0}^{2}-p^{2}\sin^{2}\theta\Big)
\end{split}  
\end{equation}
Note that $k^{\mu}=(0,0,0,k)$. Keeping this in mind, the expression can be simplified as follows:
\begin{equation}
\Delta(k)\Delta(k-p)=(2\pi)^{2}\delta(p_{0}^{2}-p^{2})\delta(p_{0}^{2}-(p^{2}+k^{2}-2pk\cos\theta))(n_{B}(p_{0})+n_{B}(p_{0})^{2})
\end{equation}
First we integrate with respect to $p_{0}$ which removes the $\delta(p_{0}^{2}-p^{2})$ function. This gives 
\begin{equation}
    \begin{split}
        \rho_{0}(0,k)=\dfrac{1}{32\pi}k^{2}\int_{0}^{\pi}d\theta\sin\theta\int_{0}^{\infty} dp p^{3}\cos^{2}\theta(n_{B}(p)+n_{B}(p)^{2})\delta(k^{2}-2pk\cos\theta)~~.
    \end{split}
\end{equation}
Substituting $\cos\theta=x$, the above equation simplifies to
\begin{equation}
  \rho_{0}(0,k)=\dfrac{1}{32\pi}k^{2}\int_{0}^{\infty} dp p^{3}\int_{-1}^{1}dx~x^{2}(n_{B}(p)+n_{B}(p)^{2})\delta(k^{2}-2pkx)~~.
\end{equation}
As long as $\vert k/2p \vert<1$, this integral is not zero as the $\delta$ function would have it's peak in the range $[-1,1]$. So, this tells us that for any function $f(x)$ and a non-zero $k$
\begin{equation}
    \int_{-1}^{1}dx f(x)\delta(k^{2}-2pkx)= f\left(\dfrac{k}{2p}\right)\dfrac{1}{2\vert k\vert p}\Theta\left(p-\dfrac{\vert k\vert}{2}\right)
\end{equation}
Using this we get
\begin{equation}
    \rho_{0}(0,k)=\dfrac{1}{256\pi}k^{2}\vert k\vert\int_{\vert k\vert/2}^{\infty}dp  (n_{B}(p)+n_{B}(p)^{2})~~.
\end{equation}
Substituting
\begin{equation}
    n_{B}(x)=\dfrac{1}{e^{\beta\vert x\vert}-1},~~\beta=T^{-1}
\end{equation}
we arrive at
\begin{equation}
\rho_{0}(0,k)= \dfrac{1}{256\pi} k^{2}\vert k\vert \dfrac{1}{\beta(e^{\beta \vert k\vert/2}-1)}~~.
\end{equation}
If we look at the Fourier transform $g(k)$ in Fig.(\ref{fig:Wall_derivative_Fourier}), it looks like a Gaussian function peaked at zero with a width. The width is proportional to $\gamma m_{a}$. So a simple approximation is to replace $\vert k\vert \approx \gamma m_{a}$ in the above equation. This can also be seen in Fig.(\ref{full_integral_vs_approximation}). Hence, we can simplify the exponential function in $\rho_{0}(0,k)$ in the limit $\beta  \gamma m_{a}\lesssim 1 $ as follows
    \begin{equation}
        \rho_{0}(0,k)=\dfrac{1}{128\pi}T^{2}k^{2}~~.
    \end{equation}
In this limit, using \eqref{eqn:pressure}, we get
    \begin{equation}
        P=\Big(\dfrac{\alpha}{\pi}\Big)^{2}\dfrac{ Tv}{128}\int_{-\infty}^{\infty} dk~ k^{2}\text{Sech}^{2}\Big(\dfrac{k \pi}{2 \gamma m_{a}}\Big)~~.
    \end{equation}
By rearranging a few terms, this integral simplifies to 
\begin{equation}
       P=\Big(\dfrac{\alpha}{\pi}\Big)^{2}\dfrac{1}{192 \pi} Tv(\gamma m_{a})^{3}~~.  
\end{equation}

One interesting feature of this expression is that it matches the thin wall answer up to the $\gamma^{3}$ factor. The reason for this is as follows: when only considering purely U(1) gauge fields and no fermions, there is no interaction between the particles in the plasma\footnote{Here, what we mean is that if there was no wall, there would not be any collisions is the plasma as there is no self-interaction in a pure U(1) gauge theory.}. So the mean free path is infinite. In fact, there is no thermal scale smaller than the wall thickness. Hence, the correct thing to do here is to integrate the wall effects first and then take the thermal average. In this way, we have already made the wall thin. On the other hand, the above calculations take into account the length contraction of the wall in the plasma frame. This is a relevant effect to consider when there are interactions in the plasma and makes sense only when the wall thickness is not the smallest scale. Hence, the $\gamma^{3}$ factor here is an excess factor. The remaining expression of the friction pressure matches the parametric dependence on $m_{a}$ and $T$ as one gets from the thin wall calculation. This parametric form is solely dependent on the nature of interaction and at small velocities $(\gamma\sim 1)$, in the absence of any thermal scales like the Debye length and mean-free path, the linear response method coincides with the particle scattering as shown in \cite{Arnold:1993wc}. 
\subsubsection{Effects of Fermions charged under the U(1) in the plasma}
When evaluating $\rho_{0}(0,k)$ above, we ignored the fact that there are fermions charged under the U(1) photon in the plasma. Let us now take one step forward and account for the fermions. These affect the $D_{21}$ propagators such that instead of $\delta(p_{0}^{2}-p^{2})$ in Eqn.\eqref{eqn:D_12_free}, one has to use the thermally corrected $\rho^{T/L}(p)$ (the spectral function) derived from the imaginary part of the thermally corrected retarded propagators \cite{Carrington:1997sq}. To elaborate, the spectral function of a free photon used in \eqref{eqn:D_12_free} takes the form
\begin{equation*}
    \rho(p_{0},p)=(2\pi)\text{sgn}(p_{0})\delta(p_{0}^{2}-p^{2})~.
\end{equation*}
This is nothing but the imaginary part of the retarded Green's function that takes the form
\begin{equation*}
    \mathcal{G}_{R}(p_{0},p)\sim \dfrac{1}{p_{0}^{2}-p^{2}+ip_{0}\epsilon}
\end{equation*}
Note that $\rho$ satisfies a couple of general properties: (a) it is an odd function of $p_{0}$, (b) it satisfies positivity condition, i.e., $\text{sgn}(p_{0})\rho>0$ \cite{Bellac:2011kqa}. Due to the thermal corrections, the retarded propagator picks up non-trivial terms in the denominator, changing the structure of $\rho$. For gauge fields, the thermal corrections due to fermions are different for the transverse and longitudinal components and are momentum-dependent. The exact equations of these corrections can be found in \cite{ Bellac:2011kqa, Kapusta:1989tk, Carrington:1997sq}. Using these corrections, one can get a full frequency-momentum dependent expression for $\rho$. These equations are complicated to simplify. Using these exact equations will make it difficult to extract the analytical behavior of the friction. Instead of solving the exact problem, we model this problem by taking two toy examples. 
\subsubsection*{Toy Model 1}
The thermal corrections give rise to a photon mass. This comes from the fact that the long-range force law for the photon is exponentially suppressed \cite{Kapusta:1989tk, Bellac:2011kqa}. This thermal mass is exactly $m_D = \lambda_D^{-1}$. Hence, we first\footnote{This model is actually a limiting case of the toy example to follow. Specifically, this is the small damping rate limit (i.e. $\Gamma \rightarrow 0$) of the toy model we show next.} consider a toy model where
\begin{equation}
    \rho(p)^{T/L}=(2\pi)\text{sgn}(p_{0})\delta(p_{0}^{2}-p^{2}-m_{D}^{2}),~~m_{D}\sim \lambda_{d}^{-1}\sim eT~~.
\end{equation}
This might not be the most accurate choice to get the exact numerical factors but, qualitatively, it is a reasonable approximation to give photons a temperature-dependent mass. Then, similar to the earlier case, one gets
\begin{equation}
    \begin{split}
        \rho_{0}(0,k)=\dfrac{1}{32\pi}k^{2}\int_{0}^{\pi}d\theta\sin\theta\int_{0}^{\infty} dp\dfrac{p^{2}}{\sqrt{p^{2}+m_{D}^{2}}}(p^{2}\cos^{2}\theta+m_{D}^{2})(n_{B}(p)+n_{B}(p)^{2})\\\delta(k^{2}-2pk\cos\theta)~~.
    \end{split}
\end{equation}
This simplifies to
\begin{equation}
  \rho_{0}(0,k)=\dfrac{1}{64 \pi}k^{2}\Big(\dfrac{k^{2}}{4}+m_{D}^{2}\Big)\dfrac{1}{\vert k\vert}\int_{\vert k\vert/2}^{\infty} dp \dfrac{p}{\sqrt{p^{2}+m_{D}^{2}}}(n_{B}(p)+n_{B}(p)^{2})
\end{equation}
with
\begin{equation}
    n_{B}(p)=\dfrac{1}{e^{\beta \sqrt{p^{2}+m_{D}^{2}}}-1},~~\beta=T^{-1}.
\end{equation}
As we saw earlier, integrating $\rho_{0}(0,k)$ with the Fourier transform of $\partial_{z}\varphi$ is equivalent to replacing each $\vert k\vert$ with $\gamma m_{a}$ (up to small numerical factors). Hence, we can look at the behavior of the integral in different regimes
\begin{itemize}
    \item $\underline{m_{D}=0:}$ \\
    This case is purely photon radiation without HTL corrections as we saw before. This gives
    \begin{equation}
    \rho_{0}(0,k)=\dfrac{1}{256\pi}k^{2}\dfrac{\vert k\vert}{\beta(e^{\beta\vert k\vert/2}-1)}
    \end{equation}
    
    \item $\underline{m_{D}\neq0, \text{ with } \gamma m_{a}<m_{D}:}$\\
    In this case
    \begin{equation}
  \rho_{0}(0,k)\approx \dfrac{1}{64\pi}\vert k\vert m_{D}^{2}\int_{0}^{\infty} dp \dfrac{p}{\sqrt{p^{2}+m_{D}^{2}}}(n_{B}(p)+n_{B}(p)^{2})
\end{equation}
We can write 
\begin{equation}
\begin{split}
   \int_{0}^{\infty} dp \dfrac{p}{\sqrt{p^{2}+m_{D}^{2}}}(n_{B}(p)+n_{B}(p)^{2})=\dfrac{1}{\beta}\int_{0}^{\infty}dx~\dfrac{x}{\sqrt{x^{2}+\beta^{2}m_{D}^{2}}}\Big(\dfrac{1}{e^{\sqrt{x^{2}+\beta^{2}m_{D}^{2}}}-1}\\+\dfrac{1}{(e^{\sqrt{x^{2}+\beta^{2}m_{D}^{2}}}-1)^{2}}\Big)   
\end{split}
\end{equation}
Now, $\beta m_{D}\sim e\sim 0.3$ and the integral is 
\begin{equation*}
    \int_{0}^{\infty}dx~\dfrac{x}{\sqrt{x^{2}+\beta^{2}m_{D}^{2}}}\Big(\dfrac{1}{e^{\sqrt{x^{2}+\beta^{2}m_{D}^{2}}}-1}\\+\dfrac{1}{(e^{\sqrt{x^{2}+\beta^{2}m_{D}^{2}}}-1)^{2}}\Big)\approx 2.85.
\end{equation*}
This gives us
\begin{equation}
\rho_{0}(0,k)\approx \dfrac{3}{64 \pi}\vert k\vert e^{2}T^{3}~~.
\end{equation}
Hence, up to a numerical constant
\begin{equation}
    P\propto \Big(\dfrac{\alpha}{\pi}\Big)^{2}\alpha (\gamma m_{a})^{2}T^{2}v.
\end{equation}
This is different from the expression for pressure in the thin wall case~\eqref{eqn:scatteringFriction}. The difference is because this toy model captures collective effects rather than just particle scattering. 
\item $\underline{m_{D}\neq 0 \text{ with } m_{D}\lesssim \gamma m_{a}:}$, \\
This case is similar to that of $m_{D}\approx 0$ as only the very high momenta contribute to the integral. This can be interpreted as the self-crossing time is fast compared to the response time of the charged particles in the plasma to neutralize the effect of the wall. Hence, the only effect on the wall is the purely high momentum modes of the photons which act like almost free photons. 
\end{itemize}
\subsubsection*{Toy Model 2 --- Including the Decay Width}
In the calculations above, we considered the collective modes to be some particles with mass $m_{D}$. This assumption included the fact that the thermal corrections are real. But this is a very crude approximation because thermal corrections do have an imaginary part \cite{Bellac:2011kqa, Bodeker:2022ihg}. The imaginary part is usually interpreted as the decay width of the modes which says that excited modes are not stable and will eventually decay. Now, the approximation we made is a good approximation as long as the wall self-crossing time $\sim (\gamma m_{a})^{-1}$  is small compared to the decay time of the collective modes $\Gamma^{-1}$. In general, the spectral function takes the form
\begin{equation*}
    \rho(p_{0},p)\sim \dfrac{f_{2}(p_{0},p)}{(p_{0}^{2}-p^{2}-f_{1}(p_{0},p))^{2}+f_{2}^{2}(p_{0},p)}~~.
\end{equation*}
Here, $f_{1}$ and $f_{2}$ contain the information about the thermally corrected part. As mentioned before, using the exact functions will make it difficult to take out any analytical information from the expression. Hence, we use a very similar toy model as follows. 

Consider a scalar field $\psi$ which has a damping factor associated with it\footnote{Here, we are considering $\Gamma<m$. This is true in a thermal plasma where the damping rate is much smaller than the plasma frequency.}. Then, the classical equation of motion looks like, 
\begin{equation}
    (\Box+m_{0}^{2}+\Gamma\partial_{t})\psi=0~~.
\end{equation}
The Green's function of this field $\psi$ takes the form,
\begin{equation}
    G_{\psi}(\omega,\mathbf{k})\sim\dfrac{1}{\omega^{2}-k^{2}-m_{0}^{2}+i\Gamma\omega}~~.
\end{equation}
The spectral function is $\rho=-\dfrac{1}{\pi}\text{Im}(G_{\psi})$ and hence takes the form
\begin{equation}
    \rho_{\psi}(\omega,k) =\dfrac{1}{\pi}\dfrac{\Gamma\omega}{(\omega^{2}-k^{2}-m_{0}^{2})^{2}+\Gamma^{2}\omega^{2}}~~.
\end{equation}
This is also a typical type of functional form a spectral function takes in a \emph{thermal bath} during the thermalization process and is also called an Ornstein-Zernike approximation\cite{Calzetta:2008iqa}. Note that if the particle would have been stable or metastable for a long time compared to any other physical process we are interested in then one can approximate $\Gamma\rightarrow 0$. In this case, the limit we recover is
\begin{equation}
    \lim_{\Gamma\rightarrow 0}\rho_{\psi}(\omega,k)=\text{sgn}(\omega)\delta(\omega^{2}-k^{2}-m_{0}^{2})~~,
\end{equation}
which has the form of an `on-shell' condition. Various studies have been done to understand the damping factor of collective excitations in plasma both in QED and QCD\footnote{The modes that follow bosonic distribution are usually called collective excitations and those that follow fermionic distribution are called quasi-particles in the literature.}. In QCD, the damping factor associated with the gluons is proportional to $g^{2}T$ (or, $\alpha_{s}T$) and is large. On the other hand, the damping factor associated with soft photons (modes with $p\lesssim eT$) is of the order $e^{4}T\sim \alpha^{2}T$ \cite{Abada:2011cc}. Further, even for hard photons ($p\sim T$), at leading order, the rate is proportional to $\alpha^{2}T$\cite{Thoma:1994fd, Blaizot:1996az}. Apart from that the photons also undergo collisional damping at a rate proportional to $\alpha^{2}T\ln(1/e)$\cite{Smilga:1996cm}. This is again different in the case of QED and QCD as mentioned in the reference. Hence, we model the spectral function in such a way that it captures these scales in a very simplistic manner. This can be done by thinking of the thermal effects on the photon as collective excitation with a decay width $\Gamma\sim \alpha^{2}T$. If we use this approximation to describe \emph{decaying collective modes}, this translates to writing the propagator as 
\begin{equation}
    \Delta(p)=(2\pi)(\theta(p_{0})+n_{B}(p_{0}))\dfrac{1}{\pi}\dfrac{2\Gamma \vert p_{0}\vert}{(p_{0}^{2}-p^{2}-m_{D}^{2})^{2}+4\Gamma^{2}p_{0}^{2}}~~.
\end{equation}
Again, we emphasize the fact that this is a toy model of the spectral function $\rho$ motivated by physical reasoning that we are using to get an analytical understaning of the nature of friction one would get in different regimes. Now, substituting this type of propagator in the equation gives
\begin{equation}
\begin{split}
    \rho_{0}(0,k)=\dfrac{1}{32\pi}k^{2}\int d^{3}p\int_{-\infty}^{\infty} dp_{0} (p_{0}^{2}-p^{2}\sin^{2}\theta)(n_{B}(p_{0})+n_{B}(p_{0})^{2})\\\times \dfrac{1}{\pi}\dfrac{2\Gamma \vert p_{0}\vert}{(p_{0}^{2}-p^{2}-m_{D}^{2})^{2}+4\Gamma^{2}p_{0}^{2}}\times \dfrac{1}{\pi}\dfrac{2\Gamma \vert p_{0}\vert}{(p_{0}^{2}-p^{2}-k^{2}+2pk\cos\theta -m_{D}^{2})^{2}+4\Gamma^{2}p_{0}^{2}}   
\end{split}
\end{equation}
This can be simplified as
\begin{equation}
\begin{split}
  \rho_{0}(0,k)\approx \dfrac{1}{32\pi^{2}} k^{2}\int d^{3}p(p^{2}\cos^{2}\theta+m_{D}^{2})(n_{B}(\sqrt{p^{2}+m_{D}^{2}})+n^{2}_{B}(\sqrt{p^{2}+m_{D}^{2}}))\\\times \dfrac{1}{\sqrt{p^{2}+m_{D}^{2}}}\dfrac{2\Gamma \sqrt{p^{2}+m_{D}^{2}}}{(k^{2}-2pk\cos\theta)^{2}+4\Gamma^{2}(p^{2}+m_{D}^{2})}~~.
\end{split}
\end{equation}
Let us now integrate with respect to $\theta$, the angular coordinate\footnote{For integrating with respect to $\theta$ one can substitute $\cos\theta=x$. On this substitution, one ends up with the integrals of the form:
\begin{equation*}
\begin{split}
 \mathcal{J}_{1}=\int_{-1}^{1}dx~\dfrac{1}{(a-bx)^{2}+c}=\dfrac{1}{b\sqrt{c}}\Big(\arctan\Big(\dfrac{b-a}{\sqrt{c}}\Big)+\arctan\Big(\dfrac{b+a}{\sqrt{c}}\Big)\Big)\\
\mathcal{J}_{2} = \int_{-1}^{1}dx~\dfrac{x^{2}}{(a-bx)^{2}+c}=\dfrac{1}{b^{3}\sqrt{c}}\Big[(a^{2}-c)\Big(\arctan\Big(\dfrac{b-a}{\sqrt{c}}\Big)+\arctan\Big(\dfrac{b+a}{\sqrt{c}}\Big)\Big)\\
  +\sqrt{c}\Big(2b+a\ln\Big(\dfrac{(a-b)^{2}+c}{(a+b)^{2}+c}\Big)\Big]
\end{split}
\end{equation*}}.
Then, we can write
\begin{equation}
\rho_{0}(0,k)\approx \dfrac{1}{8\pi}k^{2}\Gamma \int_{0}^{\infty}dp~p^{2}(n_{B}(p)+n_{B}(p)^{2})(p^{2}\mathcal{J}_{2}+m_{D}^{2}\mathcal{J}_{1})
\label{eqn:simplfied_rho_0}~~~.
\end{equation}
Here, the integrals $\mathcal{J}_{1}$ and $\mathcal{J}_{2}$ are mentioned below in the footnote with parameter values $a=k^{2},~b=2pk,~c=4\Gamma^{2}(p^{2}+m_{D}^{2})$\footnote{In Eqn.\eqref{eqn:simplfied_rho_0}, $n_{B}(p)$ is actually $n_{B}(\sqrt{p^{2}+m_{D}^{2}})$ but in the expression we write just $n_{B}(p)$ for a convenient notation}. Now, we can expand $\mathcal{J}_{1}$ and $\mathcal{J}_{2}$ in the limit where $\sqrt{c}>\vert a \pm b\vert$. This expansion will capture the leading contribution when $k\sim \gamma m_{a}$ is small compared to both $\Gamma$ and $m_{D}$. In this limit
\begin{equation}
    \begin{split}
        \mathcal{J}_{1}=\dfrac{2}{c}=\dfrac{1}{2\Gamma^{2}(p^{2}+m_{D}^{2})}\\
        \mathcal{J}_{2}=\dfrac{2a^{2}}{cb^{2}}=\dfrac{1}{8}\dfrac{k^{2}}{p^{2}(p^{2}+m_{D}^{2})\Gamma^{2}}
    \end{split}
\end{equation}
The quantity $\mathcal{J}_{2}$ is very small in the limit in which we have expanded these integrals when the wall is extremely thick. Hence, we can write the integral as 
\begin{equation}
    \rho_{0}(0,k)\approx \dfrac{k^{2}m_{D}^{2}}{16\pi\Gamma}\int_{0}^{\infty}dp~\dfrac{p^{2}}{p^{2}+m_{D}^{2}}(n_{B}(p)+n_{B}(p)^{2})~~.
\end{equation}
The final integral is just a number proportional to $T$ up to a numerical factor which is roughly 3 as shown earlier:
\begin{equation}
    \rho_{0}(0,k)\approx \alpha\dfrac{3}{4\Gamma}k^{2} T^{3}~~.
\end{equation}
From this, the pressure expression can be derived as
\begin{equation}
    P\propto \Big(\dfrac{\alpha}{\pi}\Big)^{2}\alpha(\gamma m_{a})^{2} T^{2}v\Big(\dfrac{\gamma m_{a}}{\Gamma}\Big)~~.
\end{equation}
In the limit where $\Gamma \ll k $, i.e., $\gamma m_{a}$ is large, in the leading order, one can take $c\rightarrow 0$ and one recovers toy model 1. 

Note that this model works according to our physical intuition. If $k\sim \gamma m_{a}$ is very small compared to any of the thermal scales, which include the Debye length and the decay rate of the collective modes, then the integral in Eqn.\eqref{eqn:spectral_function_rho_0} should be roughly $k$ independent at leading order, which is exactly what we get with this toy spectral function.\footnote{The second thing to note here is that the extra $\alpha$ in the expression for pressure comes from $m_{D}\sim eT$, the Debye scale. This is a property of the plasma and hence we have mentioned it separately rather than writing a $\alpha^{3}$ term.}

\section{Comparison with Hubble friction}
\label{Comparison_with_Hubble}

In the previous section, we have estimated the frictional pressure on domain walls due to the Chern-Simons interaction term using a simple toy model. There are two types of domain walls that are of interest in cosmology: (i) superhorizon walls which typically form cosmic networks and (ii) sub-horizon walls which can be either topologically closed surfaces or have an axion string boundary. Chern-Simons friction can be important for the evolution of both classes of domain walls. In the latter case, it has been demonstrated that collapsing walls attain a terminal velocity even in the absence of friction. We do not expect friction to change this qualitative conclusion but it may affect quantitative details such as the energy loss, terminal velocity, etc. These questions have to be studied with numerical simulations and we will not pursue this direction here. Instead, we will focus on analysing superhorizon walls. 

Superhorizon walls already experience friction due to the expansion of the universe. This Hubble friction can be seen from the axion equation of motion in an FLRW metric which contains a term of the form
\begin{equation}
    3H\dfrac{\partial\varphi}{\partial t}
\end{equation}
which gives a frictional pressure
\begin{equation}
    P_{H}\sim (\gamma m_{a})f_{a}^{2}\dfrac{T^{2}}{M_{pl}}v.
\end{equation}
Our goal will be to compare $P_H$ to the Chern-Simons friction and determine which one dominates as a function of axion parameters and temperature. In particular, the temperature plays an important in determining thermal scales and the functional form of the thermal friction. To apply our results from the last section, we assume that $m_{a}<T$ and focus on thermal friction given by:

\begin{numcases}{P\sim \Big(\dfrac{\alpha}{\pi}\Big)^{2}}
   \alpha m_{a}^{2}T^{2}v,\quad & $\alpha^{2}T<m_{a}<\sqrt{\alpha}T$ \label{eqn:thermalFriction_1}\\
   \vspace{-1.2em} \nonumber \\
   m_{a}^{3}Tv,\quad & $m_{a}>\sqrt{\alpha}T$ \label{eqn:thermalFriction_2}
\end{numcases}

In writing the above expressions, we have ignored factors of $\gamma$. This is justified following~\cite{Huang:1985tt, Blasi:2022ayo} where it was demonstrated that superhorizon walls are not ultra-relativistic and hence have $\gamma\sim 1$. 

The condition that thermal friction dominates is given by:
\begin{numcases}{g_{a\gamma\gamma}^2 \gtrsim}
   \dfrac{1}{\alpha m_a M_{\rm Pl}},\quad & $\alpha^{2}T<m_{a}<\sqrt{\alpha}T$ \label{eqn:thermalDominates_1}\\
   \nonumber \\
   \dfrac{T}{m_a^2 M_{\rm Pl}},\quad & $m_{a}>\sqrt{\alpha}T$ \label{eqn:thermalDominates_2}
\end{numcases}

The second expression depends on the temperature explicitly so it cannot be shown on the $(g_{a\gamma\gamma}, m_a)$-plot. However, note that in our problem the temperature is required to satisfy
\begin{equation}
    T \lesssim \sqrt{m_{a}f_{a}} = \sqrt{ \frac{\alpha m_a}{\pi g_{a\gamma\gamma}}}.
\end{equation}
for the axion mass terms, necessary for the formation of domain walls, to be generated from the confinement of a non-Abelian group. In this case, the condition
\begin{equation}
    \label{eqn:thermalDominates_3}
    g_{a\gamma\gamma}>\dfrac{1}{M_{pl}^{2/5}m_{a}^{3/5}}\Big(\dfrac{\alpha}{\pi}\Big)^{1/5}~~.
\end{equation}
guarantees that~\eqref{eqn:thermalDominates_2} is satisfied. We will use this stronger condition instead since it can be illustrated easily on the axion parameter space.

\begin{figure}[h!]
    \centering
    \includegraphics[height=8cm,width=12cm]{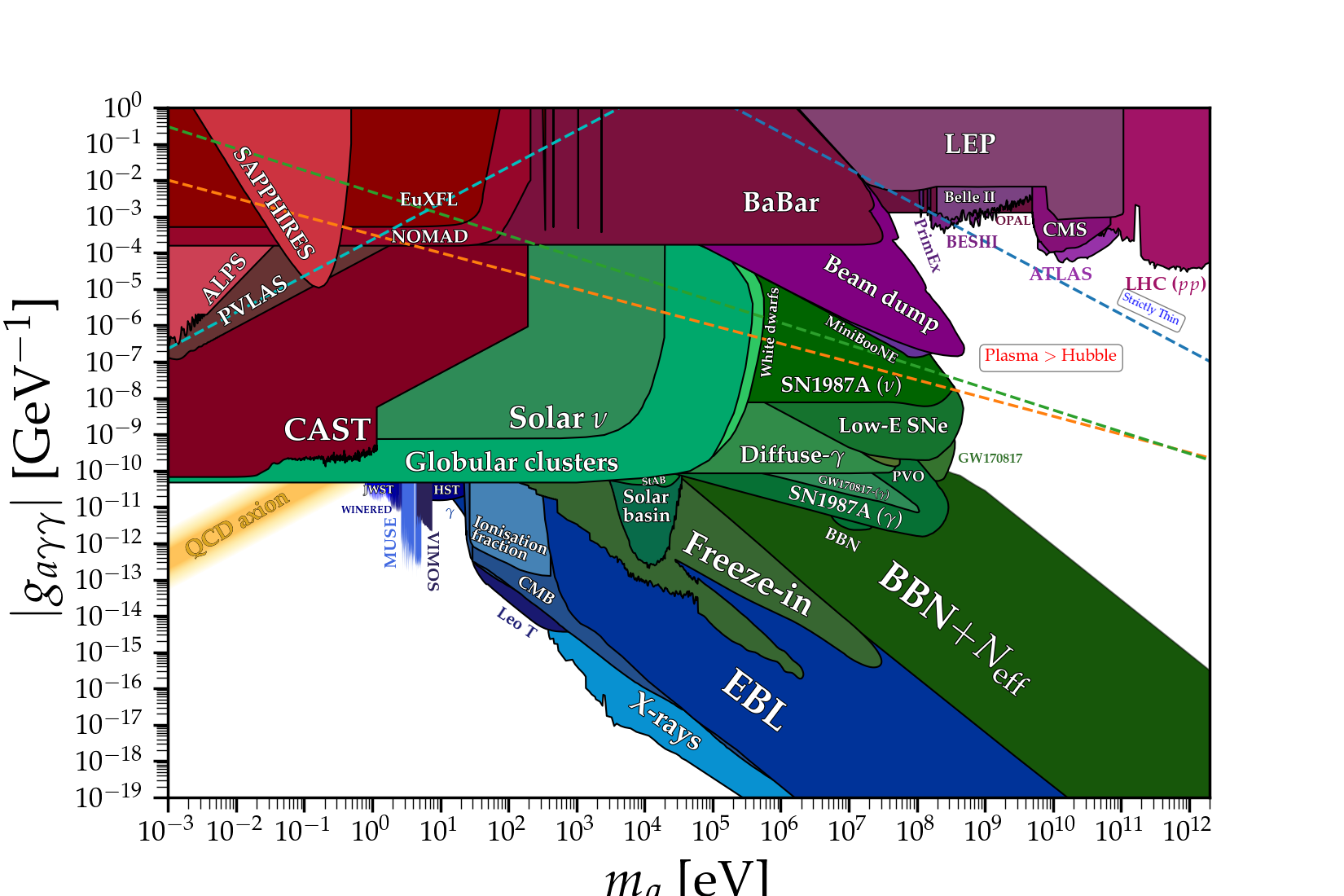}
    \caption{For superhorizon walls where mildly relativistic velocities are expected ($\gamma\sim 1$), regions above the orange and green lines indicate axion parameters where thermal friction dominates Hubble friction. (Detailed explanation is given in the text.) The region below the blue line corresponds $m_{a}<e\sqrt{m_{a} f_{a}}$, with $e=0.3$. The region below the sky-blue colored line satisfies $10\; {\rm MeV}<\sqrt{m_{a}f_{a}}$. In this region of parameter space, domain walls could have formed and annihilated before BBN. We remind the reader that this plot comes with the following caveats: (i) the axion couples only to the SM thermal plasma and not to extra or hidden degrees of freedom which might have very different properties (temperature, interaction strength, etc); (ii) the expressions for friction we used are not justified for \emph{QCD} axion domain walls which have further substructure due to the axion mixing with $\pi^0, \eta, \eta'$; and, (iii) the SM magnetic field is dominated by the usual thermal contribution and there is does not exist ano additional large, coherent, long-range magnetic field (a possibility that is
    phenomenologically allowed).  Figure generated using the open source data and code \cite{AxionLimits}.}
    \label{axion_test_plot}
\end{figure}

We have shown these conditions in the axion exclusion plot in Fig.~(\ref{axion_test_plot}). The relevant regions are above the orange and green lines and they have the following meaning.
\begin{enumerate}
    \item The region above the orange line satisfies inequality~\eqref{eqn:thermalDominates_1}. This is an exact line in the sense that the inequality is temperature-independent. Hence, as long as the plasma friction takes the functional form in~\eqref{eqn:thermalFriction_1}, it dominates over Hubble friction in this region.
    \item The region above the green line satisfies the inequality~\eqref{eqn:thermalDominates_3}. As explained above, this is stronger than~\eqref{eqn:thermalDominates_2}. What this line says is that if the plasma takes the functional form~\eqref{eqn:thermalFriction_2}, then, for the axion parameters above the green line, it is guaranteed that the plasma friction is dominant. Below the green line, the plasma friction can still dominate but one has to check this using the inequality~\eqref{eqn:thermalDominates_2}. 
\end{enumerate}
The above analysis tells us that there are regions in the axion parameter space where, as long as $m_{a}\lesssim T$, the plasma friction dominates over Hubble friction and can be important for domain wall evolution. There is plenty of opportunity for this to happen in the early universe. In particular, domain walls form at temperatures $T_{form} \sim \sqrt{m_a f_a}$ and need to annihilate before $T_{dom}$ given by~\cite{Blasi:2022ayo}:
\begin{equation}
    T_{dom}=\left(16 G\sigma\left(\sqrt{\dfrac{10}{8\pi G g_{*}}}\right)\right)^{1/2} \simeq \,\, 7 \sqrt{\frac{f_a}{M_{\rm Pl}}}\,\, T_{form}~
\end{equation}
where we have used $\sigma=8 m_{a}f_{a}^{2}$ and $g_* \sim 10$. As such, domain wall evolution can take place over temperatures spanning roughly $\log \sqrt{f_a/M_{\rm Pl}}$ orders of magnitude. This large scale separation means that domain walls can have rich dynamics in the early universe where thermal friction plays a crucial role. A complete understanding of these effects requires numerical simulations that include thermal effects calculated in this work. Finally, we stress that axions coming from GUTs and heterotic theory compactifications lie to the right of the QCD axion line where plasma friction can play an important role~\cite{Agrawal:2022lsp,Agrawal:2024ejr}.

\section{Summary and comments}
\label{Discussions}

In this work, we discussed the method of using linear response to calculate plasma friction on the axion domain walls. In particular, we were interested in the effect of the Chern-Simons interaction term. We started by describing some standard ALP domain wall preliminaries and also summarized some crucial aspects of calculating friction that have been discussed in the literature. Following this, we proceeded to work out the friction from the linear response perspective, which captures the collective processes in plasma happening faster and at smaller scales compared to the wall motion and size. In doing this, we found that the friction can be written in terms of the spectral functions of the thermal operators. Furthermore, this method allowed us to calculate the friction in the case of background magnetic fields. To extract analytical expressions, we considered simple toy models for the spectral function that capture important properties of this function. Finally, we compared this friction to the Hubble friction and qualitatively commented on how including the full time-dependent plasma friction can lead to rich wall dynamics. In the calculation above, we considered the case where $N=1$ in the axion wall potential Eqn.\eqref{eqn:simple_potential}. For $N>1$, one can always do field redefinitions to rewrite the axion field, so that in this case, the final effect is that there is an overall $1/N^{2}$ factor that multiplies the friction.

The main result of this work is summarized in Fig.\ref{thermal friction_ranges} where we show that it is essential to include collective thermal processes to get the correct functional dependence for the frictional force on a wall thicker than the Debye length $(eT)^{-1}$. For thinner walls, we correct previous expressions in the literature. This thermal contribution can dominate Hubble friction for domain walls of heavy axion (i.e. with masses larger than about a GeV) as we discussed in Section~\ref{Comparison_with_Hubble} though, as mentioned in the caption to Fig.\ref{axion_test_plot}, a broader range of axion masses allow for large friction effects if we relax the assumptions, including, e.g, long range coherent magnetic fields. We will return to this and related issues in a companion paper.

Notwithstanding these generalised early cosmology scenarios, we note that heavy axions have anyway been argued to provide natural candidate solutions to the strong CP problem in grand unification scenarios~\cite{Rubakov:1997vp} and also in more general settings like~\cite{Agrawal:2017ksf}. Domain walls of heavy axions have been invoked in~\cite{Lee:2024xjb}, for example, as a mechanism to produce QCD axion dark matter and gravitational wave signals within reach of detectors such as NANOGrav~\cite{Ferreira:2022zzo, NANOGrav:2023gor}. The idea behind this mechanism is to introduce a mixing between the QCD axion and the heavy axion which leads to the formation of QCD axion domain walls that overlap with the heavy axion domain walls. The collapse of the heavy axion domain wall also causes the QCD axion domain wall to collapse, producing a relic density of QCD axion dark matter as well as gravitational wave signals. This can happen even if the QCD axion has pre-inflationary initial conditions. More generally, domain walls of heavy axions can form in the early universe and for masses and couplings in the allowed region of Fig.\ref{axion_test_plot} we get gravitational wave signals that peak at frequencies anywhere between $10^{-14}$ to $10^{-2}$ Hz corresponding to the annihilation of these walls and these gravitational waves constitute potential targets for detectors (see for example~\cite{Kuroda:2015owv} for a survey of these detectors and signals). 

There are various simple extensions of the scenario we described above, of which we mention two. First, one can consider a monodromic axion coupling~\cite{Agrawal:2023sbp} whereby the axion couples to the U(1) gauge fields via a more general function:
\begin{align}
    \mathcal{L} \supset \frac{h(a)}{16 \pi^2} F \tilde{F}
\end{align}
where $h(a)$ has a monodromic property: $h(a + 2\pi) = h(a) + 2\pi n$ for some integer $n$. With this coupling, our treatment above is slightly modified and the pressure on the domain wall is obtained by changing the kernel in Eqn.\eqref{eqn:friction_in_terms_of_g}. Specifically, one has to replace the function $g(k_z)$ (i.e. the Fourier transform of $\partial\varphi/\partial z$) by the Fourier transform of $\partial h/\partial z$. 

Another simple extension of the above result involves considering a non-Abelian deconfined thermal plasma coupled to the wall (instead of the Abelian case we studied in this paper). This system is similar to \cite{McLerran:1990de}. The thermal effects in a non-Abelian plasma are quite different compared to the Abelian case. In the non-Abelian case, the damping of collective modes is anomalously large \cite{Bellac:2011kqa, Blaizot:1996az} which changes our treatment above. Furthermore, as shown in \cite{McLerran:1990de, Moore:2010jd}, there are non-perturbative ``sphaleron" processes that contribute to the friction. In particular, the friction is proportional to the sphaleron rate $\Gamma_{sph}$ which is given by:
 \begin{equation}
        \Gamma_{sph}\propto \alpha_{s}^{4}T^{3}/f_{a}^{2}~~.
    \end{equation}
This then gives the frictional pressure on the ALP domain wall to be,
    \begin{equation}
        P\sim \alpha_{s}^{4}(\gamma m_{a})T^{3}v~~,
    \end{equation}
The above contribution accounts for thermal gluons only and including the fermions charged under these gluons further modifies the result. In this case, the friction becomes proportional to the chirality flipping rate $\Gamma_{ch}$ which, in turn is related to $\Gamma_{sph}$. More details on this can be found in \cite{McLerran:1990de, Moore:2010jd} and references within.

In Section~\ref{Comparison_with_Hubble} we studied the effect of thermal friction on super-horizon walls and neglected subhorizon closed walls which are also affected by this friction. This system is more complicated. As shown in \cite{Widrow:1989fe, Widrow:1989vj}, the dynamics of closed spherical walls is different and these can reach high $\gamma$ factors. In this case, the inclusion of thermal friction is important for understanding wall evolution. For example, it affects primordial black hole formation \cite{Dunsky:2024zdo}. A simple case to study initially is the effect of long-range magnetic fields on such closed walls in the absence of a plasma. This can be followed by a full thermal treatment. This system can be realized if the axion is coupled to a hidden sector including heavy charged matter that is annihilated to leave only large-scale coherent electromagnetic fields. This is a phenomenologically rich system which we return to in our companion paper.

One further aspect to study is how axion particles produce friction on ALP domain walls. For sine-Gordon models this vanishes as shown in \cite{Vilenkin:2000jqa, Jackiw:1977yn}. However, the presence of non-renormalizable terms (e.g. Planck-suppressed operators that arise in a UV quantum gravity to forbid global symmetries) causes a tilt in the potential of the walls. As argued in \cite{ZambujalFerreira:2021cte, Beyer:2022ywc}, this tilt mechanism can be an effective way to solve the QCD axion domain wall problem which also gives new bounds on the QCD axion. These arguments are quite general and are applicable to an ALP as well. If such terms are included, then, ALPs in the environment will change mass across the domain wall and give rise to a different kind of friction, similar to the one discussed recently in \cite{GarciaGarcia:2024dfx}. To calculate this friction, one also has to study the abundance of ALP particles which depends on their production mechanism. This is one of the open directions to pursue.

Our results were derived for an axion domain wall neglecting mixing with QCD mesons (such as $\pi^0, \eta, \eta'$) or analogous effects from hidden groups. A more realistic scenario includes these mixing effects, in the presence of which, the wall has structure on different length scales. Typically, these meson domain walls are thinner (see e.g.~\cite{Huang:1985tt}) and one may be able to naively compute friction on them using thin-wall methods and add that contribution to the Chern-Simons friction calculated in this paper. Lastly, we stress that degrees of freedom on the domain wall can also play a crucial role and provide new scattering channels that impact the friction these walls experience. These world-volume field theories can be interesting to study in their own right. For instance, they can certainly include theories as complicated as the FQHE, and there can also be inhomogeneous or even tachyonic (e.g.~\cite{Garriga:1991ts}) field theories that require new techniques to analyse. 

We hope that our work and the discussion above shows that domain walls have very rich phenomenological aspects to be studied in relation to early universe physics. These depend on various factors including temperature, wall motion, type of interaction (e.g. Abelian vs. non-Abelian) and many more. In conclusion, there are likely many phenomenological insights hidden within the intricate details of domain wall dynamics that warrant further exploration.

\section*{Acknowledgements}
We are grateful for discussions with Prateek Agrawal, Isabel Garcia Garcia,  Ed Hardy, Marius Kongsore, Rudin Petrossian-Byrne, Arthur Platschorre, Mario Reig, Subir Sarkar, and Giovanni Villadoro. SH is thankful to the particle theory group of the physics department at University of Oxford, and is also grateful for the support of Christ Church College, University of Oxford, for a Junior Research Fellowship. GRK would like to express gratitude towards Somerville College, Oxford, and The Clarendon Fund, Oxford for supporting the research via the Oxford Ryniker Lloyd Graduate Scholarship jointly with the Clarendon Fund Scholarship. JMR is supported in part by the ``Quantum Sensors for the Hidden Sector" consortium funded by the UK Science \& Technology Facilities Council grant ST/T006277/1, and by the ``QuestDMC" consortium funded by the UK Science \& Technology Facilities Council grant ST/T006773/1.  JMR also thanks the ICTP Particle Theory Group and the University of Washington Physics Department for kind hospitality during the early and late stages of this work. The work of GO is supported by a Leverhulme Trust International Professorship grant number LIP-202-014. For the purpose of Open Access, the author has applied a CC BY public copyright licence to any Author Accepted Manuscript version arising from this submission. 
\appendix


\section{Thin wall friction - leading approximation}
\label{appendix:thin_wall_wkb}
In this section, we derive the thin wall particle scattering friction for photons. Initially, consider a particle with mass $m$ following the Bose-Einstein statistics and scattering off the wall. Consider that the reflection coefficient corresponding to this scattering is given by $\mathcal{R}(p)$. Along with this, consider that the wall is like a flat sheet in the $x-y$ plane and moving in the positive $z$ direction. Then, in the wall frame, the pressure exerted by particles on the two sides of the wall is,
\begin{equation}
\begin{split}
P_{R}=2\int\dfrac{d^{2}p}{(2\pi)^{2}}\int_{-\infty}^{0}\dfrac{dp_{z}}{2\pi}\mathcal{R}(p_{z})\dfrac{p_{z}^{2}}{E}\dfrac{1}{e^{\gamma \beta (E+vp_{z})}-1}\\
P_{L}=2\int\dfrac{d^{2}p}{(2\pi)^{2}}\int^{\infty}_{0}\dfrac{dp_{z}}{2\pi}\mathcal{R}(p_{z})\dfrac{p_{z}^{2}}{E}\dfrac{1}{e^{\gamma \beta  (E+vp_{z})}-1}
\end{split}
\end{equation}
Now, we take the difference between these two. Note that $\mathcal{R}(p_{z})=\mathcal{R}(-p_{z})$. This will give
\begin{equation}
    P=P_{R}-P_{L}=2\int \dfrac{d^{2}p}{(2\pi)^{2}}\int_{0}^{\infty}\dfrac{dp_{z}}{2\pi}\dfrac{p_{z}^{2}}{E}\mathcal{R}(p_{z})\Big(\dfrac{1}{e^{\gamma \beta (E-vp_{z})}-1}-\dfrac{1}{e^{\gamma \beta (E+vp_{z})}-1}\Big)~~.
\end{equation}
To evaluate this integral, we can first integrate the $p_{x},p_{y}$ momenta. In the equation above,
\begin{equation} E=\sqrt{\rho^{2}+p_{z}^{2}+m^{2}},~~\rho=\sqrt{p_{x}^{2}+p_{y}^{2}}.
\end{equation}
By appropriately changing the variables, the above integral can be written as, 
\begin{equation}
\begin{split}
   P=\dfrac{2}{(2\pi)^{2}}\int_{0}^{\infty} d p_{z}~p_{z}^{2}\mathcal{R}(p_{z})\int_{0}^{\infty}d\rho\Big\{\dfrac{\rho}{\sqrt{p_{z}^{2}+\rho^{2}+m^{2}}}\\\times \Big(\dfrac{1}{e^{\gamma \beta (\sqrt{p_{z}^{2}+\rho^{2}+m^{2}}-vp_{z})}-1}-\dfrac{1}{e^{\gamma \beta (\sqrt{p_{z}^{2}+\rho^{2}+m^{2}}+vp_{z})}-1}\Big)\Big\}~~.   
\end{split}
\end{equation}
Substituting $\sqrt{\rho^{2}+p_{z}^{2}+m^{2}}=x$ simplifies the integral above to, 
\begin{equation}
     P=\dfrac{2}{(2\pi)^{2}}\int_{0}^{\infty} d p_{z}~p_{z}^{2}\mathcal{R}(p_{z})\int_{\sqrt{p_{z}^{2}+m^{2}}}^{\infty}dx\Big(\dfrac{1}{e^{\gamma \beta (x-vp_{z})}-1}-\dfrac{1}{e^{\gamma \beta (x+vp_{z})}-1}\Big)~~.
     \label{eqn:pressure_thin_wall_general}
\end{equation}
We can do the integrals individually. Let us first look at the first term in the integral:
\begin{equation}
\begin{split}
\int_{\sqrt{p_{z}+m^{2}}}^{\infty}dx\dfrac{1}{e^{\gamma \beta (x-vp_{z})}-1}=\int_{\sqrt{p_{z}+m^{2}}-vp_{z}}^{\infty}dx\dfrac{1}{e^{\gamma \beta y}-1}=\dfrac{1}{\beta\gamma}\int_{\gamma \beta(\sqrt{p^{2}_{z}+m^{2}}-vp_{z})}^{\infty}dx\dfrac{1}{e^{y}-1}\\
=\dfrac{1}{\beta\gamma}\Big(\beta\gamma(\sqrt{p^{2}_{z}+m^{2}}-vp_{z})-\ln(e^{\beta\gamma(\sqrt{p^{2}_{z}+m^{2}}-vp_{z})}-1)\Big)
\end{split}
\end{equation}
Similarly doing the other integral and taking the difference gives,
\begin{equation}
    P=\dfrac{2}{(2\pi)^{2}\beta \gamma}\int_{0}^{\infty}dp_{z}~p_{z}^{2}\mathcal{R}(p_{z})\left(\ln\left(\dfrac{e^{\gamma \beta (E_{z}+vp_{z})}-1}{e^{\gamma \beta (E_{z}-vp_{z})}-1}\right)-2\gamma \beta vp_{z}\right),~~E_{z}=\sqrt{p_{z}^{2}+m^{2}}
\end{equation}

Let us now look at the reflection coefficient. The WKB approximation can be done by writing the equations of motion of the vector potential in the presence of a wall and treating the vector potential as a wave function. We direct the readers to \cite{Huang:1985tt} for more details. This equation takes the form,
\begin{equation}
    \omega^{2}A_{\pm}=\Big(-\dfrac{d^{2}}{dz^{2}}\pm\dfrac{\alpha}{\pi}\omega\dfrac{d\varphi}{dz}\Big)A_{\pm}~~.
\end{equation}
Substituting the free wall solution gives in the equation above will give 
\begin{equation*}
    \dfrac{d\varphi}{dz}=\dfrac{2m_{a}}{\cosh(m_{a}z)}~~.
\end{equation*}
Solving this equation analytically in closed form is not possible. We can estimate the order of friction by modeling the wall by replacing this potential by, 
\begin{equation}
    V(z)=\begin{cases}
        2\pi m_{a}~~~~0<z<m_{a}^{-1}\\
        0~~~~~~~~~\vert z\vert >m_{a}^{-1}~~.
    \end{cases}
\end{equation}
This potential gives introduces a \emph{wall} of the thickness $m_{a}^{-1}$. The height of the wall is chosen such that the area under the curve is preserved. This is an exactly solved problem in non-relativistic quantum mechanics, see for example \cite{Landau:1975pou}. The reflection coefficient for this then takes the form,
\begin{equation}
R=\dfrac{(k_{1}^{2}-k_{2}^{2})^{2}\sin^{2}(k_{2}/m_{a})}{(k_{1}^{2}-k_{2}^{2})^{2}\sin^{2}(k_{2}/m_{a})+4k_{1}^{2}k_{2}^{2}},~~k_{1}=\omega,~k_{2}=\sqrt{\omega^{2}-2\alpha m_{a}\omega}~~.
\end{equation}
If $\omega^{2}<2\alpha m_{a}\omega$, one can still analytically continue the formula given above. This can then be written as,
\begin{equation}
R=\begin{cases}
    \dfrac{\alpha ^{2}\sin^{2}(\sqrt{x^{2}-2\alpha x})}{\alpha^{2}\sin^{2}(\sqrt{x^{2}-2\alpha x})+(x^{2}-2\alpha x)}~~~~~x>2\alpha \\
    \\
    \dfrac{\alpha ^{2}\sinh^{2}(\sqrt{2\alpha x-x^{2})}}{\alpha^{2}\sinh^{2}(\sqrt{2\alpha x-x^{2}})+(2\alpha x-x^{2})}~~~~x<2\alpha
\end{cases},~~x=\dfrac{\omega}{m_{a}}
\end{equation}
Then, it the limit $\omega< m_{a}$, the reflection coefficient in the leading order is, 
\begin{equation}
    R\approx \alpha^{2}~~.
\end{equation}
In the limit $\omega>m_{a}$, the reflection coefficient is suppressed. 
If we substitute this in Eqn.\eqref{eqn:pressureThermalAverage}, the integral simplifies to,
\begin{equation}
    P\sim \dfrac{2 \alpha^{2}}{(2\pi)^{2}}\dfrac{1}{\beta\gamma}\int_{0}^{m_{a}}dp~p^{2}\mathcal{R}(p)\left(\log\left(\dfrac{f(-v)}{f(v)}\right)-2\beta\gamma vp\right)~~.
\end{equation}
When $\gamma m_{a}<T$, the above integral can be further simplified by taking the leading order expansion in $\gamma m_{a}/T$. This gives us,
\begin{equation}
    P\sim \dfrac{1}{2}\Big(\dfrac{\alpha}{\pi}\Big)^{2}\dfrac{1}{\gamma}m_{a}^{3}T\log\left(\dfrac{1+v}{1-v}\right)\sim\Big(\dfrac{\alpha}{\pi}\Big)^{2}\dfrac{1}{\gamma}m_{a}^{3}Tv
\end{equation}
We direct the readers to \cite{Favitta:2023hlx, Ganoulis:1986rd} for a detailed treatment of a similar model. For QCD axion domain wall, as described in \cite{Huang:1985tt}, the reflection coefficient takes the form,
\begin{equation}
    \mathcal{R}(p)\approx \alpha^{2}m_{a}\delta(p-m_{a})~~.
\end{equation}
If one uses this reflection coefficient, one still ends up with the same expression for pressure as above in the leading order. 
\section{Comments on friction due ALPs in the environment}
One important point to mention when discussing the friction experienced by domain walls is the friction of ALP particles with their domain walls. These ALP particles are usually cold ALP particles (non-thermal) sitting in the environment and are produced by some different mechanisms such as emission from axion strings. This has been studied in \cite{Huang:1985tt, Vilenkin:2000jqa, Jackiw:1977yn}. The analysis can be done in a similar fashion by finding the effective potential that the free ALPs in the environment experience in the vicinity of the wall. This can be done by expanding the Lagrangian of the field with $\varphi=\varphi_{0}+\delta\varphi$, with $\varphi_{0}$ being the domain wall solution. In this case, by substituting this expression in the Lagrangian, the equation of motion for the $\delta\varphi$ is,
\begin{equation}
\partial_{\mu}\partial^{\mu}\delta\varphi=-m_{a}^{2}\delta\varphi\cos(\varphi_{0})~~.
\label{eqn:alp_scattering_wall_eom}
\end{equation}
We can consider the wall in the $x-y$ plane. The friction due to free ALPs in the environment can then be calculated by looking at the scattering of $\delta\varphi$ in the $z$ direction. This can be done with WKB analysis. Hence, we can express the $\delta\varphi$ in the ansatz given by,
\begin{equation}
    \delta\varphi=\chi(z)e^{-i\omega t+ik_{x}x+ik_{y}y}
\end{equation}
If we work in the frame of the wall, and substitute Eqn.\eqref{eqn:static_planar_pure_wall} in Eqn.\eqref{eqn:alp_scattering_wall_eom}, we arrive at,
\begin{equation}
    \chi''(z)+k_{z}^{2}\chi(z)-U(z)\chi(z)=0~,
\end{equation}
where,
\begin{equation}
\begin{split}
k_{z}^{2}=\omega^{2}-m_{a}^{2}-k_{x}^{2}-k_{y}^{2},\\
U(z)=-\dfrac{2}{\cosh^{2}(m_{a}z)},~~.
\end{split}
\end{equation}
The reflection coefficient for a general potential of the form $U(x)=U_{0}\cosh^{-2}(\alpha x)$ has been worked out in detail in \cite{Landau:1975pou}. Substituting the above equations in the result from \cite{Landau:1975pou} gives a vanishing reflection coefficient. This is a consequence of the fact that the Sine-Gordon model is an Integrable model \cite{PhysRevD.109.085019}.

One may think that in this case, there is no friction from the ALPs sitting around in the environment. This is true for ALP models that use the Sine-Gordon model. Further, this comes out to be true for domain walls formed out of the $\mathbb{Z}_{2}$ symmetry vacuum \cite{Vilenkin:2000jqa, Jackiw:1977yn}.  However as shown in \cite{Huang:1985tt, GrillidiCortona:2015jxo}\footnote{Also, one can refer to Appendix of \cite{Blasi:2022ayo} to see the thin wall approach where the reflection coefficient is calculated using the WKB approach.}, a more realistic model in the case of QCD axion is not a pure Sine-Gordon model, but a slight deviation from it \footnote{This is due to the fact that the up and down quarks have different masses that leads to neutral pion axion mixing leading to a coupled differential equation for the two fields. If one substitutes $m_{u}=m_{d}$ in the potential derived in \cite{GrillidiCortona:2015jxo}, one recovers Sine-Gordon potential.}. Even though there is a deviation from the exact Sine-Gordon model, the effective potential that the ALPs in the environment experience is qualitatively the same.
\begin{figure}[h!]
    \centering
    \includegraphics[width=8cm,height=6cm]{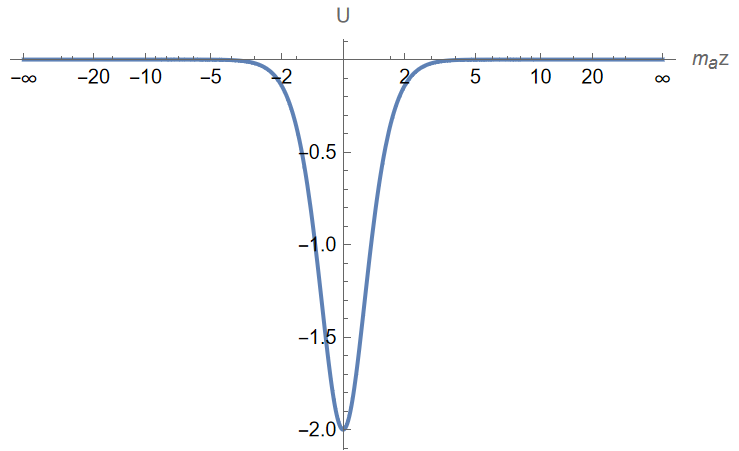}
    \caption{Effective potential experienced by ALPs in the environment around the domain wall, $U(z)$ as a function of $m_{a}z$. This is qualitatively similar to the effective potential derived in \cite{Huang:1985tt}.}
    \label{fig:scattering_potential}
\end{figure}
\\
In such a case, the WKB analysis shows that the reflection coefficient drops down exponentially as a function of $\gamma$, the relative boost factor with the ALP wall and particle. As the authors of \cite{Huang:1985tt} mention, there is a large population of cold ALPs (non-thermal) sitting in the environment, and these can have significant frictional force. But this is true only when the wall is non-relativistic. Even a $\gamma$ factor of the order 10 can make this friction negligible \cite{Blasi:2022ayo}\footnote{Please Refer to Appendix B.1 of \cite{Blasi:2022ayo}.}. Especially, when the walls are closed and have a spherical topology and are completely contained in the horizon, the tension force of the wall eventually dominates and these configurations become highly relativistic. Hence, the relative $\gamma$ factor of the ALPs in the environment with the wall increases, and this friction becomes negligible.


\section{Useful thermal field theory results}
In this Appendix we re-derive the linear response relations and the general expressions for retarded Green's functions in terms of the Spectral Density Function for the sake of completeness. Similar derivations can be found in canonical texts and references dealing with thermal field theory such as \cite{Bellac:2011kqa, Kapusta:1989tk, Laine:2016hma}.
\subsection{Average in the linear response limit}
\label{appendix:Apx_linear_resp}
Consider a general Hamiltonian of the form,
\begin{equation}
    H=H_{0}+H_{int}~~.
\end{equation}
Here, $H_{0}$ is the Hamiltonian without any perturbation, and $H_{int}$ is the perturbation part. Consider an operator $\mathcal{O}(t,\mathbf{x})$. Then the average of this operator is given by,
\begin{equation}
\langle \mathcal{O}(t,\mathbf{x})\rangle=\Tr(\mathcal{O}(t,\mathbf{x})\rho)=\Tr(\mathcal{O}_{I}(t,\mathbf{x})\rho_{I})~~,
\end{equation}
where $\rho$ is the density matrix of the full system. The second equality comes from the fact that the transformation from the Heisenberg picture to the interaction picture is a unitary transformation that leaves the trace invariant. In the interaction picture, the density matrix satisfies, 
\begin{equation}
    \dfrac{d\rho_{I}}{dt}=-i[(H_{int})_{I},\rho_{I}]~~.
\end{equation}
Hence, in the linear limit, this gives us,
\begin{equation}
\rho_{I}(t)=\rho_{I}(t_{0})-i\int_{t_{0}}^{t}dt' [(H_{int})_{I}(t'),\rho_{I}(t_{0})]+O(2)~~.
\end{equation}
In the above equation $t_{0}$ is the time when the interaction Hamiltonian is turned on. One can take the limit $t_{0}\rightarrow -\infty$. Also, $\rho_{I}(t_{0})=\rho_{0}(t_{0})$, where $\rho_{0}$ is the density matrix calculated with respect to $H_{0}$ at time $t_{0}$. This gives us,
\begin{equation}
 \rho_{I}(t)=\rho_{0}-i\int_{-\infty}^{\infty}dt' \theta(t-t')[(H_{int})_{I}(t'),\rho_{0}]~~.   
\end{equation}
Using this, we get, 
\begin{equation}
    \langle \mathcal{O}(t,\mathbf{x})\rangle=\langle \mathcal{O}_{I}(t,\mathbf{x})\rangle_{0}-i\int_{-\infty}^{\infty}dt'\theta(t-t')\langle [\mathcal{O}_{I}(t,\mathbf{x}),(H_{int})_{I}(t')]\rangle_{0}~~,
\end{equation}
where, the average is taken with respect to the $\rho_{0}$ matrix. 

\subsection{Retarded Green's function}
\label{appendix:retarded_greens_function_general_operator}
Consider an operator $\mathcal{O}(t,\mathbf{x})$. The retarded Green's function of this operator is given by,
\begin{equation}
    \mathcal{G}^{R}(t-t',\mathbf{x}-\mathbf{x}')=i\theta(t-t')\langle [\mathcal{O}(t,\mathbf{x}),\mathcal{O}(t',\mathbf{x}')]\rangle~~.
    \label{eqn:green_function_retarded_general}
\end{equation}
We define the Fourier transform of the left-hand side expression as,
\begin{equation}
\mathcal{G}^{R}(t-t,\mathbf{x}-\mathbf{x}')=\int_{-\infty}^{\infty}\dfrac{d\omega}{2\pi} e^{-i\omega(t-t')}\int\dfrac{d^{3}k}{(2\pi)^{3}}e^{i\mathbf{k}\cdot(\mathbf{x}-\mathbf{x}')}\widetilde{\mathcal{G}}^{R}(\omega,\mathbf{k})~~. 
\label{eqn:green_function_fourier_1}
\end{equation}
On the other hand, we can derive the same by expressing the right-hand side of Eqn.\eqref{eqn:green_function_retarded_general}. We can express the thermal average as follows,
\begin{equation}
    \langle \mathcal{O}(t,\mathbf{x})\mathcal{O}(t',\mathbf{x}')\rangle=\sum_{n,m} \bra{n}\mathcal{O}(t,\mathbf{x})\ket{m}\bra{m}\mathcal{O}(t',\mathbf{x}')\ket{n}e^{-\beta E_{n}}~~.
\end{equation}
We can use the spatial translation operator $\mathbf{P}$ and the Hamiltonian as follows,
\begin{equation*}
    \mathcal{O}(t,x)=e^{iHt-i\mathbf{P}\cdot\mathbf{x}}\mathcal{O}(0,0)e^{-iHt+i\mathbf{P}\cdot\mathbf{x}}~~.
\end{equation*}
This when substituted in the equation above gives us,
\begin{equation}
 \langle \mathcal{O}(t,\mathbf{x})\mathcal{O}(t',\mathbf{x}')\rangle=\sum_{n,m} e^{-i (E_{m}-E_{n})(t-t')+i(\mathbf{k}_{m}-\mathbf{k}_{n})\cdot(\mathbf{x}-\mathbf{x}')}e^{-\beta E_{n}}\vert\bra{n}\mathcal{O}(0,0)\ket{m}\vert^{2} 
\end{equation}
Similarly, we can exchange $(t',\mathbf{x}')\leftrightarrow (t,\mathbf{x})$ 
    \begin{equation}
 \langle \mathcal{O}(t,\mathbf{x})\mathcal{O}(t',\mathbf{x}')\rangle=\sum_{n,m} e^{-i (E_{m}-E_{n})(t-t')+i(\mathbf{k}_{m}-\mathbf{k}_{n})\cdot(\mathbf{x}-\mathbf{x}')}e^{-\beta E_{m}}\vert\bra{m}\mathcal{O}(0,0)\ket{n}\vert^{2} 
\end{equation}
Taking the difference then gives us,
\begin{equation}
\begin{split}
\langle [\mathcal{O}(t,\mathbf{x}),\mathcal{O}(t',\mathbf{x}')]\rangle=\sum_{n,m} e^{-i(E_{m}-E_{n})(t-t')+i(\mathbf{k}_{m}-\mathbf{k}_{n})\cdot(\mathbf{x}-\mathbf{x}')} e^{-\beta E_{n}}\\\times (1-e^{-\beta(E_{m}-E_{n})})\vert\bra{m}\mathcal{O}(0,0)\ket{n}\vert^{2}    
\end{split} 
\end{equation}
We define $E_{m}-E_{n}=\omega$. This gives us,
\begin{equation}
\begin{split}
\langle [\mathcal{O}(t,\mathbf{x}),\mathcal{O}(t',\mathbf{x}')]\rangle=\sum_{n,m} e^{-i\omega(t-t')+i(\mathbf{k}_{m}-\mathbf{k}_{n})\cdot(\mathbf{x}-\mathbf{x}')} e^{-\beta E_{n}}\\\times (1-e^{-\beta\omega})\vert\bra{m}\mathcal{O}(0,0)\ket{n}\vert^{2}    
\end{split} 
\end{equation}
Hence, this tells us that the Fourier transform of the commutator is 
\begin{equation}
    \mathsf{FT}\Big(\langle [\mathcal{O}(t,\mathbf{x}),\mathcal{O}(t',\mathbf{x}')]\rangle\Big)=(1-e^{-\beta\omega})\mathsf{FT}\Big(\langle \mathcal{O}(t,\mathbf{x})\mathcal{O}(t',\mathbf{x}')\rangle\Big)~~.
\end{equation}
Let us denote, 
\begin{equation}
\mathsf{FT}\Big(\langle \mathcal{O}(t,\mathbf{x})\mathcal{O}(t',\mathbf{x}')\rangle\Big)= \rho_{0}(\omega,\mathbf{k})~~.
\end{equation}
Furthermore, we can write the Heaviside function $\theta(t-t')$ as,
\begin{equation}
    \theta(t-t')=\dfrac{i}{2\pi}\int_{-\infty}^{\infty}\dfrac{d\omega}{\omega +i\varepsilon}e^{-i\omega(t-t')}~~.
\end{equation}
Using this, we can write the RHS of Eqn.\eqref{eqn:green_function_retarded_general} as,
\begin{equation}
    -\int_{-\infty}^{\infty}\dfrac{d\omega}{2\pi}\dfrac{e^{-i\omega(t-t')}}{\omega+i\varepsilon}\int_{-\infty}^{\infty}\dfrac{d\omega'}{2\pi}e^{-i\omega'(t-t')}\int e^{i\mathbf{k}\cdot(\mathbf{x}-\mathbf{x}')}\dfrac{d^{3}k}{(2\pi)^{3}}(1-e^{-\beta\omega'})\rho_{0}(\omega',\mathbf{k})~~.
\end{equation}
Rearranging this gives us,
\begin{equation}
-\int_{-\infty}^{\infty}\dfrac{d\omega}{2\pi}e^{-i\omega(t-t')}\int e^{i\mathbf{k}\cdot(\mathbf{x}-\mathbf{x}')}\dfrac{d^{3}k}{(2\pi)^{3}}\int_{-\infty}^{\infty}\dfrac{d\omega'}{2\pi}\dfrac{1}{\omega-\omega'+i\varepsilon}(1-e^{-\beta\omega'})\rho_{0}(\omega',\mathbf{k})~~.
\label{eqn:green_function_fourier_2}
\end{equation}
Comparing Eqn.\eqref{eqn:green_function_fourier_1} and Eqn.\eqref{eqn:green_function_fourier_2} gives us,
\begin{equation}
    \widetilde{\mathcal{G}}^{R}(\omega,\mathbf{k})=-\int_{-\infty}^{\infty}\dfrac{d\omega'}{2\pi}\dfrac{1}{\omega-\omega'+i\varepsilon}(1-e^{-\beta\omega'})\rho_{0}(\omega',\mathbf{k})~~.
\end{equation}
using the identity, 
\begin{equation*}
    \dfrac{1}{\Delta \pm i\varepsilon}=\mathbb{P}\Big(\dfrac{1}{\Delta}\Big)\mp i\pi \delta(\Delta)~~,
\end{equation*}
gives us,
\begin{equation}
    \widetilde{\mathcal{G}}^{R}(\omega,\mathbf{k})=\dfrac{i}{2}(1-e^{-\beta\omega})\rho_{0}(\omega,\mathbf{k})-\mathbb{P}\int_{-\infty}^{\infty}\dfrac{d\omega'}{2\pi}\dfrac{1}{\omega-\omega'}(1-e^{-\beta\omega'})\rho_{0}(\omega',\mathbf{k})
\end{equation}
From this, we can write,
\begin{equation}
    \mathcal{\widetilde{G}^{R}}(0,\mathbf{k})=\mathbb{P}\int_{-\infty}^{\infty}\dfrac{d\omega'}{2\pi}\dfrac{1}{\omega'}(1-e^{-\beta\omega'})\rho_{0}(\omega',\mathbf{k}),~~.
\end{equation}
and 
\begin{equation}
    \dfrac{\partial\mathcal{\widetilde{G}^{R}}}{\partial \omega}(0,\mathbf{k})=\dfrac{i}{2T}\rho_{0}(0,\mathbf{k})~~.
\end{equation}
In the above equation, we have used the fact that the spectral density function $\rho(\omega,\mathbf{k})=(1-e^{-\beta\omega})\rho_{0}(\omega,\mathbf{k})$ is an odd function of $\omega$ \cite{Bellac:2011kqa}.


\section{Simplification of integral $\mathcal{I}_{1}$}
\subsection{Breaking the integral in two parts}
\label{breaking the integrals in two parts}
Let us start with the integral (dropping the $\alpha/\pi$ factor)
\begin{equation}
    \mathcal{I}_{1}=i\int_{-\infty}^{\infty}dt'\theta(t-t')\int d^{3}x'\langle[E_{i}(t,\mathbf{x}),E_{j}(t',\mathbf{x}')]\rangle_{0}\xi_{ij}(t-t',\mathbf{x}-\mathbf{x}')\varphi(t',\mathbf{x}')~. 
\end{equation} 
One can express the commutator in terms of the vector potential as follows \cite{Kapusta:1989tk},
\begin{equation}
    	\begin{split}
		\langle[E_{i}(t,\mathbf{x}),E_{j}(t',\mathbf{x}')]\rangle_{0}\theta(t-t')=\partial_{i}\partial'_{j}(\langle [A_{0}(t,\mathbf{x}),A_{0}(t'\mathbf{x}')]\rangle\theta(t-t'))\\-\partial_{i}\partial'_{0}(\langle [A_{0}(t,\mathbf{x}),A_{j}(t'\mathbf{x}')]\rangle\theta(t-t'))\\ -\partial_{0}\partial'_{j}(\langle [A_{i}(t,\mathbf{x}),A_{0}(t'\mathbf{x}')]\rangle\theta(t-t'))\\+\partial_{0}\partial'_{0}(\langle [A_{i}(t,\mathbf{x}),A_{j}(t'\mathbf{x}')]\rangle\theta(t-t'))\\
		-i\delta_{ij}\delta^{3}(\mathbf{x}-\mathbf{x}')\delta(t-t')
	\end{split}
 \label{eqn:electric_field_thermal_corrected_retarded_green}
\end{equation}
Let us include the $\delta$ function at the end. In the standard notation of \cite{Kapusta:1989tk} where, 
\begin{equation}
D^{R}_{ij}=i\theta(t-t')\langle [A_{i}(t,\mathbf{x}),A_{j}(t',\mathbf{x}')]\rangle_{0}~~,
\end{equation}
we can express the quantity of interest as, 
\begin{equation}
\mathcal{G}_{ij}(t-t',\mathbf{x}-\mathbf{x}')=\partial_{i}\partial'_{j}D^{R}_{00}-\partial_{i}\partial'_{0}D^{R}_{0j}-\partial_{0}\partial'{j}D^{R}_{i0}+\partial_{0}\partial'_{0}D^{R}_{ij}
\end{equation}
In the Temporal Gauge, the only non-vanishing propagator is $D_{ij}$, hence, this reduces our expression to,
\begin{equation}
\mathcal{G}_{ij}(t-t',\mathbf{x}-\mathbf{x}')=\partial_{0}\partial'_{0}D^{R}_{ij}
\end{equation}
In the Fourier modes, this looks like,
\begin{equation}
\widetilde{\mathcal{G}}_{ij}(\omega,\mathbf{k})=\omega^{2}\widetilde{D}^{R}_{ij}(\omega,\mathbf{k})~~.
\end{equation}
We have a standard expression for $\widetilde{D}^{R}_{ij}$ given by \cite{Kapusta:1989tk}
\begin{equation}
\widetilde{D}^{R}_{ij}(\omega,\mathbf{k})=
	-\Big(\dfrac{1}{\omega^{2}-k^{2}-F}P_{ij}^{L}+\dfrac{1}{\omega^{2}-k^{2}-G}P_{ij}^{T}\Big)
\label{eqn:thermal_propagator}
\end{equation}
where, $\vert\mathbf{k}\vert=k$, and
\begin{equation}
\begin{split}
P_{ij}^{L}=\dfrac{\omega^{2}-k^{2}}{\omega^{2}}\dfrac{k_{i}k_{j}}{k^{2}}\\
P^{T}_{ij}=\delta_{ij}-\dfrac{k_{i}k_{j}}{k^{2}}\\
F=-2\dfrac{m^{2}(\omega^{2}-k^{2})}{k^{2}}\Big(1-\dfrac{\omega}{k}Q_{0}\Big(\dfrac{\omega}{k}\Big)\Big)\\
		G=m^{2}\Big(\dfrac{\omega}{k}\Big)\Big[\dfrac{\omega}{k}+\Big(1-\Big(\dfrac{\omega}{k}\Big)^{2}\Big)Q_{0}\Big(\dfrac{\omega}{k}\Big)\Big]\\
		Q_{0}\Big(\dfrac{\omega}{k}\Big)=\dfrac{1}{2}\ln\Big(\dfrac{\omega+k}{\omega-k}\Big)-\dfrac{i\pi}{2}\theta(k^{2}-\omega^{2})
\end{split}
\end{equation}
The propagator can be expressed in other gauges but as shown in \cite{Kapusta:1989tk}, will leave us with the same result. Here, the quantity $m$ is related to the plasma temperature as $m=eT/\sqrt{6}$. 

As we are interested in the regime where we expand near $\omega=0$, in this limit, the terms that are relevant to us are:
\begin{equation}
    \mathsf{Re}(\widetilde{\mathcal{G}}_{ij})(0,k)=-\dfrac{k^{2}}{2m^{2}+k^{2}}
\end{equation}
and,
\begin{equation}
\mathsf{Im}\Big(\dfrac{\partial\widetilde{\mathcal{G}}_{ij}}{\partial\omega}(0,k)\Big)=\dfrac{m^{2}\pi k}{(2m^{2}+k^{2})^{2}}
\end{equation}
The imaginary part of $\widetilde{\mathcal{G}}_{ij}(\omega,\mathbf{k})$ and the real part of $\partial_{\omega}\widetilde{\mathcal{G}}_{ij}(\omega,\mathbf{k})$ vanishes in the limit $\omega\rightarrow 0$. If these parts were non-zero, they would correspond to microscopic \emph{oscillations} of the axion profile. However, as we look at the coherent state, these microscopic oscillations are of less importance when looking at macroscopic motion. The integral then looks like,
\begin{equation}
    \begin{split}
        \mathcal{I}_{1}=\int_{-\infty}^{\infty}dt'\int d^{3}x'\int_{-\infty}^{\infty}\dfrac{d\omega}{2\pi}\int\dfrac{d^{3}k}{(2\pi)^{3}}e^{-i\omega(t-t')+i\mathbf{k}\cdot(\mathbf{x}-\mathbf{x}')}\Big(-\dfrac{k^{2}}{2m^{2}+k^{2}}\\+i\omega\dfrac{m^{2}\pi k}{(2m^{2}+k^{2})^{2}}\Big)P^{L}_{ij}\xi_{ij}\varphi(t',\mathbf{x}')~~.
    \end{split}
\end{equation}
\subsection{Simplifying the two parts}
\label{Appendix:Simplification of integral I_1}
Let us now first concentrate on the second integral, i.e.,
\begin{equation}
    \mathcal{I}_{1,2}=-\int d^{3}x'\int\dfrac{d^{3}k}{(2\pi)^{3}}e^{i\mathbf{k}\cdot(\mathbf{x}-\mathbf{x}')}\dfrac{m^{2}\pi k}{(2m^{2}+k^{2})^{2}}P^{L}_{ij}\xi_{ij}\partial_{t}(\varphi(t,\mathbf{x}'))~~.
\end{equation}
We consider that the wall's radius of curvature is large compared to its thickness. This is a valid assumption to make as usually, the walls are of the horizon size or a few orders smaller, which is still much larger than the thickness. This allows us to treat the wall flat locally.  We consider that $\varphi$ is a function of $x_{\perp}$. In this case, 
\begin{equation}
    \mathcal{I}_{1,2}=-m^{2}\pi\int d^{3}x' \int \dfrac{d^{3}k}{(2\pi)^{3}}e^{i\mathbf{k}\cdot(\mathbf{x}-\mathbf{x}')}\dfrac{k}{(2m^{2}+k^{2})^{2}}\xi_{ij}(0,\vert\mathbf{x}-\mathbf{x}'\vert)P^{L}_{ij}\partial_{t}(\varphi(t,x'_{\perp}))~~.
\end{equation}
Let us Fourier transform both $\xi_{ij}$ and $\partial_{t}\varphi$. This is given by,
\begin{equation}
\begin{split}
  \mathcal{I}_{1,2}=-m^{2}\pi\int d^{3}x'\int\dfrac{d^{3}k}{(2\pi)^{3}}\int\dfrac{d^{3}p}{(2\pi)^{3}}e^{i(\mathbf{k}+\mathbf{p})\cdot(\mathbf{x}-\mathbf{x}')}\dfrac{k}{(2m^{2}+k^{2})^{2}}\eta(p)\Big(\delta_{ij}-\dfrac{p_{i}p_{j}}{p^{2}}\Big)\dfrac{k_{i}k_{j}}{k^{2}} \\\times \int_{-\infty}^{\infty}\dfrac{dp_{2}}{2\pi}  e^{i p_{2}(x'_{\perp}-vt)}g(p_{2}) 
\end{split}
\end{equation}
Here we have considered that the wall depends on a single coordinate given by $x'_{\perp}$. As mentioned before, the tensor structure of $\xi_{ij}$ comes from the $\nabla\cdot \mathbf{B}_{bg}=0$ condition. 
 The tensor structure is,
\begin{equation}
    \Big(\delta_{ij}-\dfrac{p_{i}p_{j}}{p^{2}}\Big)\dfrac{k_{i}k_{j}}{k^{2}} =1-\dfrac{(\mathbf{p}\cdot\mathbf{k})^{2}}{p^{2}k^{2}}=1-\dfrac{(p_{\perp}k_{\perp}+p_{\parallel}\cdot k_{\parallel})^{2}}{(p_{\perp}^{2}+p_{\parallel}^{2})(k_{\perp}^{2}+k_{\parallel}^{2})}
\end{equation}
The coordinates parallel to the wall are denoted by $x'_{\parallel}$. We can integrate with respect to them. This gives, $(2\pi)^{2}\delta^{(2)}(k_{\parallel}+p_{\parallel})$.
This can then be integrated with respect to $d^{2}p_{\parallel}$. This then leaves us with, 
\begin{equation}
\begin{split}
  \mathcal{I}_{1,2}=-m^{2}\pi\int dx'_{\perp}\int\dfrac{d^{3}k}{(2\pi)^{3}}\int\dfrac{dp_{\perp}}{2\pi}e^{i(k_{\perp}+p_{\perp})\cdot(x_{\perp}-x'_{\perp})}\dfrac{k}{(2m^{2}+k^{2})^{2}}\eta(\sqrt{p_{\perp}^{2}+k_{\parallel}^{2}})\\\times \Big(1-\dfrac{(p_{\perp}k_{\perp}-k_{\parallel}^{2})^{2}}{(k_{\parallel}^{2}+p_{\perp}^{2})(k_{\parallel}^{2}+k_{\perp}^{2})}\Big)\int_{-\infty}^{\infty}\dfrac{dp_{2}}{2\pi}  e^{i p_{2}(x'_{\perp}-vt)}g(p_{2}) 
\end{split}
\end{equation}
Now, one can integrate with respect to $x_{\perp}'$. This will gives $2\pi \delta(k_{\perp}+p_{\perp}-p_{2})$. This object can then be integrated with respect to $p_{2}$ variable. This gives,
\begin{equation}
  \begin{split}
  \mathcal{I}_{1,2}=-m^{2}\pi\int\dfrac{d^{3}k}{(2\pi)^{3}}\int\dfrac{dp_{\perp}}{2\pi}e^{i(k_{\perp}+p_{\perp})\cdot(x_{\perp}-vt)}\dfrac{k}{(2m^{2}+k^{2})^{2}}\eta(\sqrt{p_{\perp}^{2}+k_{\parallel}^{2}})\\ \times \Big(1-\dfrac{(p_{\perp}k_{\perp}-k_{\parallel}^{2})^{2}}{(k_{\parallel}^{2}+p_{\perp}^{2})(k_{\parallel}^{2}+k_{\perp}^{2})}\Big)g(k_{\perp}+p_{\perp})
\end{split}  
\end{equation}
We can simplify the bracket inside the integral as:
\begin{equation}
    1-\dfrac{(p_{\perp}k_{\perp}-k_{\parallel}^{2})^{2}}{(k_{\parallel}^{2}+p_{\perp}^{2})(k_{\parallel}^{2}+k_{\perp}^{2})}=\dfrac{k_{\parallel}^{2}(p_{\perp}+k_{\perp})^{2}}{(k_{\parallel}^{2}+p_{\perp}^{2})(k_{\parallel}^{2}+k_{\perp}^{2})}
\end{equation}
Hence, the integral looks like,
\begin{equation}
 \begin{split}
  \mathcal{I}_{1,2}=-m^{2}\pi\int\dfrac{d^{3}k}{(2\pi)^{3}}\int\dfrac{dp_{\perp}}{2\pi}e^{i(k_{\perp}+p_{\perp})\cdot(x_{\perp}-vt)}\dfrac{k}{(2m^{2}+k^{2})^{2}}\eta(\sqrt{p_{\perp}^{2}+k_{\parallel}^{2}})\\ \times \Big(\dfrac{k_{\parallel}^{2}(p_{\perp}+k_{\perp})^{2}}{(k_{\parallel}^{2}+p_{\perp}^{2})(k_{\parallel}^{2}+k_{\perp}^{2})}\Big)g(k_{\perp}+p_{\perp})
\end{split}  
\end{equation}
One can substitute $p_{\perp}+k_{\perp}=q$. This gives, 
\begin{equation}
 \begin{split}
  \mathcal{I}_{1,2}=-m^{2}\pi\int\dfrac{d^{3}k}{(2\pi)^{3}}\int\dfrac{dq}{2\pi}e^{iq(x_{\perp}-vt)}\dfrac{k}{(2m^{2}+k^{2})^{2}}\eta(\sqrt{(q-k_{\perp}^{2}+k_{\parallel}^{2}})\\ \times \Big(\dfrac{k_{\parallel}^{2}q^{2}}{(k_{\parallel}^{2}+(q-k_{\perp})^{2})(k_{\parallel}^{2}+k_{\perp}^{2})}\Big)g(q)
\end{split}  
\end{equation}
Rearranging this gives us,
\begin{equation}
  \begin{split}
  \mathcal{I}_{1,2}=-m^{2}\pi\int\dfrac{dq}{2\pi}e^{iq(x_{\perp}-vt)}g(q)q^{2}\int\dfrac{d^{3}k}{(2\pi)^{3}}\dfrac{k}{(2m^{2}+k^{2})^{2}}\eta(\sqrt{(q-k_{\perp})^{2}+k_{\parallel}^{2}})\\\times \Big(\dfrac{k_{\parallel}^{2}}{(k_{\parallel}^{2}+(q-k_{\perp})^{2})(k_{\parallel}^{2}+k_{\perp}^{2})}\Big)
\end{split}    
\end{equation}
Now, we can shift the vector $\mathbf{k}\rightarrow \mathbf{k}+q\hat{k}_{\perp}$. This will leave the measure invariant. This will give us,
\begin{equation}
  \begin{split}
  \mathcal{I}_{1,2}=-m^{2}\pi\int\dfrac{dq}{2\pi}e^{iq(x_{\perp}-vt)}g(q)q^{2}\int\dfrac{d^{3}k}{(2\pi)^{3}}\dfrac{1}{(2m^{2}+k_{\parallel}^{2}+(k_{\perp}+q)^{2})^{2}}\eta(k)\\\times \Big(\dfrac{k_{\parallel}^{2}}{k^{2}\sqrt{(k_{\parallel}^{2}+(k_{\perp}+q)^{2})}}\Big)
\end{split}    
\end{equation}
Here, $k_{\parallel}=k\sin\theta,~k_{\perp}=k\cos\theta$. Substituting this gives us,
\begin{equation}
 \begin{split}
  \mathcal{I}_{1,2}=-\dfrac{1}{4\pi^{2}}m^{2}\pi\int\dfrac{dq}{2\pi}e^{iq(x_{\perp}-vt)}g(q)q^{2}\int_{0}^{\infty}dk\int_{0}^{\pi}d\theta  \dfrac{\sin\theta}{(2m^{2}+k^{2}\sin^{2}\theta+(k\cos\theta+q)^{2})^{2}}\\\times \eta(k) \Big(\dfrac{k^{2}\sin^{2}\theta}{\sqrt{(k^{2}\sin^{2}\theta+(k\cos\theta+q)^{2})}}\Big)
\end{split}    
\end{equation}
Further, if write $\cos\theta=y$, then,
\begin{equation}
   \begin{split}
  \mathcal{I}_{1,2}=-\dfrac{1}{4\pi^{2}}m^{2}\pi\int\dfrac{dq}{2\pi}e^{iq(x_{\perp}-vt)}g(q)q^{2}\int_{0}^{\infty}dk k^{2}\eta(k)\\\times \int_{-1}^{1}dx \dfrac{1-x^{2}}{(2m^{2}+k^{2}+q^{2}+2kqx)^{2}}\dfrac{1}{\sqrt{k^{2}+q^{2}+2kqx}}
\end{split} 
\end{equation}
The integral with respect to $x$ is done by Mathematica,
\begin{equation}
\begin{split}
 \int_{-1}^{1}dx \dfrac{1-x^{2}}{(2m^{2}+k^{2}+q^{2}+2kqx)^{2}}\dfrac{1}{\sqrt{k^{2}+q^{2}+2kqx}}=\\\dfrac{1}{8(\sqrt{2}m)^{3}k^{3}q^{3}}\Big(\sqrt{2}m((\sqrt{(k-q)^{2}}-\sqrt{(k+q)^{2}})(3(\sqrt{2}m)^{2}+k^{2}+q^{2})\\+2kq(\sqrt{(k-q)^{2}}+\sqrt{(k+q)^{2}}))
 +3(\sqrt{2}m)^{4}-(k^{2}-q^{2})^{2}\\+2(\sqrt{2}m)^{2}(k^{2}+q^{2})\Big(\text{arccot}\Big(\dfrac{\sqrt{2}m}{\sqrt{(k+q)^{2}}}\Big)-\text{arccot}\Big(\dfrac{\sqrt{2}m}{\sqrt{(k-q)^{2}}}\Big)\Big)\Big)
\end{split}
\label{eqn:horrible_eqn_1}
\end{equation}
Note that the function $g(q)$, i.e., the Fourier transform of $\partial_{t}\varphi$ peaks around $q=0$. It has a characteristic width of $\gamma m_{a}$. The Linear response with initial simplification of expanding the electric field correlator around $\omega=0$ is correct when $\gamma m_{a}< eT$. Otherwise, for a relativistic wall, the self-crossing time is approximately around $(\gamma m_{a})^{-1}\sim (eT)^{-1}$ where the $\omega=0$ approximation is not a good one. In the limit where $\gamma m_{a}<eT$, the leading order contribution of the integral in $q$ can be calculated by setting \eqref{eqn:horrible_eqn_1} near $q=0$. This is given by,
\begin{equation}
    \dfrac{4}{3}\dfrac{1}{k(2m^{2}+k^{2})^{2}}
\end{equation}
Hence, what we have is,
\begin{equation}
 \begin{split}
  \mathcal{I}_{1,2}=-\dfrac{1}{3\pi^{2}}m^{2}\pi\int\dfrac{dq}{2\pi}e^{iq(x_{\perp}-vt)}g(q)q^{2}\int_{0}^{\infty}dk \dfrac{k}{(2m^{2}+k^{2})^{2}}\eta(k)
\end{split} 
\end{equation}
The $q$ integral is nothing by,
\begin{equation}
    \int\dfrac{dq}{2\pi}e^{iq(x_{\perp}-vt)}g(q)q^{2}=-\partial_{x_{\perp}}^{2}\partial_{t}\varphi(t,x_{\perp})
\end{equation}
Hence, we get,
\begin{equation}
    \mathcal{I}_{1,2}=\dfrac{m^{2}}{3\pi}\partial_{t}\partial_{x_{\perp}}^{2}\varphi(t,x_{\perp})\int_{0}^{\infty}dk \dfrac{k}{(2m^{2}+k^{2})^{2}}\eta(k)~~.
    \label{eqn:I_1_2_full_appendix}
\end{equation}
The same procedure can be repeated for the first term in Eqn.\eqref{eqn:I_1_integral_full}. This simplifies the equation to,
\begin{equation}
    \mathcal{I}_{1,1}=\dfrac{1}{3\pi^{3}}\partial_{x_{\perp}}^{2}\varphi(t,x_{\perp})\int_{0}^{\infty}\dfrac{k^{2}}{2m^{2}+k^{2}}\eta(k)~~.
\end{equation}
\section{Calculating the spectral density}
\label{Appendix:Calculating the Spectral density}
Let us start with the equation at hand, 
\begin{equation}
\begin{split}
\mathcal{I}_{3}=\epsilon^{\mu\nu\sigma\tau}\epsilon^{\alpha\beta\gamma\delta}\langle \partial_{\mu}A_{\nu}\partial_{\sigma}A_{\tau}\partial'_{\alpha}A_{\beta}\partial'_{\gamma}A_{\delta}\rangle=\epsilon^{\mu\nu\sigma\tau}\epsilon^{\alpha\beta\gamma\delta}\int\prod\dfrac{d^{4}p_{i}}{(2\pi)^{4}}e^{-i(\omega_{1}+\omega_{2})t}e^{i(\mathbf{p}_{1}+\mathbf{p}_{2})\cdot\mathbf{x})}\\ \times e^{-i(\omega_{3}+\omega_{4})t'}e^{i(\mathbf{p}_{3}+\mathbf{p}_{4})\cdot\mathbf{x})}(p_{1})_{\mu}(p_{2})_{\sigma}(p_{3})_{\alpha}(p_{4})_{\gamma} \langle \widetilde{A}_{\nu}(p_{1})\widetilde{A}_{\tau}(p_{2})\widetilde{A}_{\beta}(p_{3})\widetilde{A}_{\delta}(p_{4})\rangle    
\end{split}
\end{equation}
We look at the \emph{tree level diagram} without considering any fermion loops between the four-photon operators. We can contract the fields in two ways: $\{(\nu,\beta),(\tau,\delta)\},~\{(\nu,\delta),(\tau,\beta)\}$. The remaining contraction vanishes due to the $\epsilon$-tensor sitting outside the integral. We write:
\begin{equation}
    \langle \widetilde{A}_{\nu}(p_{1})\widetilde{A}_{\beta}(p_{3})\rangle= (2\pi)^{4}\delta^{(4)}(p_{1}+p_{3})(P^{T}_{\nu\beta}\Delta^{T}(p_{1})+P^{L}_{\nu\beta}\Delta^{L}(p_{1}))
\end{equation}
Here, $P^{T/L}$ are the transverse and longitudinal projection operators respectively and $\Delta^{T/L}$ are the longitudinal and transverse propagators. Note that these are not time-ordered thermal averages. Hence, in the standard thermal field theory notation, these are $D_{21}$ type propagators\cite{Bellac:2011kqa}. We can substitute this in the equation above. This will give us,
\begin{equation}
  \begin{split}
\mathcal{I}_{3}=\epsilon^{\mu\nu\sigma\tau}\epsilon^{\alpha\beta\gamma\delta}\int\prod\dfrac{d^{4}p_{i}}{(2\pi)^{4}}e^{-i(\omega_{1}+\omega_{2})t}e^{i(\mathbf{p}_{1}+\mathbf{p}_{2})\cdot\mathbf{x})}e^{-i(\omega_{3}+\omega_{4})t'}e^{i(\mathbf{p}_{3}+\mathbf{p}_{4})\cdot\mathbf{x})}(p_{1})_{\mu}(p_{2})_{\sigma}(p_{3})_{\alpha}(p_{4})_{\gamma}\\
\times (2\pi)^{8}(P^{T}_{\nu\beta}\Delta^{T}(p_{1})+P^{L}_{\nu\beta}\Delta^{L}(p_{1}))(P^{T}_{\tau\delta}\Delta^{T}(p_{2})+P^{L}_{\tau\delta}\Delta^{L}(p_{2}))\\
(\delta^{(4)}(p_{1}+p_{3})\delta^{(4)}(p_{2}+p_{4})-\delta^{(4)}(p_{1}+p_{4})\delta^{(4)}(p_{2}+p_{3}))
  \end{split}  
\end{equation}
The negative sign in the last bracket comes from exchanging $\beta,\delta$ in the $\epsilon$ tensor. This then gives us,
\begin{equation}
\begin{split}
  \mathcal{I}_{3}  =\epsilon^{\mu\nu\sigma\tau}\epsilon^{\alpha\beta\gamma\delta}\int\prod\dfrac{d^{4}p_{i}}{(2\pi)^{4}}e^{-i(\omega_{1}+\omega_{2})(t-t')}e^{i(\mathbf{p}_{1}+\mathbf{p}_{2})\cdot(\mathbf{x}-\mathbf{x}')}\\
\times (P^{T}_{\nu\beta}\Delta^{T}(p_{1})+P^{L}_{\nu\beta}\Delta^{L}(p_{1}))(P^{T}_{\tau\delta}\Delta^{T}(p_{2})+P^{L}_{\tau\delta}\Delta^{L}(p_{2}))\\
((p_{1})_{\mu}(p_{2})_{\sigma}(p_{1})_{\alpha}(p_{2})_{\gamma}-(p_{1})_{\mu}(p_{2})_{\sigma}(p_{2})_{\alpha}(p_{1})_{\gamma})
\end{split}
\end{equation}
The term in the bracket can be rearranged using the property of $\epsilon$ tensor to get,
\begin{equation}
\begin{split}
  \mathcal{I}_{3}  =2\epsilon^{\mu\nu\sigma\tau}\epsilon^{\alpha\beta\gamma\delta}\int\prod\dfrac{d^{4}p_{i}}{(2\pi)^{4}}e^{-i(\omega_{1}+\omega_{2})(t-t')}e^{i(\mathbf{p}_{1}+\mathbf{p}_{2})\cdot(\mathbf{x}-\mathbf{x}')}\\
\times (P^{T}_{\nu\beta}\Delta^{T}(p_{1})+P^{L}_{\nu\beta}\Delta^{L}(p_{1}))(P^{T}_{\tau\delta}\Delta^{T}(p_{2})+P^{L}_{\tau\delta}\Delta^{L}(p_{2}))\\
((p_{1})_{\mu}(p_{2})_{\sigma}(p_{1})_{\alpha}(p_{2})_{\gamma})
\end{split}
\end{equation}
We can now define $p_{1}+p_{2}=k$. This gives us,
\begin{equation}
  \begin{split}
  \mathcal{I}_{3}  =2\epsilon^{\mu\nu\sigma\tau}\epsilon^{\alpha\beta\gamma\delta}\int\dfrac{d^{4}k}{(2\pi)^{4}}e^{-i\omega(t-t')}e^{i\mathbf{k}\cdot(\mathbf{x}-\mathbf{x}')}\int\dfrac{d^{4}p_{1}}{(2\pi)^{4}}\\
\times (P^{T}_{\nu\beta}\Delta^{T}(p_{1})+P^{L}_{\nu\beta}\Delta^{L}(p_{1}))(P^{T}_{\tau\delta}\Delta^{T}(k-p_{1})+P^{L}_{\tau\delta}\Delta^{L}(k-p_{1}))\\
((p_{1})_{\mu}(k-p_{1})_{\sigma}(p_{1})_{\alpha}(k-p_{1})_{\gamma})
\end{split}  
\end{equation}
One can think of this as the following diagram, where one operator injects momentum $k$ and the other one collects it. 
\begin{figure}[h!]
    \centering
  \begin{tikzpicture}
\begin{feynman}
    \node [crossed dot] (a);
    \node [crossed dot, right=3cm of a] (b);
    \node [left=0.5cm of a] {$\mathcal{O}$};
    \node [right=0.5cm of b] {$\mathcal{O}$};
    \diagram* {
      (a) -- [boson, half left, looseness=1,  edge label'=\(p_{1}\)] (b),
      (b) -- [boson, half left, looseness=1, edge label'=\(k-p_{1}\)] (a),
    };
  \end{feynman}
  \end{tikzpicture}
    \caption{The correlator can be thought of as an exchange at one point where the operator takes momentum $k$ from the corresponding mode of the wall and the other operator collects it. Note that as mentioned in the text, these are the $D_{12}$ correlators.}
    \label{fig:enter-label}
\end{figure}
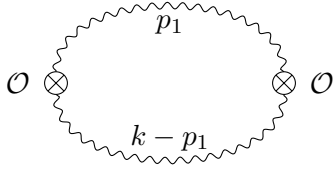
\\
This tells us that the quantity we want is,
\begin{equation}
\begin{split}
 \rho_{0}(\omega,\mathbf{k})=  \dfrac{1}{32}\epsilon^{\mu\nu\sigma\tau}\epsilon^{\alpha\beta\gamma\delta}\int\dfrac{d^{4}p_{1}}{(2\pi)^{4}}
\times (P^{T}_{\nu\beta}\Delta^{T}(p_{1})+P^{L}_{\nu\beta}\Delta^{L}(p_{1}))\\(P^{T}_{\tau\delta}\Delta^{T}(k-p_{1})+P^{L}_{\tau\delta}\Delta^{L}(k-p_{1}))
((p_{1})_{\mu}(k)_{\sigma}(p_{1})_{\alpha}(k)_{\gamma}) 
\end{split}
\end{equation}
Now, we can choose a gauge to work with. The choice of gauge only affects the projectors and not the $\Delta$ propagators. Let us work in the Coulomb Gauge. In this gauge, the projection tensors take the form,
\begin{equation}
    P^{T}_{ij}=\delta_{ij}-\dfrac{p_{i}p_{j}}{p^{2}},~~P^{L}_{\mu\nu}=\dfrac{p_{0}^{2}-p^{2}}{p^{2}}u_{\mu}u_{\nu},~~p=\vert\mathbf{p}\vert
\end{equation}
where $u_{\mu}$ is frame four velocity. Here, we are considering everything in the plasma frame which is at rest hence $u=(1,0,0,0)$. From the structure of these, we can directly see that $\Delta^{L}(p)^{2}$ terms won't contribute. For the transverse part to be non-zero, $\nu,\beta,\tau,\delta\in\{1,2\}$. This means $\mu=\alpha=0$. Hence, what we get is,
\begin{equation}
    \epsilon^{\nu\tau 03}\epsilon^{\beta\delta03}\Big(\delta_{\nu\beta}-\dfrac{p_{\nu}p_{\beta}}{p^{2}}\Big)\Big(\delta_{\tau\delta}-\dfrac{p_{\tau}p_{\delta}}{p^{2}}\Big)p_{0}^{2}\Delta^{T}(p)\Delta^{T}(k-p)
\end{equation}
This is nothing but,
\begin{equation}
    2\Big(1-\dfrac{p_{1}^{2}+p_{2}^{2}}{p^{2}}\Big)p_{0}^{2}\Delta^{T}(p)\Delta^{T}(k-p)=2\cos^{2}\theta p_{0}^{2}\Delta^{T}(p)\Delta^{T}(k-p)
\end{equation}
Similarly, the $\Delta^{T}\Delta^{L}$ also has a non-zero coefficient. It is given by, 
\begin{equation}
    \epsilon^{\mu\nu0 3}\epsilon^{\alpha\beta0 3}\Big(\delta_{\nu\beta}-\dfrac{p_{\nu}p_{\beta}}{p^{2}}\Big)p_{\mu}p_{\alpha}=(p_{1}^{2}+p_{2}^{2})=p^{2}\sin^{2}\theta
\end{equation}
So, the integral looks like:
\begin{equation}
\begin{split}
   \rho_{0}(0,k)=\dfrac{1}{32}k^{2}\int\dfrac{d^{4}p}{(2\pi)^{4}}\Big(2\cos^{2}\theta p_{0}^{2}\Delta^{T}(p)^{2}+\sin^{2}\theta (p_{0}^{2}-p^{2})(\Delta^{T}(p)\Delta^{L}(k-p)\\+\Delta^{T}(k-p)\Delta^{L}(p))\Big)
\end{split}
\end{equation}
We can integrate with respect to the azimuthal angle to get an additional factor of $2\pi$


\bibliography{references}
\bibliographystyle{JHEP}

\end{document}